\documentclass[twocolumn,english,showpacs,showkeys,preprintnumbers,amsmath,amssymb,nofootinbib,superscriptaddress]{revtex4}
\usepackage{ae,aecompl}
\usepackage[T1]{fontenc}
\usepackage[latin1]{inputenc}
\usepackage{color}
\definecolor{shadecolor}{rgb}{1, 0, 0}
\usepackage{framed}
\usepackage{mathrsfs}
\usepackage{bm}
\usepackage{amsmath}
\usepackage{amssymb}
\usepackage{graphicx}
\usepackage{esint}

\makeatletter

\providecommand{\tabularnewline}{\\}

\@ifundefined{textcolor}{}
{%
 \definecolor{BLACK}{gray}{0}
 \definecolor{WHITE}{gray}{1}
 \definecolor{RED}{rgb}{1,0,0}
 \definecolor{GREEN}{rgb}{0,1,0}
 \definecolor{BLUE}{rgb}{0,0,1}
 \definecolor{CYAN}{cmyk}{1,0,0,0}
 \definecolor{MAGENTA}{cmyk}{0,1,0,0}
 \definecolor{YELLOW}{cmyk}{0,0,1,0}
}

\@ifundefined{definecolor}{\@ifundefined{definecolor}
 {\usepackage{color}}{}
}{}
\@ifundefined{definecolor}{\@ifundefined{definecolor}
 {\@ifundefined{definecolor}
 {\usepackage{color}}{}
}{}
}{}
\@ifundefined{definecolor}{\@ifundefined{definecolor}
 {\@ifundefined{definecolor}
 {\@ifundefined{definecolor}
 {\usepackage{color}}{}
}{}
}{}
}{}
\@ifundefined{definecolor}{\@ifundefined{definecolor}
 {\@ifundefined{definecolor}
 {\@ifundefined{definecolor}
 {\@ifundefined{definecolor}
 {\usepackage{color}}{}
}{}
}{}
}{}
}{}
\@ifundefined{definecolor}{\@ifundefined{definecolor}
 {\@ifundefined{definecolor}
 {\@ifundefined{definecolor}
 {\@ifundefined{definecolor}
 {\usepackage{color}}{}
}{}
}{}
}{}
}{}
\@ifundefined{definecolor}{\@ifundefined{definecolor}
 {\@ifundefined{definecolor}
 {\@ifundefined{definecolor}
 {\@ifundefined{definecolor}
 {\@ifundefined{definecolor}
 {\@ifundefined{definecolor}
 {\usepackage{color}}{}
}{}
}{}
}{}
}{}
}{}
}{}
\@ifundefined{definecolor}{\@ifundefined{definecolor}
 {\@ifundefined{definecolor}
 {\@ifundefined{definecolor}
 {\@ifundefined{definecolor}
 {\@ifundefined{definecolor}
 {\@ifundefined{definecolor}
 {\@ifundefined{definecolor}
 {\usepackage{color}}{}
}{}
}{}
}{}
}{}
}{}
}{}
}{}
\@ifundefined{definecolor}{\@ifundefined{definecolor}
 {\@ifundefined{definecolor}
 {\@ifundefined{definecolor}
 {\@ifundefined{definecolor}
 {\@ifundefined{definecolor}
 {\@ifundefined{definecolor}
 {\@ifundefined{definecolor}
 {\usepackage{color}}{}
}{}
}{}
}{}
}{}
}{}
}{}
}{}
\@ifundefined{definecolor}{\@ifundefined{definecolor}
 {\@ifundefined{definecolor}
 {\@ifundefined{definecolor}
 {\@ifundefined{definecolor}
 {\@ifundefined{definecolor}
 {\@ifundefined{definecolor}
 {\@ifundefined{definecolor}
 {\usepackage{color}}{}
}{}
}{}
}{}
}{}
}{}
}{}
}{}
\@ifundefined{definecolor}{\@ifundefined{definecolor}
 {\@ifundefined{definecolor}
 {\@ifundefined{definecolor}
 {\@ifundefined{definecolor}
 {\@ifundefined{definecolor}
 {\@ifundefined{definecolor}
 {\@ifundefined{definecolor}
 {\usepackage{color}}{}
}{}
}{}
}{}
}{}
}{}
}{}
}{}
\@ifundefined{definecolor}{\@ifundefined{definecolor}
 {\@ifundefined{definecolor}
 {\@ifundefined{definecolor}
 {\@ifundefined{definecolor}
 {\@ifundefined{definecolor}
 {\@ifundefined{definecolor}
 {\@ifundefined{definecolor}
 {\usepackage{color}}{}
}{}
}{}
}{}
}{}
}{}
}{}
}{}
\@ifundefined{definecolor}{\@ifundefined{definecolor}
 {\@ifundefined{definecolor}
 {\@ifundefined{definecolor}
 {\@ifundefined{definecolor}
 {\@ifundefined{definecolor}
 {\@ifundefined{definecolor}
 {\@ifundefined{definecolor}
 {\usepackage{color}}{}
}{}
}{}
}{}
}{}
}{}
}{}
}{}
\@ifundefined{definecolor}{\@ifundefined{definecolor}
 {\@ifundefined{definecolor}
 {\@ifundefined{definecolor}
 {\@ifundefined{definecolor}
 {\@ifundefined{definecolor}
 {\@ifundefined{definecolor}
 {\@ifundefined{definecolor}
 {\usepackage{color}}{}
}{}
}{}
}{}
}{}
}{}
}{}
}{}
\@ifundefined{definecolor}{\@ifundefined{definecolor}
 {\@ifundefined{definecolor}
 {\@ifundefined{definecolor}
 {\@ifundefined{definecolor}
 {\@ifundefined{definecolor}
 {\@ifundefined{definecolor}
 {\@ifundefined{definecolor}
 {\usepackage{color}}{}
}{}
}{}
}{}
}{}
}{}
}{}
}{}
\@ifundefined{definecolor}{\@ifundefined{definecolor}
 {\@ifundefined{definecolor}
 {\@ifundefined{definecolor}
 {\@ifundefined{definecolor}
 {\@ifundefined{definecolor}
 {\@ifundefined{definecolor}
 {\@ifundefined{definecolor}
 {\usepackage{color}}{}
}{}
}{}
}{}
}{}
}{}
}{}
}{}
\@ifundefined{definecolor}{\@ifundefined{definecolor}
 {\@ifundefined{definecolor}
 {\@ifundefined{definecolor}
 {\@ifundefined{definecolor}
 {\@ifundefined{definecolor}
 {\@ifundefined{definecolor}
 {\@ifundefined{definecolor}
 {\usepackage{color}}{}
}{}
}{}
}{}
}{}
}{}
}{}
}{}
\@ifundefined{definecolor}{\@ifundefined{definecolor}
 {\@ifundefined{definecolor}
 {\@ifundefined{definecolor}
 {\@ifundefined{definecolor}
 {\@ifundefined{definecolor}
 {\@ifundefined{definecolor}
 {\@ifundefined{definecolor}
 {\usepackage{color}}{}
}{}
}{}
}{}
}{}
}{}
}{}
}{}
\@ifundefined{definecolor}{\@ifundefined{definecolor}
 {\@ifundefined{definecolor}
 {\@ifundefined{definecolor}
 {\@ifundefined{definecolor}
 {\@ifundefined{definecolor}
 {\@ifundefined{definecolor}
 {\@ifundefined{definecolor}
 {\usepackage{color}}{}
}{}
}{}
}{}
}{}
}{}
}{}
}{}

\usepackage{babel}

\usepackage{babel}

\usepackage{babel}

\usepackage{babel}

\usepackage{babel}

\usepackage{babel}

\usepackage{babel}
\usepackage{appendix}

\usepackage{babel}

\usepackage{babel}

\usepackage{babel}

\usepackage{babel}

\usepackage{babel}

\usepackage{babel}

\usepackage{babel}

\makeatother

\usepackage{babel}
\begin{document}

\title{Spacetime dynamics of spinning particles --- exact electromagnetic
analogies}

\author{L.~Filipe~O.~Costa}

\email{lfpocosta@math.ist.utl.pt}

\affiliation{Centro de Física do Porto --- CFP, Departamento de Física e Astronomia,
Faculdade de Ciências da Universidade do Porto --- FCUP, Rua do Campo
Alegre 687, 4169-007 Porto, Portugal}

\affiliation{Center for Mathematical Analysis, Geometry and Dynamical Systems,
Instituto Superior Técnico, Universidade de Lisboa, Portugal}

\author{José~Natário}

\email{jnatar@math.ist.utl.pt}

\affiliation{Center for Mathematical Analysis, Geometry and Dynamical Systems,
Instituto Superior Técnico, Universidade de Lisboa, Portugal}

\author{Miguel~Zilhão}

\email{mzilhao@ffn.ub.es}

\affiliation{Centro de Física do Porto --- CFP, Departamento de Física e Astronomia,
Faculdade de Ciências da Universidade do Porto --- FCUP, Rua do Campo
Alegre 687, 4169-007 Porto, Portugal}

\affiliation{Departamento de Física da Universidade de Aveiro and I3N, Campus
de Santiago, 3810-183 Aveiro, Portugal}

\affiliation{Departament de Física Fonamental \& Institut de Ciències del Cosmos,
Universitat de Barcelona, Martí i Franquès 1, E-08028 Barcelona, Spain}

\date{\today}
\begin{abstract}
We compare the rigorous equations describing the motion of spinning
test particles in gravitational and electromagnetic fields, and show
that if the Mathisson-Pirani spin condition holds then exact gravito-electromagnetic
analogies emerge. These analogies provide a familiar formalism to
treat gravitational problems, as well as a means for comparing the
two interactions. Fundamental differences are manifest in the symmetries
and time projections of the electromagnetic and gravitational tidal
tensors. The physical consequences of the symmetries of the tidal
tensors are explored comparing the following analogous setups: magnetic
dipoles in the field of non-spinning/spinning charges, and gyroscopes
in the Schwarzschild, Kerr, and Kerr-de Sitter spacetimes. The implications
of the time projections of the tidal tensors are illustrated by the
work done on the particle in various frames; in particular, a reciprocity
is found to exist: in a frame comoving with the particle, the electromagnetic
(but not the gravitational) field does work on it, causing a variation
of its proper mass; conversely, for ``static observers\textquotedblright{},
a stationary gravitomagnetic (but not a magnetic) field does work
on the particle, and the associated potential energy is seen to embody
the Hawking-Wald spin-spin interaction energy. The issue of hidden
momentum, and its counterintuitive dynamical implications, is also
analyzed. Finally, a number of issues regarding the electromagnetic
interaction and the physical meaning of Dixon's equations are clarified. 
\end{abstract}



\maketitle
\tableofcontents{}

\section{Introduction\label{sec:Introduction}}

Analogies between the equations of motion for gyroscopes in a gravitational
field and magnetic dipoles in an electromagnetic field have been known
for a long time, and were presented in many different forms throughout
the years. This is the case for both the force and the spin evolution
equations for these test particles in external fields. The former
was first found by Wald~\cite{Wald} in the framework of linearized
theory: he showed that the gravitational force exerted on a spinning
pole-dipole test particle (hereafter a gyroscope), whose center of
mass is \emph{at rest} in a \emph{stationary field}, takes the form
$\vec{F}_{{\rm G}}=K\nabla(\vec{H}\cdot\vec{S})$, where $\vec{H}$
is the so-called ``gravitomagnetic field'', $K$ is some constant
(depending on the precise definition of $\vec{H}$, e.g.~\cite{Ruggiero:2002hz,Gravitation and Inertia,Black Holes,Harris1991}),
and $\vec{S}$ is the particle's angular momentum. This formula is
similar to the formula for the electromagnetic force on a magnetic
dipole, $\vec{F}_{{\rm EM}}=\nabla(\vec{B}\cdot\vec{\mu})$, where
$\vec{B}$ is the magnetic field and $\vec{\mu}$ is the dipole's
magnetic moment. The analogy was later cast in an \emph{exact} form
by one of the authors in~\cite{Natario}, using the \emph{exact}
``gravitoelectromagnetic'' (GEM) inertial fields from the so-called
1+3 ``quasi-Maxwell'' formalism. The force was seen therein to consist
of an electromagnetic-like term of the form above plus a term interpreted
as the ``weight of the energy'' of the gravitomagnetic dipole, and
the limit of validity of the analogy was extended to arbitrarily strong
\emph{stationary} fields \emph{and} when the gyroscope's worldline
\emph{is tangent to any time-like Killing vector field} (which comprehends
e.g.~circular trajectories with arbitrary speed in axisymmetric spacetimes).
In a different framework, it was later shown that there is actually
an exact, \emph{covariant}, and \emph{fully general} analogy relating
the two forces; such analogy is made explicit \emph{not} in the framework
of the GEM inertial fields, but by using instead the tidal tensors
of both theories, introduced in~\cite{CHPRD}.

The analogy between the so-called ``precession'' of a gyroscope
in a gravitational field and the precession of a magnetic dipole under
the action of a magnetic field was noticed long ago, in the framework
of linearized theory, by a number of authors, e.g. \cite{Black Holes,Near Zero,Gravitation and Inertia,O'Connel Spin Rotation 1974,lense,Harris1991,Tucker Clark},
who pointed out that the spin vector of a gyroscope at rest in a stationary
field evolves as $d\vec{S}/dt=K\vec{S}\times\vec{H}$. This formula
is similar to the formula for the precession of a magnetic dipole
in a magnetic field, $d\vec{S}/dt=\vec{\mu}\times\vec{B}$. The analogy
was later cast in \emph{exact} forms in e.g.~\cite{Black Holes,Natario,The many faces,GEM User Manual};
these are not covariant, holding only in specific frames, but (in
the more general formulations in~\cite{The many faces,GEM User Manual,PaperAnalogies}
and herein) the test particle can be moving with arbitrary velocity
in an \emph{arbitrary field.}

These analogies provide a familiar formalism to treat otherwise complicated
gravitational effects, as well as a means to compare the two interactions.
In this work we explore them, exemplifying their usefulness in some
applications, and the insight they provide into fundamental, yet not
well known aspects of both interactions.

We will also make use of a third exact gravito-electromagnetic analogy
(see e.g.~\cite{matte,bel,Maartens:1997fg}), this one a \emph{purely
formal} one (see~\cite{PaperAnalogies}), relating the quadratic
scalar invariants of the Maxwell and Weyl tensors~\cite{matte,bel,PaperInvariantes},
which proves useful in some applications.

\subsection{The equations of motion}

In this work we start, in Sec.~\ref{sec:Eqns-Motion}, by writing
the general relativistic equations describing the motion of spinning
test particles with gravitational and electromagnetic pole-dipole
moments, subject to gravitational and electromagnetic external fields,
in terms of quantities with a clear physical meaning. This turns out
not to be a straightforward task, as the covariant equations for this
problem are still not generally well understood, with different methods
and derivations leading to different versions of the equations, the
relationship between them not being clear. Perhaps more surprising
is the fact that it is the electromagnetic sector that has been posing
more difficulties, with a number of misconceptions arising in the
physical interpretation of the quantities involved. These issues are
clarified in Appendix \ref{sec:DixonEqs}, where the relation between
the different versions of the equations and their physical interpretation
are discussed in detail.

In order to form a determined system, the equations of motion need
to be supplemented by a spin condition; the latter is even today still
regarded as an open question, with a long history of debates concerning
which one is the ``best'' condition (see \cite{Semerak I} for a
review and list of references). In Sec.~\ref{sub:Center-of-mass}
we briefly discuss its meaning and the problem of the relativistic
definition of center of mass. This is of relevance here because the
two physical analogies mentioned above (for the force and for the
spin precession) rely on a specific choice --- the Mathisson-Pirani
spin condition.

Also related with the spin supplementary condition is an issue central
to the understanding of the dynamics of a spinning particle: the decoupling
of the 4-velocity $U^{\alpha}$ from the 4-momentum $P^{\alpha}$,
discussed in Sec.~\ref{sub:Momentum-of-the-Particle}. In general,
$U^{\alpha}$ is not parallel to $P^{\alpha}$; the particle is said
to possess ``hidden momentum'', for which another exact analogy
is seen to emerge. The hidden momentum is known to lead to counter-intuitive
behaviors of the spinning particles; examples are the bobbings studied
in~\cite{Wald et al 2010}, and the Mathisson helical motions themselves,
where a particle accelerates without the action of any force \cite{Helical}.
Herein (Sec.~\ref{sub:Radial-Schwa}) we present another, perhaps
even more surprising consequence: a magnetic dipole with radial initial
velocity in the field of a point charge accelerates in approximately
the \emph{opposite} direction to the force.

\subsection{The main realizations}

Most of our applications, Secs.~\ref{sub:Symmetries}-\ref{sub:Weak-field-regime:}
of this paper, will deal with the tidal tensor formalism introduced
in~\cite{CHPRD}, and the exact analogy it unveils: both the electromagnetic
force on a magnetic dipole and the gravitational force on a gyroscope
are given by a contraction of a rank 2 magnetic type tidal tensor
($B_{\alpha\beta}$, $\mathbb{H}_{\alpha\beta}$), with the dipole/spin
4-vector. Here $B_{\alpha\beta}$ gives the tidal effects of the magnetic
field and $\mathbb{H}_{\alpha\beta}$ is the magnetic part of the
Riemann tensor, both measured in the particle's rest frame. This makes
this formalism specially suited to comparing the two forces --- it
amounts to simply comparing the two tidal tensors. Such comparison
is done through Einstein's and Maxwell's equations, as they also can
be written in terms of tidal tensors. Apart from the non-linearity
of $\mathbb{H}_{\alpha\beta}$, the tensorial structure differs when
the fields vary along the test particle's worldline, since this endows
$B_{\alpha\beta}$ with an antisymmetric part, and a non-vanishing
time projection along that worldline, whereas its gravitational counterpart
is spatial relative to that worldline and, in vacuum, symmetric. Both
these aspects are related with the laws of electromagnetic induction
(and the absence of a counterpart in the gravitational tidal effects);
we discuss them in two separate sections, as described below.

In Sec.~\ref{sub:Symmetries} we explore the physical consequences
of the different symmetries of the gravitational and electromagnetic
tidal tensors. They are seen to imply e.g.~that particles moving
in a non-homogeneous electromagnetic field always measure a non-vanishing
$B_{\alpha\beta}$ (thus feel a force), which is not necessarily the
case in gravity. The following analogous setups are compared: magnetic
dipoles in the field of non-spinning/spinning charges, and gyroscopes
in the Schwarzschild, Kerr, and Kerr-dS spacetimes. It is seen that
in the cases where $B_{\alpha\beta}$ reduces to $B_{[\alpha\beta]}$,
we have $\mathbb{H}_{\alpha\beta}=0$ (thus no force) in the gravitational
analogue. Geodesic motions for spinning particles are even found to
exist in the Schwarzschild (radial geodesics) and in the Kerr-dS (circular
equatorial geodesics) spacetimes.

In Sec.~\ref{sub:Time-Components} we explore the physical content
of the time projections of the forces in different frames, which are
related with the rate of work done on the test particle by the external
fields. In order to obtain the relationship, we start by deriving
the general equation yielding the variation of energy of a particle
with multipole structure with respect to an arbitrary congruence of
observers. We then show that the electromagnetic force has a non-vanishing
time projection along $U^{\alpha}$, which is the power transferred
to the dipole by Faraday's induction, reflected in a variation of
its proper mass $m$. The projection of the gravitational force along
$U^{\alpha}$, by contrast, vanishes (as $\mathbb{H}_{\alpha\beta}$
is spatial relative to $U^{\alpha}$), leading to the conservation
of the gyroscope's mass. Also of particular interest in this context
are the time projections as measured by ``static observers'', analyzed
in Sec.~\ref{sub:Time-components-Static}. For these observers, the
time projection of the electromagnetic force vanishes, meaning that
the total work done on the magnetic dipole is zero. \textcolor{black}{This
reflects the well known fact that the work done by the stationary
magnetic field 
is zero; in this framework, it is seen to arise from an exchange of
energy between three forms, translational kinetic energy, proper mass
$m$, and ``hidden energy'', occurring in a way such that their
variations cancel out, keeping the total energy constant.} \textcolor{black}{In
the gravitational case, since $m$ is constant, such cancellation}
does not occur and (by contrast with its electromagnetic counterpart)
a stationary field \emph{does work} on mass currents, so that there
exists an associated potential spin-curvature potential energy, of
which the Wald-Hawking spin-spin interaction energy \cite{Wald,Hawking}
is seen to be a special case.

In Sec.~\ref{sub:Weak-field-regime:} we study the weak field and
slow motion regime, and show that the above mentioned differences
between the two interactions appear at leading order (thus are \emph{not}
negligible) therein, which is commonly overlooked in the literature
concerning this regime.

\subsection{Beyond the pole-dipole}

In Sec.~\ref{sec:Beyond-pole-dipole} we go beyond the pole-dipole
approximation, including the moments of quadrupole order, to clarify
the mechanism by which the proper mass of a spinning particle in an
electromagnetic field varies, and solve an apparent contradiction
of the former approximation: on the one hand, as stated above, the
mass $m$ of a particle with magnetic moment varies due (from the
point of view of the particle's frame) to the work done on it by the
induced electric field (which, by having a curl, should torque the
body). On the other hand, the associated torque is not manifest in
the dipole order equations. In Sec.~\ref{sub:Electromagnetic-torque}
we show that such torque is indeed exerted on the particle (governed
by the time projection of the magnetic tidal tensor, $B_{\ \beta}^{\alpha}U^{\beta}$),
but it involves quadrupole order moments of the charge, which is why
it does not show up to dipole order. The subtlety here is that \emph{the
work it does}, and the associated variation of mass/kinetic energy
of rotation, \emph{is of dipole order} (yielding indeed the time projection
of the dipole force along its worldline, as obtained in Sec.~\ref{sub:Time-components-in-CM-frame }).
Then in Sec.~\ref{sub:Gravitational-torque} we study the analogous
gravitational problem, showing that, as expected (as $\mathbb{H}_{\alpha\beta}$
is spatial with respect to $U^{\alpha}$), no analogous torque exists.

\subsection{Notation and conventions\label{sub:Notation-and-conventions}}
\begin{enumerate}
\item \emph{Signature and signs}. We use the signature $-+++$; $\epsilon_{\alpha\beta\gamma\delta}\equiv\sqrt{-g}[\alpha\beta\gamma\delta]$
denotes the Levi-Civita tensor, and we follow the orientation $[1230]=1$
(i.e., in flat spacetime, $\epsilon_{1230}=1$). $\epsilon_{ijk}\equiv\epsilon_{ijk0}$
is the 3-D alternating tensor. Our convention for the Riemann tensor
is $R_{\ \beta\mu\nu}^{\alpha}=\Gamma_{\beta\nu,\mu}^{\alpha}-\Gamma_{\beta\mu,\nu}^{\alpha}+...$. 
\item \textcolor{black}{We use bold fonts to denote tensors $\mathbf{T}$
(including 4-vectors $\mathbf{P}$), and arrows for the spatial components
$\vec{P}$ of a 4-vector $\mathbf{P}$.} \textcolor{black}{Greek letters
$\alpha$, $\beta$, $\gamma$, ... denote 4-D spacetime indices,
running 0-3; Roman letters $i,j,k,...$ denote spatial indices, running
1-3. Following the usual practice, we sometimes use component notation
$T^{\alpha\beta}$ to refer to a tensor $\mathbf{T}$.} 
\item By $u^{\alpha}$ we denote a generic unit time-like vector, which
can be interpreted as the instantaneous 4-velocity of some observer.
$U^{\alpha}\equiv dz^{\alpha}/d\tau$ is the tangent vector to the
test body's representative worldline $z^{\alpha}(\tau)$, taken to
be a suitably defined center of mass (CM). $U^{\alpha}$ is thus the
CM 4-velocity. 
\item \emph{Time and space projectors}. $(\top^{u})_{\ \beta}^{\alpha}\equiv-u^{\alpha}u_{\beta}$
and $(h^{u})_{\ \beta}^{\alpha}\equiv\delta_{\ \beta}^{\alpha}+u^{\alpha}u_{\beta}$
denote, respectively, the projectors parallel and orthogonal to a
unit time-like vector $u^{\alpha}$; they may be interpreted as the
time and space projectors in the local rest frame of an observer with
4-velocity $u^{\alpha}$. 
\item \emph{Tensors resulting from a measurement process}. $(A^{u})_{\ \ }^{\alpha_{1}..\alpha_{n}}$
denotes the tensor $\mathbf{A}$ as measured by an observer $\mathcal{O}(u)$
of 4-velocity $u^{\alpha}$. For example, $(E^{u})^{\alpha}\equiv F_{\ \beta}^{\alpha}u^{\beta}$,
$(E^{u})_{\alpha\beta}\equiv F_{\alpha\gamma;\beta}u^{\gamma}$ and
$(\mathbb{E}^{u})_{\alpha\beta}\equiv R_{\alpha\mu\beta\nu}u^{\mu}u^{\nu}$
denote, respectively, the electric field, electric tidal tensor, and
gravitoelectric tidal tensor as measured by $\mathcal{O}(u)$. For
the space components of a vector in a given frame we use the notation
$\vec{A}(u)$; for example, $\vec{E}(u)$ denotes the space components
of $(E^{u})^{\alpha}$. When $u^{\alpha}=U^{\alpha}$ (i.e., the particle's
CM 4-velocity) we drop the superscript (e.g.~$(E^{U})^{\alpha}\equiv E^{\alpha}$),
or the argument of the vector: $\vec{E}(U)\equiv\vec{E}$. 
\item \emph{Electromagnetic field.} The Maxwell tensor $F^{\alpha\beta}$
and its dual $\star F^{\alpha\beta}$ decompose in terms of the electric
$(E^{u})^{\alpha}\equiv F_{\ \beta}^{\alpha}u^{\beta}$ and magnetic
$(B^{u})^{\alpha}\equiv\star F_{\ \beta}^{\alpha}u^{\beta}$ fields
measured by an observer of 4-velocity $u^{\alpha}$ as 
\begin{eqnarray}
F_{\alpha\beta} & = & 2u_{[\alpha}(E^{u})_{\beta]}+\epsilon_{\alpha\beta\gamma\delta}u^{\delta}(B^{u})^{\gamma}\ ;\label{eq:Fdecomp}\\
\star F_{\alpha\beta} & = & 2u_{[\alpha}(B^{u})_{\beta]}-\epsilon_{\alpha\beta\gamma\delta}u^{\delta}(E^{u})^{\gamma}\ .\label{eq:FstarDecomp}
\end{eqnarray}

\item \emph{\label{enu:Static-observers}Static observers.} In stationary,
asymptotically flat spacetimes, we dub ``static observers'' the
\emph{rigid} congruence of observers whose worldlines are tangent
to the temporal Killing vector field $\mathbf{\xi}=\partial/\partial t$;
they may be interpreted as the set of points rigidly fixed to the
``distant stars'' (the asymptotic inertial rest frame of the source).
In the Kerr spacetime, these correspond to the observers of zero 3-velocity
in Boyer-Lindquist coordinates. This agrees with the convention in
e.g.~\cite{Gravitation,BiniStaticObs}. (The denomination ``static
observers'' has, however, a different meaning in some literature,
e.g.~\cite{WyllemanBeke}, where it designates rigid, \emph{vorticity-free}
congruences tangent to a time-like Killing vector field, existing
only in \emph{static spacetimes}). In the case of the electromagnetic
systems in flat spacetime, by static observers we mean the globally
inertial rest frame of the sources. 
\item \emph{GEM}. This is the acronym for ``gravitoelectromagnetism''.
By ``inertial GEM fields'', we mean the fields of inertial forces
that arise from the 1+3 splitting of spacetime: the gravitoelectric
field $\vec{G}$, which plays in this framework a role analogous to
the electric field of electromagnetism, and the gravitomagnetic field
$\vec{H}$, analogous to the magnetic field. We discuss these fields
in detail in \cite{PaperAnalogies}. 
\end{enumerate}

\section{Equations of motion for spinning pole-dipole particles\label{sec:Eqns-Motion}}

In most of this work we will be dealing with the dynamics of the so-called
pole-dipole spinning test particles. We consider systems composed
of a test body plus background gravitational and electromagnetic fields.
Let $(T_{{\rm tot}})_{\ }^{\alpha\beta}=\Theta^{\alpha\beta}+(T_{{\rm matter}})^{\alpha\beta}$
denote the total energy-momentum tensor, which splits into the electromagnetic
stress-energy tensor $\Theta^{\alpha\beta}$ and the energy-momentum
of the matter $(T_{{\rm matter}})^{\alpha\beta}$. Moreover, let $T^{\alpha\beta}$
and $j^{\alpha}$ denote, respectively, the energy-momentum tensor
and the current density 4-vector \emph{of the test body}. We also
consider that the only matter and currents present are the ones arising
from the test body: $(T_{{\rm matter}})^{\alpha\beta}=$ $T^{\alpha\beta}$,
$j_{{\rm tot}}^{\alpha}=j^{\alpha}$. In this case (see \cite{EPAPS}
for details) the conservation of total energy-momentum tensor yields
(cf. e.g. \cite{Dixon1970I,Wald et al 2010,Wald et al}): 
\begin{equation}
(T_{{\rm tot}})_{\ \ ;\beta}^{\alpha\beta}=0\ \Rightarrow\ T_{\ \ ;\beta}^{\alpha\beta}=-\Theta_{\ \ ;\beta}^{\alpha\beta}\ \Leftrightarrow\ T_{\ \ ;\beta}^{\alpha\beta}=F^{\alpha\beta}j_{\beta}\ ,\label{eq:ConservT}
\end{equation}
where $F^{\alpha\beta}$ is the Maxwell tensor of the \emph{external}
(background) electromagnetic field.

In a multipole expansion the body is represented by the moments of
$j^{\alpha}$ (its ``electromagnetic skeleton\textquotedbl{}) and
a set of moments of $T^{\alpha\beta}$, called ``inertial'' or ``gravitational''
moments (forming the so called~\cite{MathissonNeueMechanik} ``gravitational
skeleton''). Truncating the expansion at dipole order, the equations
of motion for such a particle involve only two moments of $T^{\alpha\beta}$,
\begin{eqnarray}
P^{\hat{\alpha}} & \equiv & \int_{\Sigma(\tau,U)}T^{\hat{\alpha}\hat{\beta}}d\Sigma_{\hat{\beta}}\ ,\label{eq:Pgeneral}\\
S^{\hat{\alpha}\hat{\beta}} & \equiv & 2\int_{\Sigma(\tau,U)}x^{[\hat{\alpha}}T^{\hat{\beta}]\hat{\gamma}}d\Sigma_{\hat{\gamma}}\ ,\label{eq:Sab}
\end{eqnarray}
and the electromagnetic moments~\cite{Dixon1967}: 
\begin{eqnarray}
q & \equiv & \int_{\Sigma}j^{\alpha}d\Sigma_{\alpha}\ ,\label{eq:}\\
d^{\hat{\alpha}} & \equiv & \int_{\Sigma(\tau,U)}x^{\hat{\alpha}}j^{\hat{\beta}}d\Sigma_{\hat{\beta}}\ ,\label{eq:d_a}\\
\mu^{\hat{\alpha}} & \equiv & \frac{1}{2}\epsilon_{\ \hat{\beta}\hat{\gamma}\hat{\delta}}^{\hat{\alpha}}U^{\hat{\delta}}\int_{\Sigma(\tau,U)}x^{\hat{\beta}}j^{\hat{\gamma}}d\Sigma\ .\label{eq:mu_a}
\end{eqnarray}
These are taken with respect to a reference worldline $z^{\alpha}(\tau)$,
of proper time $\tau$ and (unit) tangent vector $U^{\alpha}\equiv dz^{\alpha}/d\tau$,
and to a hypersurface of integration $\Sigma(\tau,u)$, which is the
spacelike hypersurface generated by all geodesics orthogonal to some
time-like vector $u^{\alpha}$ at the point $z^{\alpha}(\tau)$; following
\cite{Dixon1967} we take $u^{\alpha}=U^{\alpha}$. Also, 
\begin{equation}
d\Sigma_{\gamma}\equiv-n_{\gamma}d\Sigma\quad{\rm (at}\ z^{\alpha}:\ n^{\alpha}=U^{\alpha}{\rm )},\label{eq:dSigma}
\end{equation}
where $n^{\alpha}$ is the (future-pointing) unit vector normal do
$\Sigma(\tau,U)$, and $d\Sigma$ is the 3-volume element of this
hypersurface. The integrations are performed in a system of Riemann
\emph{normal coordinates} $\{x^{\hat{\alpha}}\}$ (e.g.~\cite{Gravitation,Madore:1969})
centered at the point $z^{\alpha}$ of the reference worldline (i.e.,
$z^{\hat{\alpha}}=0$). The resulting expressions, however, are \emph{tensors}
(see below), which can be expressed in any frame. $P^{\alpha}(\tau)$
is the 4-momentum of the test particle, $q$ its total charge (an
invariant, independent of $\Sigma$), and $S^{\alpha\beta}(\tau)$,
$d^{\alpha}(\tau)$ and $\mu^{\alpha}(\tau)$ are, respectively, the
angular momentum, and the \emph{intrinsic} electric and magnetic dipole
moments about the point $z^{\alpha}(\tau)$ of the reference worldline.
It is useful to introduce also the magnetic dipole 2-form $\mu_{\alpha\beta}$
by 
\begin{equation}
\mu_{\alpha\beta}\equiv\epsilon_{\alpha\beta\gamma\delta}\mu^{\gamma}U^{\delta};\qquad\mu^{\alpha}=\frac{1}{2}\epsilon_{\ \beta\gamma\delta}^{\alpha}U^{\beta}\mu^{\gamma\delta}\ .\label{eq:mu_ab}
\end{equation}
In some applications we will assume $\mu^{\alpha\beta}$ to be proportional
to the spin tensor: $\mu^{\alpha\beta}=\sigma S^{\alpha\beta}$, where
$\sigma$ is the gyromagnetic ratio. The moments $d^{\alpha}$ and
$\mu^{\alpha}$ are dubbed ``intrinsic'' because they are evaluated
in a frame comoving with the particle's representative point $z^{\alpha}(\tau)$
(where $U^{i}=0$). If this frame is inertial, they take the forms
$d^{\alpha}=(0,\vec{d})$ and $\mu^{\alpha}=(0,\vec{\mu})$, where
$\vec{d}$ and $\vec{\mu}$ are given by the usual textbook definitions
(e.g.~\cite{Jackson}): $\vec{d}=\int\rho_{c}\vec{x}d^{3}x$, $\vec{\mu}=\int\vec{x}\times\vec{j}d^{3}x/2$.

Expressions (\ref{eq:Pgeneral}), (\ref{eq:Sab}), (\ref{eq:d_a})
and (\ref{eq:mu_a}) are integrals of tensors over $\Sigma$ (i.e.,
they add tensor components at different points in a curved spacetime),
which requires a justification. By using Riemann normal coordinates,
one is implicitly using the exponential map to pull back the integrands
from the spacetime manifold to the tangent space at $z^{\alpha}$,
and integrating therein, which is a well defined tensor operation,
see \cite{Madore:1969,CostaNatario2014}. (Note also that, by being
associated to the exponential map, such coordinates are naturally
adapted to integrations over geodesic hypersurfaces $\Sigma$). Other
schemes to perform such integrations were proposed in the literature,
based on bitensors in \cite{Dixon1964,Dixon1970I,Dixon1970II,Dixon1974III,Wald et al 2010},
or less sophisticated ones in e.g.~\cite{Papapetrou I}. In the pole-dipole
approximation (where $T^{\alpha\beta}$ and $j^{\alpha}$ are non-vanishing
only in a very small region around $z^{\alpha}(\tau)$, so that only
terms linear in $x^{\hat{\alpha}}$ are kept) they are \emph{all}
\emph{equivalent} (see Appendix \ref{sub:Relation-with-the-Eqs-Herein}
and \cite{CostaNatario2014,Dixon1974III}).

The motion of the test particle is described by the reference worldline
$z^{\alpha}(\tau)$; its choice will be discussed below. The equations
of motion that follow from \eqref{eq:ConservT} are~\cite{Dixon1964,Dixon1965,Dixon1967}
(see Appendix~\ref{sec:DixonEqs} for a discussion): 
\begin{eqnarray}
\frac{DP^{\alpha}}{d\tau} & = & qF_{\ \beta}^{\alpha}U^{\beta}+\frac{1}{2}F^{\mu\nu;\alpha}\mu_{\mu\nu}-\frac{1}{2}R_{\ \beta\mu\nu}^{\alpha}S^{\mu\nu}U^{\beta}\nonumber \\
 &  & +F_{\ \gamma;\beta}^{\alpha}U^{\gamma}d^{\beta}+F_{\ \beta}^{\alpha}\frac{Dd^{\beta}}{d\tau}\ ,\label{eq:ForceDS0}\\
\frac{DS^{\alpha\beta}}{d\tau} & = & 2P^{[\alpha}U^{\beta]}+2\mu^{\theta[\beta}F_{\ \ \theta}^{\alpha]}+2d^{[\alpha}F_{\ \ \gamma}^{\beta]}U^{\gamma}\ ,\label{eq:SpinDS0}
\end{eqnarray}
where $F^{\alpha\beta}$ is the background Maxwell tensor.

The first term in (\ref{eq:ForceDS0}) is the Lorentz force; the second
term, $\frac{1}{2}F^{\mu\nu;\alpha}\mu_{\mu\nu}\equiv F_{{\rm EM}}^{\alpha}$,
is the force due to the tidal coupling of the electromagnetic field
to the magnetic dipole moment; and the third, $-\frac{1}{2}R_{\ \beta\mu\nu}^{\alpha}S^{\mu\nu}U^{\beta}\equiv F_{{\rm G}}^{\alpha}$,
is the Mathisson-Papapetrou spin-curvature force. The last two terms
are the force exerted on the electric dipole, consisting of a tidal
term $F_{\ \gamma;\beta}^{\alpha}U^{\gamma}d^{\beta}$ and of a \emph{non
tidal} term $F_{\ \beta}^{\alpha}Dd^{\beta}/d\tau$. Note that the
terms involving $\mu^{\alpha}$ and $d^{\alpha}$ are substantially
different; this can be traced back to the intrinsic difference between
the two types of dipole --- $\mu^{\alpha}$ being the dipole moment
of the spatial current density $(h^{U})_{\ \beta}^{\alpha}j^{\beta}$,
and $d^{\alpha}$ the dipole moment of the charge density $\rho_{c}\equiv-j^{\alpha}U_{\alpha}$,
cf.~Eqs.~(\ref{eq:d_a})-(\ref{eq:mu_a}). The former can be modeled
by a current loop, the latter by a pair of oppositely charged monopoles,
and these two types of objects behave differently as test particles;
in this work we shall discuss some dynamical implications.

Up until now, the reference worldline $z^{\alpha}(\tau)$, relative
to which the moments in Eqs.~(\ref{eq:ForceDS0})-(\ref{eq:SpinDS0})
are taken, is still undefined. Had we made an exact expansion keeping
all the infinite multipole moments as in~\cite{Dixon1970II,Dixon1974III},
such worldline would be \emph{arbitrary.} Herein, however, it must
be assumed that it passes through the body (or close enough), so that
the pole-dipole approximation is valid; it will be chosen as being
prescribed by a suitably defined center of mass of the test particle.
As discussed in the next section, that is done \textcolor{black}{through
a supplementary condition} $S^{\alpha\beta}u_{\beta}=0$, \textcolor{black}{for
some time-like unit vector field $u^{\alpha}$}. \textcolor{black}{If
$F^{\alpha\beta}=0$ there are 13 unknowns} in Eqs.~(\ref{eq:ForceDS0})-(\ref{eq:SpinDS0})
\textcolor{black}{($P^{\alpha}$, 3 independent components of $U^{\alpha}$,
and 6 independent components of $S^{\alpha\beta}$) for only 10 equations.
The condition} $S^{\alpha\beta}u_{\beta}=0$, \textcolor{black}{for
a} \textcolor{black}{\emph{definite~}} \textcolor{black}{$u^{\alpha}$,
closes the system as it kills off 3 components of $S^{\alpha\beta}$.
In the general case where $F^{\alpha\beta}\ne0$ one also needs to
give the laws of evolution for $\mu^{\alpha\beta}$} and $d^{\alpha}$
\textcolor{black}{in order for the system to be determined, cf.~\cite{Wald et al}.}

\subsection{Center of mass (CM) and spin supplementary condition\label{sub:Center-of-mass}}

In relativistic physics, the center of mass of a spinning body is
observer dependent. This is illustrated in Fig.~1 of~\cite{Helical}.
Thus one needs to specify the frame where it is to be evaluated. That
amounts to supplementing Eqs.~(\ref{eq:ForceDS0})-(\ref{eq:SpinDS0})
(which, as discussed above, would otherwise be undetermined) by the
spin supplementary condition $S^{\alpha\beta}u_{\beta}=0$, as we
will show next. The vector $(d_{G}^{u})^{\alpha}\equiv-S^{\alpha\beta}u_{\beta}$
yields the ``mass dipole moment'' (i.e.~the mass times the displacement
of the reference worldline relative to the center of mass) as measured
by the observer $\mathcal{O}$ of 4-velocity $u^{\alpha}$. In order
to see this consider, at the point $z^{\alpha}$ of the reference
worldline, a system of Riemann normal coordinates $\{x^{\hat{\alpha}}\}$
momentarily comoving with $\mathcal{O}$ (i.e., $\partial_{\hat{0}}=\mathbf{u}$
\emph{at} $z^{\alpha}$). In this frame, $u^{\hat{i}}=0$ and $S^{\hat{i}\hat{\beta}}u_{\hat{\beta}}=S^{\hat{i}\hat{0}}u_{\hat{0}}=-S^{\hat{i}\hat{0}}$.
From Eq.~(\ref{eq:Sab}) we have: 
\begin{equation}
S^{\hat{i}\hat{0}}=\int_{\Sigma(\tau,u)}x^{\hat{i}}T^{\hat{0}\hat{\gamma}}d\Sigma_{\hat{\gamma}}\equiv m(u)x_{{\rm CM}}^{\hat{i}}(u),\label{eq:Massdipole}
\end{equation}
where $m(u)=-P^{\alpha}u_{\alpha}$ denotes the mass as measured by
$\mathcal{O}$, and we used the fact that $\Sigma(\tau,u)$ coincides
with the spatial hypersurface $x^{\hat{0}}=0$. We see that $S^{\hat{i}\hat{0}}$
is by definition the mass dipole in the frame $\{x^{\hat{\alpha}}\}$:
$S^{\hat{i}\hat{0}}=m(u)x_{{\rm CM}}^{\hat{i}}(u)\equiv(d_{G}^{u})^{\hat{i}}$,
and $x_{{\rm CM}}^{\hat{i}}(u)=S^{\hat{i}\hat{0}}/m(u)$ is the position
of the center of mass. Thus the condition 
\begin{equation}
S^{\alpha\beta}u_{\beta}=0\ ,\label{eq:Spin_condition}
\end{equation}
implying in this frame $S^{\hat{i}\hat{0}}=0\Rightarrow x_{{\rm CM}}^{\hat{i}}(u)=0$,
is precisely the condition that the reference worldline is the center
of mass (or ``centroid'') as measured in this frame (or, equivalently,
that the mass dipole vanishes for such an observer). For details on
how the center of mass position changes in a change of observer, we
refer the reader to~\cite{Helical,CostaNatario2014}. \textcolor{black}{The
set of all the possible positions of the center of mass, as measured
by every possible observer, forms a worldtube (the ``minimal worldtube''
\cite{Semerak II}), which is typically very narrow, and always contained
within the convex hull of the body's worldtube (see \cite{CostaNatario2014,MollerAIP,MollerBook}).}

Usually one prefers equations of motion that do not depend on a CM
measured by some ``external'' observer, but instead the field $u^{\alpha}$
to be defined in terms of the time-like vectors ($P^{\alpha}$ or
$U^{\alpha}$) already present in Eqs.~\eqref{eq:ForceDS0}-\eqref{eq:SpinDS0}.
This is the case of the two most common \cite{Semerak I} conditions
in the literature: the Frenkel-Mathisson-Pirani \cite{Frenkel,MathissonNeueMechanik,Mathisson Zitterbewegung,Pirani 1956}
condition $S^{\alpha\beta}U_{\beta}=0$ (hereafter the Mathisson-Pirani
condition, as it is best known) and the Tulczyjew-Dixon \cite{TulczyjewII,Dixon1964}
condition $S^{\alpha\beta}P_{\beta}=0$. The former seems the most
natural choice, as it amounts to computing the center of mass in its
proper frame, i.e., \emph{in the frame where it has zero 3-velocity}.
It also arises in a natural fashion in some derivations \cite{Taub,EulerTop}
(see also~\cite{Plyatsko Non-Oscillatory}), and has been argued
\cite{BaylinMassless,BaylinMassless II,MashhoonMassless} to be the
only one that can be applied in the case of massless particles. It
turns out, however, that it does not determine the worldline uniquely.
For instance, in the case of a free particle in flat spacetime, it
is known to lead, in addition to the expected straightline motion,
to an infinite set of helical motions, notably found by Mathisson~\cite{Mathisson Zitterbewegung},
and which have been poorly understood and subject of some misconceptions
in the literature. These were clarified in~\cite{Helical}, where
it was shown that the different worldlines compatible with this condition
are just equivalent descriptions of the same physical motion.

The Tulczyjew-Dixon condition $S^{\alpha\beta}P_{\beta}=0$ amounts
to computing the CM \emph{in the frame where it has zero 3-momentum}.
This condition determines uniquely the CM worldline~\cite{Beigblock,Dixon1970I,DixonGRG1973};
there is no ambiguity in this case, since $P^{\alpha}$ is given in
advance by (\ref{eq:Pgeneral}), and for this reason it is preferred
by a number of authors. \textcolor{black}{For a free particle in flat
spacetime, the worldline specified by} $S^{\alpha\beta}P_{\beta}=0$
\textcolor{black}{corresponds to Mathisson's non-helical solution;
but in the presence of gravitational/electromagnetic field, $P^{\alpha}$
cannot in general be parallel to $U^{\alpha}$ under these spin conditions}
\textcolor{black}{(cf.~Eqs.~\eqref{eq:Momentum}, \eqref{eq:MomentumMP},
below), so the solutions do not coincide.}

The fundamental point to be emphasized here is that these two conditions,
as well as other reasonable conditions in the literature (such as
the Corinaldesi-Papapetrou condition~\cite{Corinaldesi Papapetrou},
the ``parallel'' condition in \cite{Semerak II}, or the Newton-Wigner
condition \cite{Newton-Wigner,Pryce}, used in Hamiltonian and effective
field theory approaches \cite{Buonanno2009,Hanson-Regge,Steinhoffr,HergtSteinhoffSchaefer2010,Kunst2014,Porto,BakerOConnel19741975}),
are, as shown explicitly in \cite{CostaNatario2014}, \emph{equivalent}
descriptions of the motion of the test particle, the choice between
them being a matter of convenience.

In most of this work we will adopt the Mathisson-Pirani condition,
since it is the one that leads to the exact gravito-electromagnetic
analogies we use. If the Tulczyjew-Dixon condition is chosen instead,
one still recovers the same analogies to a good approximation. The
spin conditions, their effective differences and their suitability
for the applications in this work, as well as their impact on the
gravito-electromagnetic analogies, are discussed in detail in Appendix
\ref{sec:Spin Conditions}. Therein we show that exact analogies turn
out to exist for an arbitrary spin condition, only the corresponding
equations are slightly more complicated.

With the Mathisson-Pirani condition, we have $S^{\mu\nu}=\epsilon^{\mu\nu\tau\lambda}S_{\tau}U_{\lambda}$,
where $S^{\alpha}$ is the spin 4-vector, which has components $(0,\vec{S})$
in an orthonormal frame comoving with the CM. Substituting into Eqs.~(\ref{eq:ForceDS0})-(\ref{eq:SpinDS0})
(and also performing the contractions with $U^{\alpha}$) we obtain
\begin{eqnarray}
\frac{DP^{\alpha}}{d\tau} & = & qE^{\alpha}+E^{\alpha\beta}d_{\beta}+B^{\beta\alpha}\mu_{\beta}-\mathbb{H}^{\beta\alpha}S_{\beta}\ \nonumber \\
 &  & +F_{\ \beta}^{\alpha}\frac{Dd^{\beta}}{d\tau}\ ;\label{eq:ForcePirani}\\
\frac{D_{F}S^{\mu}}{d\tau} & = & \epsilon_{\ \alpha\beta\nu}^{\mu}U^{\nu}(d^{\alpha}E^{\beta}+\mu^{\alpha}B^{\beta})\ ,\label{eq:SpinPirani}
\end{eqnarray}
where $E^{\alpha}\equiv F^{\alpha\beta}U_{\beta}$ and $B^{\alpha}\equiv\star F^{\alpha\beta}U_{\beta}$
are the electric and magnetic fields \emph{as measured by the test
particle}, and $E_{\alpha\beta}\equiv F_{\alpha\mu;\beta}U^{\mu}$,
$B_{\alpha\beta}\equiv\star F_{\alpha\mu;\beta}U^{\mu}$ and $\mathbb{H}_{\alpha\beta}\equiv\star R_{\alpha\mu\beta\sigma}U^{\mu}U^{\sigma}$
are, respectively, the electric, magnetic and gravitomagnetic tidal
tensors as defined in \cite{CHPRD,PaperAnalogies}, \emph{as measured
by the test particle}. The operator $D_{F}/d\tau$ denotes the Fermi-Walker
derivative along $U^{\alpha}$ (e.g.~\cite{Gravitation,Stephani}),
which, for some vector $V^{\mu}$, reads 
\begin{equation}
\frac{D_{F}V^{\mu}}{d\tau}=\frac{DV^{\mu}}{d\tau}-2U^{[\mu}a^{\nu]}V_{\nu}\ ,\label{eq:FermiTransport}
\end{equation}
where $a^{\alpha}\equiv DU^{\alpha}/d\tau$ is the CM acceleration.

\subsection{Force on gyroscope vs. force on magnetic dipole - exact analogy based
on tidal tensors \label{sub:analogy based on tidal tensors}}

\begin{table*}
\caption{\label{tab:Analogy}Analogy between the electromagnetic force on a
magnetic dipole and the gravitational force on a gyroscope}

\begin{ruledtabular} %
\begin{tabular}{cc}
\raisebox{2.4ex}{}Electromagnetic Force~  & \raisebox{2.4ex}{}Gravitational Force\tabularnewline
\raisebox{0.4ex}{on a Magnetic Dipole}  & \raisebox{0.4ex}{on a Spinning Particle~} \tabularnewline
\hline 
\begin{tabular}{cc}
\raisebox{6.5ex}{}\raisebox{2ex}{${\displaystyle F_{{\rm EM}}^{\beta}=B_{\alpha}^{\,\,\,\beta}\mu^{\alpha}};\quad\ B_{\ \beta}^{\alpha}\equiv\star F_{\ \mu;\beta}^{\alpha}U^{\mu}$}  & \raisebox{2ex}{(\ref{tab:Analogy}.1a)}\tabularnewline
\raisebox{3ex}{}\raisebox{0.4ex}{Eqs. for the Magnetic Tidal Tensor}  & \tabularnewline
\raisebox{3ex}{}$B_{\ \alpha}^{\alpha}=0$  & (\ref{tab:Analogy}.2a)\tabularnewline
\raisebox{3ex}{}$B_{[\alpha\beta]}=\frac{1}{2}\star F_{\alpha\beta;\gamma}U^{\gamma}-2\pi\epsilon_{\alpha\beta\sigma\gamma}j^{\sigma}U^{\gamma}$  & (\ref{tab:Analogy}.3a)\tabularnewline
\raisebox{6ex}{}\raisebox{2ex}{$B_{\alpha\beta}U^{\alpha}=0;\quad\ B_{\alpha\beta}U^{\beta}=\epsilon_{\ \ \alpha\delta}^{\beta\gamma}E_{[\beta\gamma]}U^{\delta}$}  & \raisebox{2ex}{(\ref{tab:Analogy}.4a)}\tabularnewline
\end{tabular} & %
\begin{tabular}{cc}
\raisebox{6.5ex}{}\raisebox{2ex}{${\displaystyle F_{{\rm G}}^{\beta}=-\mathbb{H}_{\alpha}^{\,\,\,\beta}S^{\alpha}};\quad\ \mathbb{H}_{\ \beta}^{\alpha}\equiv\star R_{\,\,\,\mu\beta\nu}^{\alpha}U^{\mu}U^{\nu}$}  & \raisebox{2ex}{(\ref{tab:Analogy}.1b)}\tabularnewline
\raisebox{3ex}{}\raisebox{0.4ex}{Eqs.~for the Gravitomagnetic
Tidal Tensor}  & \tabularnewline
\raisebox{3ex}{}$\mathbb{H}_{\ \alpha}^{\alpha}=0$  & (\ref{tab:Analogy}.2b)\tabularnewline
\raisebox{3ex}{}$\mathbb{H}_{[\alpha\beta]}=-4\pi\epsilon_{\alpha\beta\sigma\gamma}J^{\sigma}U^{\gamma}$  & (\ref{tab:Analogy}.3b)\tabularnewline
\raisebox{6ex}{}\raisebox{2ex}{$\mathbb{H}_{\alpha\beta}U^{\alpha}=\mathbb{H}_{\alpha\beta}U^{\beta}=0$}  & \raisebox{2ex}{(\ref{tab:Analogy}.4b)}\tabularnewline
\end{tabular}\tabularnewline
\end{tabular}\end{ruledtabular} 
\end{table*}

Herein we are interested in \emph{purely magnetic} dipoles, i.e.,
dipoles whose electric moment vanishes in the CM frame; this is ensured
by the condition $d^{\alpha}=0$. In this case, Eq.~(\ref{eq:ForcePirani})
simplifies to 
\begin{equation}
\frac{DP^{\alpha}}{d\tau}=qF^{\alpha\beta}U_{\beta}+B^{\beta\alpha}\mu_{\beta}-\mathbb{H}^{\beta\alpha}S_{\beta}\ .\label{eq:ForceAnalogy}
\end{equation}
These equations manifest the physical analogy $B_{\alpha\beta}\leftrightarrow\mathbb{H}_{\alpha\beta}$,
summarized in Table \ref{tab:Analogy}: (i) both the electromagnetic
force on a magnetic dipole and the gravitational force on a gyroscope
are determined by a contraction of the spin/magnetic dipole 4-vector
with a magnetic type tidal tensor. $B_{\alpha\beta}$ may be cast
as the derivative of the magnetic field $B^{\alpha}=\star F_{\ \beta}^{\alpha}U^{\beta}$
as measured in the \emph{inertial} frame \emph{momentarily} comoving
with the test particle: $B_{\alpha\beta}=B_{\alpha;\beta}|_{U=const.}$.
For this reason it is dubbed the \emph{magnetic tidal tensor}, and
its gravitational counterpart $\mathbb{H}_{\alpha\beta}$ the \emph{gravitomagnetic
tidal tensor} \cite{CHPRD}. (ii) It turns out that $B_{\alpha\beta}$
and $\mathbb{H}_{\alpha\beta}$ obey the formally similar equations
(\ref{tab:Analogy}.2) and (\ref{tab:Analogy}.3) in Table \ref{tab:Analogy},
which in one case are Maxwell's equations, and in the other are \emph{exact}
Einstein's equations. That is: the traces (\ref{tab:Analogy}.2) are,
respectively, the time projection (with respect to $U^{\alpha}$)
of the electromagnetic Bianchi identity $\star F_{\ \ ;\beta}^{\alpha\beta}=0$
and the time-time projection of the algebraic Bianchi identities $\star R_{\ \ \ \gamma\beta}^{\gamma\alpha}=0$;
the antisymmetric parts (\ref{tab:Analogy}.3a) are, respectively,
the space projection of Maxwell's equations $F_{\ \ ;\beta}^{\alpha\beta}=4\pi j^{\alpha}$
and the time-space projection of Einstein's equations $R_{\mu\nu}=8\pi(T_{\mu\nu}-\frac{1}{2}g_{\mu\nu}T_{\,\,\,\alpha}^{\alpha})$.
The electromagnetic equations take a familiar form in an inertial
frame: Eq.~(\ref{tab:Analogy}.2a) becomes $\nabla\cdot\vec{B}=0$;
the space part of (\ref{tab:Analogy}.3a) is the Maxwell-Ampère law
$\nabla\times\vec{B}=\partial\vec{E}/\partial t+4\pi\vec{j}$. The
latter means that the space part of $B_{[\alpha\beta]}$ encodes the
curl of $B^{\alpha}$, which is actually a more general statement,
holding in arbitrarily accelerated frames: denote by $U^{\alpha}$
the 4-velocity of the rest observers in such frames; if the frame
is non-rotating and non-shearing, $U_{\alpha;\beta}=-a_{\alpha}U_{\beta}$,
cf.~Eq.~(\ref{eq:uKinDecomp}) below, and we have 
\begin{equation}
\epsilon_{\ \ \alpha\delta}^{\beta\gamma}B_{\gamma\beta}U^{\delta}=\epsilon_{\ \ \alpha\delta}^{\beta\gamma}B_{\gamma;\beta}U^{\delta}\;\Rightarrow\;\epsilon^{ikj}B_{jk}=(\nabla\times\vec{B})^{i}.\label{eq:Bab-Ba;b}
\end{equation}
Expressing also the second member of (\ref{tab:Analogy}.3a) in terms
of the electric and magnetic fields $E^{\alpha}$ and $B^{\alpha}$
measured in this frame, we obtain, in 3-vector notation, 
\begin{equation}
\nabla\times\vec{B}=\frac{D\vec{E}}{d\tau}-\vec{a}\times\vec{E}+4\pi\vec{j}\label{eq:CurlBAccel}
\end{equation}
which is the generalization of Maxwell-Ampère law for accelerated
frames (cf.~Eq.~(19) of~\cite{Maartens:1997fg}). This equation,
as well as Eq.~(\ref{eq:MaxFardayGenVector}) below, is of use in
the particle's CM frame (as it in general accelerates).

Note this important aspect of Eq.~(\ref{tab:Analogy}.3a), considering
for simplicity the vacuum case $j^{\alpha}=0$: it tells us that when
the field $F_{\alpha\beta}$ varies along the particle's worldline
(of 4-velocity $U^{\alpha}$), that endows $B_{\alpha\beta}$ with
an antisymmetric part, implying that $B_{\alpha\beta}$ itself is
non-vanishing. Hence, whenever the particle moves in a non-homogeneous
field, a force will be exerted on it (except for special orientations
of $\vec{\mu}$). From Eqs.~(\ref{eq:Bab-Ba;b})-(\ref{eq:CurlBAccel}),
this can be interpreted, taking the perspective of the inertial frame
momentarily comoving with the particle, as the time-varying electric
field inducing a curl in the magnetic field $\vec{B}$ (and thus a
non-vanishing magnetic tidal tensor). The fact that its gravitomagnetic
counterpart $\mathbb{H}_{\alpha\beta}$ is symmetric in vacuum tells
us that no analogous induction phenomenon occurs in gravity. The physical
consequences shall be explored in Sec.~\ref{sub:Symmetries} below.

There is an electric counterpart to this analogy, relating the electric
tidal tensor $E_{\alpha\beta}$ with the electric part of the Riemann
tensor: 
\[
E_{\alpha\beta}\equiv F_{\alpha\mu;\beta}U^{\mu}\,\longleftrightarrow\,\mathbb{E}_{\alpha\beta}\equiv R_{\alpha\mu\beta\nu}U^{\mu}U^{\nu},
\]
which is manifest in the worldline deviations of both theories, see
\cite{CHPRD}, and together they form the gravito-electromagnetic
analogy based on tidal tensors \cite{CHPRD,PaperAnalogies}. These
tensors obey the following equations, which will be useful in this
work: 
\begin{equation}
E_{[\alpha\beta]}=\frac{1}{2}F_{\alpha\beta;\gamma}U^{\gamma};\quad{\rm (a)}\qquad\mathbb{E}_{[\alpha\beta]}=0.\quad{\rm (b)}\label{eq:Eabanti}
\end{equation}
Eq.~(\ref{eq:Eabanti}a) results from the space projection (relative
to $U^{\alpha}$) of the identity $\star F_{\ \ ;\beta}^{\alpha\beta}=0$,
and Eq.~(\ref{eq:Eabanti}b) from the time-space projection of the
identity $\star R_{\ \ \ \gamma\beta}^{\gamma\alpha}=0$. Contracting
(\ref{eq:Eabanti}a) with the spatial 3-form $\epsilon_{\alpha\beta\gamma\delta}U^{\delta}$
yields Eq.~(\ref{tab:Analogy}.4a) of Table \ref{tab:Analogy}. Again,
for a (non-rotating and non-shearing) arbitrarily accelerated frame
we have: 
\begin{equation}
\epsilon_{\ \ \alpha\delta}^{\beta\gamma}E_{\gamma\beta}U^{\delta}=\epsilon_{\ \ \alpha\delta}^{\beta\gamma}E_{\gamma;\beta}U^{\delta}\;\Rightarrow\;\epsilon^{ikj}E_{jk}=(\nabla\times\vec{E})^{i}.\label{eq:Eab-Ea;b}
\end{equation}
Expressing also the second member of (\ref{eq:Eabanti}a) in terms
of the fields $E^{\alpha}$ and $B^{\alpha}$ measured in this frame,
we obtain, in 3-vector notation: 
\begin{equation}
\nabla\times\vec{E}=-\frac{D\vec{B}}{d\tau}-\vec{a}\times\vec{E},\label{eq:MaxFardayGenVector}
\end{equation}
which is a generalization of \emph{Maxwell-Faraday equation} $\nabla\times\vec{E}=-\partial\vec{B}/\partial t$
for accelerated frames, cf.~Eq.~(20) of~\cite{Maartens:1997fg}.

The fact that the gravitoelectric tidal tensor $\mathbb{E}_{\alpha\beta}$
is symmetric again means that there is no analogous gravitational
induction effect, and this is a key observation for the applications
in Secs.~\ref{sub:Time-components-in-CM-frame } and \ref{sec:Beyond-pole-dipole}.

Equations (\ref{tab:Analogy}.4) are the time projections of the tidal
tensors with respect to \emph{the observer} $U^{\alpha}$ \emph{measuring
them} (if $U^{\alpha}$ is the particle's CM 4-velocity, they are
the time projection in its rest frame); they are zero in the gravitational
case, as $\mathbb{H}_{\alpha\beta}$ is spatial relative to $U^{\alpha}$,
and non-zero in the electromagnetic case, which again is related to
electromagnetic induction, as the right Eq.~(\ref{tab:Analogy}.4a)
corresponds to the spatially projected Eq.~(\ref{eq:Eabanti}a).
This means that $F_{{\rm G}}^{\alpha}$ is spatial with respect to
$U^{\alpha}$, whereas $F_{{\rm EM}}^{\alpha}$ is not, which has
important implications on the work done by the fields on the particle,
as will be discussed in Sec.~\ref{sub:Time-Components}.

Finally, note that $F_{{\rm EM}}^{\alpha}=B^{\beta\alpha}\mu_{\beta}$
is the covariant generalization of the familiar textbook 3-D expression
$\vec{F}_{{\rm EM}}=\nabla(\vec{\mu}\cdot\vec{B})$, the latter being
valid only in the \emph{inertial} frame \emph{momentarily} comoving
with the particle.

\subsection{Spin ``precession'' - exact analogy based on inertial GEM fields
from the 1+3 formalism\label{sub:Spin-``precession''--}}

For \emph{purely magnetic} dipoles ($d^{\alpha}=0$), Eq.~(\ref{eq:SpinPirani})
for the spin evolution under the Mathisson-Pirani condition simplifies
to 
\begin{equation}
\frac{D_{F}S^{\mu}}{d\tau}=\epsilon_{\ \alpha\beta\nu}^{\mu}U^{\nu}\mu^{\alpha}B^{\beta},\label{eq:EqSpinFermi}
\end{equation}
or equivalently (cf. Eq. (\ref{eq:FermiTransport})) 
\begin{eqnarray}
\frac{DS^{\mu}}{d\tau} & = & S_{\nu}a^{\nu}U^{\mu}+\epsilon_{\ \alpha\beta\nu}^{\mu}U^{\nu}\mu^{\alpha}B^{\beta},\label{eq:EqSpinVector}
\end{eqnarray}
where \textbf{$B^{\beta}$} is the magnetic field \emph{as measured
by the test particle}. The first term in (\ref{eq:EqSpinVector})
embodies the Thomas precession. The second term is a covariant form
for the familiar torque $\tau=\vec{\mu}\times\vec{B}$ causing the
Larmor precession of a magnetic dipole under a magnetic field.

Consider now an orthonormal frame $\mathbf{e_{\hat{\alpha}}}$ carried
by an observer of 4-velocity $U^{\alpha}$, such that $\mathbf{U}=\mathbf{e}_{\hat{0}}$,
comoving with the test particle. In such frame, $S^{\hat{0}}=0$ and
$U^{\hat{\alpha}}=\delta_{\hat{0}}^{\hat{\alpha}}$, and equation
(\ref{eq:EqSpinVector}) reduces to (see \cite{PaperAnalogies}):
\begin{equation}
\frac{DS^{\hat{i}}}{d\tau}=(\vec{\mu}\times\vec{B})^{\hat{i}}\ \Leftrightarrow\ \frac{dS^{\hat{i}}}{d\tau}=\left(\vec{S}\times\vec{\Omega}+\vec{\mu}\times\vec{B}\right)^{\hat{i}}\ \ \label{eq:Spin3+1}
\end{equation}
where $\vec{\Omega}$ is angular velocity of rotation of the spatial
axes $\mathbf{e}_{\hat{i}}$ relative to a tetrad Fermi-Walker transported
along the center of mass worldline. If $B^{\alpha}=0$, Eqs.~(\ref{eq:EqSpinFermi})-(\ref{eq:Spin3+1})
tell us that $S^{\alpha}$ undergoes Fermi-Walker transport, i.e.,
it follows the local ``compass of inertia''~\cite{MassaZordan,Gravitation and Inertia}
(the so-called gyroscope ``precession'', of frequency $-\vec{\Omega}$,
is thus in fact just minus the rotation of the $\mathbf{e}_{\hat{i}}$
relative to a locally non-rotating frame, and therefore, \emph{locally},
an artifact of the reference frame, manifest only in the ordinary
derivative $d\vec{S}/d\tau$). Up until now $\vec{\Omega}$ is arbitrary;
of special interest is, in asymptotically flat spacetimes, the case
where the triads $\mathbf{e}_{\hat{i}}$ are locked to the so-called
``frame of the distant stars''. If the spacetime is stationary,
such frame is set up by choosing the congruence of static observers
(cf.~point \ref{enu:Static-observers} of Sec.~\ref{sub:Notation-and-conventions}),
and demanding $\vec{\Omega}$ to match their vorticity $\vec{\omega}$
(defined in Eq.~\eqref{eq:uKinDecomp} below), $\vec{\Omega}=\vec{\omega}$.
That is, demanding the triads $\mathbf{e}_{\hat{i}}$ to \emph{co-rotate}
\cite{Massa,The many faces} with the observers, relative to Fermi-Walker
transport; we dub such frame \emph{congruence adapted}. This ensures
that the axes $\mathbf{e}_{\hat{i}}$ point to fixed neighboring observers,
cf.~Eq.~(41) of \cite{PaperAnalogies}. Since the observer congruence
is rigid and, \emph{at infinity}, inertial, the axes $\mathbf{e}_{\hat{i}}$
locked to it are locked to the inertial frame at infinity (the rest
frame of the ``distant stars''), and Eq.~(\ref{eq:Spin3+1}) yields
the precession of spinning particle with respect to the distant stars.
For more details we refer to Secs.~3.1 and 3.3 of the companion paper
\textcolor{black}{\cite{PaperAnalogies}}.

Note the analogy between the two terms of the second Eq.~(\ref{eq:Spin3+1});
when the frame is congruence adapted, then $\vec{\Omega}=\vec{H}/2$,
where $\vec{H}$ is the ``gravitomagnetic'' or Coriolis field felt
in such frame, which plays in the \emph{exact} geodesic equations
(e.g.~Eq.~(58) of \cite{PaperAnalogies}) the same role as the magnetic
field $\vec{B}$ in the electromagnetic Lorentz force. Moreover, the
field equations for $\vec{H}$ exhibit striking similarities with
the Maxwell equations for $\vec{B}$ in an accelerated, rotating frame,
see Table 2 of \cite{PaperAnalogies}. And in the linear regime, for
stationary fields, they become similar to Maxwell's equations in a
Lorentz frame, as is well known \cite{Gravitation and Inertia,General Relativity,Ciufolini Nature Review,Wald et al 2010,Wald,Tucker Clark}.
That tells us that analogous setups generate fields alike. A well
known realization is the similarity between the gravitomagnetic field
produced by a spinning mass (as measured by the congruence of static
observers), and the magnetic field produced by a spinning charge,
e.g.~Eqs.~(6.1.9), (6.1.25) of \cite{Gravitation and Inertia}.

The analogy in Eq.~(\ref{eq:Spin3+1}) is valid for arbitrary fields,
unlike the case of most gravito-electromagnetic analogies%
\footnote{In the framework of the GEM inertial fields, the force on the gyroscope
\cite{Natario,PaperAnalogies} and the equation for the geodesics
of (non-spinning) test particles (e.g.~\cite{Natario,PaperAnalogies,The many faces})
can be exactly described by equations analogous to the ones from electromagnetism,
but \emph{only} if the fields are stationary \emph{and} the gyroscope
it at rest with respect to a stationary observer (i.e., its \emph{worldline}
is tangent to a time-like Killing vector), or, in the case of the
geodesic equation, if one considers a frame adapted to a rigid congruence
of stationary observers. See Secs.~3.2 and 3.6 of \cite{PaperAnalogies}. %
} based on GEM \emph{inertial fields} (not tidal tensors), which do
not hold (in the sense of a one to one correspondence) when one considers
time-dependent fields \cite{PaperAnalogies,PaperIAU} (another exception
is the hidden momentum analogy, presented in the next section).

Finally, note that, if we assume $\vec{\mu}=\sigma\vec{S}$, then
the quantity $S^{2}=S^{\alpha}S_{\alpha}=S^{\alpha\beta}S_{\alpha\beta}/2$
is a constant of the motion, which is immediately seen contracting
(\ref{eq:EqSpinVector}) with $S^{\mu}$.

\subsection{Momentum of the spinning particle - ``hidden momentum'' and exact
analogy based on inertial GEM fields from the 1+3 formalism\label{sub:Momentum-of-the-Particle}}

The momentum (\ref{eq:Pgeneral}) of a spinning particle is not in
general parallel to its center of mass 4-velocity $U^{\alpha}$. In
order to see that, let us re-write the spin evolution equation (\ref{eq:SpinDS0})
as 
\begin{equation}
\frac{DS^{\alpha\beta}}{d\tau}=2P^{[\alpha}U^{\beta]}+\tau^{\alpha\beta}\label{eq:DSabdtHidden}
\end{equation}
where we denoted 
\begin{equation}
\tau^{\alpha\beta}\equiv2\mu^{\theta[\beta}F_{\ \ \theta}^{\alpha]}+2d^{[\alpha}F_{\ \ \gamma}^{\beta]}U^{\gamma}\ ,\label{eq:taudip}
\end{equation}
which is sometimes called the dipole ``torque'' tensor (although
only its spatial part contributes to the actual torque, cf.~Eq.~(\ref{eq:FermiTorque})).
Consider a generic spin condition $S^{\alpha\beta}u_{\beta}=0$, where
$u^{\alpha}$ denotes the 4-velocity of an arbitrary observer $\mathcal{O}(u)$
(as discussed in Sec.~\ref{sub:Center-of-mass}, this condition means
that we take as reference worldline the center of mass as measured
by $\mathcal{O}(u)$). An expression for $P^{\alpha}$ can be obtained
contracting (\ref{eq:DSabdtHidden}) with $u_{\beta}$, leading to
\begin{equation}
P^{\alpha}=\frac{1}{\gamma(u,U)}\left(m(u)U^{\alpha}+S^{\alpha\beta}\frac{Du_{\beta}}{d\tau}+\tau^{\alpha\beta}u_{\beta}\right)\label{eq:Momentum}
\end{equation}
where \textcolor{black}{$\gamma(U,u)\equiv-U^{\alpha}u_{\alpha}$,
$m(u)\equiv-P^{\alpha}u_{\alpha}$, and }in the second term we used
\textcolor{black}{$S^{\alpha\beta}u_{\beta}=0$}. We split $P^{\alpha}$
in its projections parallel and orthogonal to the CM 4-velocity $U^{\alpha}$:
\begin{equation}
P^{\alpha}=P_{{\rm kin}}^{\alpha}+P_{{\rm hid}}^{\alpha};\quad P_{{\rm kin}}^{\alpha}\equiv mU^{\alpha},\; P_{{\rm hid}}^{\alpha}\equiv(h^{U})_{\ \beta}^{\alpha}P^{\beta}.\label{eq:HiddenMomentum}
\end{equation}
We dub the parallel projection $P_{{\rm kin}}^{\alpha}=mU^{\alpha}$
``kinetic momentum'' associated with the motion of the center of
mass. This is the most familiar part of $P^{\alpha}$, formally similar
to the momentum of a monopole particle. The component $P_{{\rm hid}}^{\alpha}$
orthogonal to $U^{\alpha}$ is the so-called ``hidden momentum''
\cite{Wald et al 2010}. The reason for the latter denomination is
seen taking the perspective of the particle's center of mass frame
(i.e., the frame where $\vec{U}=0$): the 3-momentum is in general
not zero therein, $\vec{P}=\vec{P}_{{\rm hid}}\ne0$; however, by
definition, the particle's CM is at rest in that frame, and so this
momentum must be somehow hidden in the spinning particle. The hidden
momentum $P_{{\rm hid}}^{\alpha}$ consists of two parts of distinct
origin: $P_{{\rm hid}}^{\alpha}=P_{{\rm hidI}}^{\alpha}+P_{{\rm hidEM}}^{\alpha}$,
where 
\begin{align}
P_{{\rm hidI}}^{\alpha} & \equiv\frac{1}{\gamma(u,U)}(h^{U})_{\ \sigma}^{\alpha}S^{\sigma\beta}\frac{Du_{\beta}}{d\tau}\ ;\label{eq:HiddenInertial}\\
P_{{\rm hidEM}}^{\alpha} & \equiv\frac{1}{\gamma(u,U)}(h^{U})_{\ \sigma}^{\alpha}\tau^{\sigma\beta}u_{\beta},\label{eq:PhidEM-0}
\end{align}
\textcolor{black}{The term $P_{{\rm hidI}}^{\alpha}$, which we dub
}``inertial'' hidden momentum \textcolor{black}{(the reason for
such denomination will be clear below)},\textcolor{black}{{} is a
gauge} term that depends only on the spin supplementary condition,
i.e., on the choice of the vector field $u^{\alpha}$ (the 4-velocity
of the observers $\mathcal{O}(u)$ relative to which the CM is being
computed). This type of \textcolor{black}{hidden momentum was first
discussed in~\cite{Wald et al 2010} (dubbed ``kinematical''} therein).
It is in general not zero when $Du^{\alpha}/d\tau\ne0$; this comes
as a natural consequence of what we discussed in Sec.~\ref{sub:Center-of-mass}:
the position of the CM of a spinning body depends on the vector $u^{\alpha}$
relative to which it is computed; if that vector varies along the
reference worldline, it is clear that this is reflected in the velocity
$U^{\alpha}$ of the CM (which in general will accelerate even without
the action of any forces; see Figs.~1 and 2 of \cite{CostaNatario2014}).
Since the momentum $P^{\alpha}$ remains the same, $U^{\alpha}$ will
in general not be parallel to $P^{\alpha}$, and so the centroid is
not at rest in the frame where $P^{i}=0$; \textcolor{black}{conversely,
the momentum is not zero in the CM frame (hidden momentum). If we
take a field $u^{\alpha}$ such that $Du^{\alpha}/d\tau=0$ (which
was proposed in~\cite{Semerak II} as one of the possible spin supplementary
conditions), i.e., if we take as reference worldline the center of
mass as measured with respect to a field $u^{\alpha}$ that} \textcolor{black}{\emph{is
parallel transported along it}}\textcolor{black}{, then $P_{{\rm hidI}}^{\alpha}$
(as well as the motion effects induced by it, such as the bobbings
studied in~\cite{Wald et al 2010}, or the helical motions discussed
in }\cite{Helical}\textcolor{black}{) is made to vanish.}

The term $P_{{\rm hidEM}}^{\alpha}$ is what we dub ``electromagnetic''
hidden momentum; it is a still not well known feature of relativistic
electrodynamics (despite its discovery \cite{Shockley} dating back
from the 60's, and having since been discussed in number of papers,
e.g.~\cite{GriffithsAmJPhys,Vaidman,Shockley,HnizdoFluid,Van Vleck,Wald et al 2010,ProcERE2011,Hartle2009,GrallaHerrmannHidden}).
It is associated with the electromagnetic torque tensor $\tau^{\alpha\beta}$,
and consists of a part which is gauge and arises, again, from the
choice of centroid (vanishing for suitable choices, see \cite{CostaNatario2014}
for details), plus a part that is not gauge, whose motion effects
(such as the bobbings in electromagnetic systems studied in \cite{Wald et al 2010})
cannot in general be made to vanish by any choice of center of mass.

With the Mathisson-Pirani condition $S^{\alpha\beta}U_{\beta}=0$,
the hidden momentum in Eqs.~(\ref{eq:HiddenMomentum})-(\ref{eq:PhidEM-0})
takes the suggestive form 
\begin{equation}
P_{{\rm hidI}}^{\alpha}\equiv-\epsilon_{\ \beta\gamma\delta}^{\alpha}S^{\beta}a^{\gamma}U^{\delta}\ ;\quad P_{{\rm hidEM}}^{\alpha}\equiv\epsilon_{\ \beta\gamma\delta}^{\alpha}\mu^{\beta}E^{\gamma}U^{\delta}\ ,\label{eq:PhidDecompMP}
\end{equation}
and so the particle's total momentum, Eq (\ref{eq:HiddenMomentum}),
reads 
\begin{equation}
P^{\alpha}=mU^{\alpha}-\epsilon_{\ \beta\gamma\delta}^{\alpha}S^{\beta}a^{\gamma}U^{\delta}+\epsilon_{\ \beta\gamma\delta}^{\alpha}\mu^{\beta}E^{\gamma}U^{\delta}\ ,\label{eq:MomentumMP}
\end{equation}
where $E^{\alpha}=F_{\ \beta}^{\alpha}U^{\beta}$ is the electric
field \emph{as measured in the particle's CM frame} (of 4-velocity
$U^{\alpha}$), and $a^{\alpha}$ its acceleration. In the particle's
CM frame (where $U^{i}=0$), and in vector notation, the space part
reads ($P_{{\rm hid}}^{0}=0$) 
\begin{equation}
\vec{P}_{{\rm hid}}=\vec{P}=-\vec{S}\times\vec{a}+\vec{\mu}\times\vec{E}=\vec{S}\times\vec{G}+\vec{\mu}\times\vec{E}\ .\label{eq:Pvec}
\end{equation}
The term $\vec{P}_{{\rm hidEM}}=\vec{\mu}\times\vec{E}$ is the most
usual form for the electromagnetic hidden momentum in the literature,
e.g. \cite{Vaidman,HnizdoFluid,Van Vleck,GriffithsBook,GriffithsAmJPhys,Jackson}.
It equals \emph{minus} the electromagnetic field momentum $\vec{P}_{\times}$
generated by a magnetic dipole when placed in an external electromagnetic
field, which, in the particle's frame, reads (see \cite{EPAPS}) 
\[
\vec{P}_{\times}=\int\vec{E}\times\vec{B}_{{\rm dipole}}=-\vec{\mu}\times\vec{E}\ =-\vec{P}_{{\rm hidEM}}.
\]
It should be noted however that $\vec{P}_{{\rm hidEM}}$ (unlike $\vec{P}_{\times}$)
is \emph{purely mechanical in nature} (\emph{not} field momentum,
even though it is ultimately originated by the action of the electromagnetic
field), as explained in \cite{GriffithsAmJPhys,GriffithsBook,Vaidman}
using simple models. This hidden momentum implies that, in the presence
of an electromagnetic field, the spatial momentum of a dipole whose
center of mass is \emph{at rest} is in general not zero. As explained
in detail in \cite{EPAPS}, this actually plays a crucial role in
the conservation laws: consider a magnetic dipole at rest in a stationary
field; it is $\vec{P}_{{\rm hidEM}}$ which allows for the total spatial
momentum $\vec{P}_{{\rm tot}}\equiv\vec{P}_{{\rm matter}}+\vec{P}_{{\rm EM}}$
to vanish, as required by the conservation equations $(T_{{\rm tot}})_{\ \ ;\beta}^{\alpha\beta}=0$
for a stationary configuration.

Equations \eqref{eq:PhidDecompMP}-\eqref{eq:Pvec} manifest an \emph{exact}
analogy: $G^{\alpha}=-a^{\alpha}$ is the gravitoelectric field (as
defined in \cite{PaperAnalogies,Natario,The many faces}) associated
to the CM frame, which is a field of ``inertial forces'', and so
$P_{{\rm hidI}}^{\alpha}$ is the ``inertial'' analogue of $P_{{\rm hidEM}}^{\alpha}$,
with $S^{\alpha}$ and $G^{\alpha}$ in the roles of $\mu^{\alpha}$
and $E^{\alpha}$. The analogy above is useful to understand the famous
helical solutions allowed by the condition $S^{\alpha\beta}U_{\beta}=0$:
we show in \cite{Helical,ProcERE2011} that they are a phenomena which
can be cast as analogous to the bobbings of a magnetic dipole in an
external electric field (studied in Sec.~III.B.1 of \cite{Wald et al 2010}),
in both cases the effect being driven not by a force but solely by
an interchange between kinetic and hidden momentum.

\subsection{Mass of the spinning particle\label{sub:Mass-of-the}}

We take the scalar $m=-P^{\alpha}U_{\alpha}$ as ``the proper mass''%
\footnote{This is the most natural definition of the body's mass if one uses
the Mathisson-Pirani spin condition, since it is the quantity which
is conserved when $F^{\alpha\beta}=0$, cf.~Eq.~(\ref{eq:dm1}).
If one uses the Tulczyjew-Dixon condition $S^{\alpha\beta}P_{\beta}=0$
instead, then the conserved quantity is $M\equiv\sqrt{-P^{\alpha}P_{\alpha}}$
(not $m$), i.e., the particle's energy as measured in the zero 3-momentum
frame (see e.g. \cite{Semerak I}).%
}~\cite{MollerBook} of the spinning particle. It is simply the time
projection of $P^{\alpha}$ in the particle's CM frame, i.e., the
particle's energy as measured in its center of mass rest frame. Whereas
for a monopole particle $m$ is a constant of the motion, for a spinning
particle with dipole moments that is not the case in general. It follows
from the definition of $m$ that 
\begin{equation}
\frac{dm}{d\tau}=-\frac{DP^{\alpha}}{d\tau}U_{\alpha}-P^{\alpha}a_{\alpha}=-\frac{D_{F}P^{\alpha}}{d\tau}U_{\alpha}\ ;\label{eq:dm/dt0}
\end{equation}
i.e., $dm/d\tau$ is the time projection, in the CM frame, of the
Fermi-Walker derivative of the momentum. Noting that $P^{\alpha}a_{\alpha}=P_{{\rm hidEM}}^{\alpha}a_{\alpha}$,
and using the orthogonality $P_{{\rm hidEM}}^{\alpha}U_{\alpha}=0$,
we can rewrite this equation as: 
\begin{equation}
\frac{dm}{d\tau}=-\left(\frac{DP^{\alpha}}{d\tau}-\frac{DP_{{\rm hidEM}}^{\alpha}}{d\tau}\right)U_{\alpha}\ .\label{eq:Phid-Contrib-m}
\end{equation}
Thus $dm/d\tau$ equals also the time projection, in the CM frame,
of the force $DP^{\alpha}/d\tau$ \emph{subtracted} by the derivative
of the ``electromagnetic'' hidden momentum $DP_{{\rm hidEM}}^{\alpha}/d\tau$.
Let us see the meaning of the first term. Contracting (\ref{eq:ForcePirani})
with $U^{\alpha}$, and noting that $B^{\beta\alpha}U_{\alpha}=U_{\gamma}D\!\star\! F^{\beta\gamma}/d\tau$,
we obtain 
\begin{eqnarray}
-\frac{DP^{\alpha}}{d\tau}U_{\alpha} & = & -\frac{D\!\star\! F^{\beta\gamma}}{d\tau}U_{\gamma}\mu_{\beta}+E_{\beta}\frac{Dd^{\beta}}{d\tau}\ ,\label{eq:ProjDP/dt}
\end{eqnarray}
showing that the force has a time projection if the Maxwell tensor
and/or the electric dipole vector vary along the CM worldline. Now,
noting from Eqs.~(\ref{eq:MomentumMP}) and (\ref{eq:FstarDecomp})
that $P^{\alpha}a_{\alpha}=\star F^{\beta\gamma}a_{\gamma}\mu_{\beta}$,
and putting Eqs.~(\ref{eq:dm/dt0}) and (\ref{eq:ProjDP/dt}) together,
we see that 
\begin{equation}
\frac{dm}{d\tau}=-\mu_{\gamma}\frac{DB^{\gamma}}{d\tau}+E_{\gamma}\frac{Dd^{\gamma}}{d\tau}\ .\label{eq:dm1}
\end{equation}
Hence the mass of a particle possessing electric and magnetic dipole
moments is not constant in general. The two contributions are \textcolor{black}{substantially}
different: the mass variation due to the coupling of the field to
the magnetic dipole occurs when the magnetic field varies along the
particle's worldline; it may be interpreted as \emph{essentially}
the rate of work done on the magnetic dipole through Faraday's law
of induction (Fig.~\ref{fig:DipoleMagnet} below), as we shall see
in detail in Sec.~\ref{sub:Time-components-in-CM-frame }. The second
term corresponds to the work done on the electric dipole by the electric
field when the dipole vector varies, e.g., when it rotates; this term
has nothing to do with induction, and is non-zero even for constant,
uniform electric fields. \textcolor{black}{The case of electric dipoles
is discussed in detail in Appendix \ref{sub:Edipole Proper-mass-and-time proj}.}

We are interested mostly in \emph{purely magnetic} dipoles, $d^{\alpha}=0$;
in this case, if we take $\mu^{\alpha}=\sigma S^{\alpha}$, \textcolor{black}{with
$\sigma$ a constant}, and, since from Eq.~(\ref{eq:EqSpinVector}),
$B^{\mu}DS_{\mu}/d\tau=0$, we have~\cite{CorbenBook,Corben,WeyssenhoffRaabe,Wald et al}
\begin{align}
 & \frac{dm}{d\tau}=-\sigma\frac{d}{d\tau}(S_{\mu}B^{\mu})\label{eq:dm2}\\
 & \Rightarrow m=m_{0}-\sigma S_{\mu}B^{\mu}\ =\ m_{0}-\sigma\vec{S}\cdot\vec{B}\ ,\label{eq:m}
\end{align}
where $m_{0}$ is a constant. \textcolor{black}{Thus, if $\vec{\mu}=\sigma\vec{S}$,
the mass $m$ is the sum of a constant plus a variable part $-\vec{\mu}\cdot\vec{B}$,
about which we would like to make some remarks. The expression $-\vec{\mu}\cdot\vec{B}$
is commonly dubbed in elementary textbooks ``magnetic potential energy'';
for this reason some authors \cite{Dixon1970I,WeyssenhoffRaabe,Dixon1965}
have interpreted this term as meaning that the potential energy contributes
to the particle's mass. We argue (in agreement with the analysis in~\cite{Coombes,Young,YoungQuestion66,Deissler}),
that the term $-\vec{\mu}\cdot\vec{B}$ is actually} \textcolor{black}{\emph{internal
}}\textcolor{black}{(not potential) energy} \textcolor{black}{of the
test particle; in fact, we shall see (Sec.~\ref{sub:Electromagnetic Torque-and-force on Spherical})
that, for a quasi-rigid body, it is essentially rotational kinetic
energy, associated with the rotation of the body around its center
of mass. What it actually does is to ensure that} \textcolor{black}{\emph{the
net work done by the magnetic field on a magnetic dipole is zero}}
\textcolor{black}{(hence no potential energy can be assigned to it).
Potential energy comes into play instead in the case of a monopole
charged particle or of an electric dipole in an electric field; but
in neither case does it contribute to the mass ($m$ is a constant
for a monopole particle, as well as for an electric dipole if $d^{\alpha}$
is parallel transported, cf.~Eq.~(\ref{eq:dm1})). These issues
are discussed in detail in Sec.~\ref{sub:TimeProj_Static_EM} and
Appendix \ref{sub:Conserved-quantities,-proper}.}

It is also important to understand that the varying mass $m$ (and
its variable part $-\vec{\mu}\cdot\vec{B}$), are real and physically
measurable, not just a matter of definition (i.e.~not an issue that
goes away by redefining $m_{0}$ in Eq.~(\ref{eq:m}) as the particle's
mass), for $m$ is the \emph{inertial} mass of the particle. In order
to see that, take for simplicity the case when \textcolor{black}{$P_{{\rm hid}}^{\alpha}=0$};
we have 
\[
\frac{DP^{\alpha}}{d\tau}=ma^{\alpha}+\frac{dm}{d\tau}U^{\alpha}\ ,
\]
i.e., the projection of the force in the orthogonal space to $U^{\alpha}$
is $ma^{\alpha}$ (thus, in the CM frame, $D\vec{P}/d\tau=m\vec{a}$).
This inertial mass is measurable, for instance in collisions. The
angular velocity of rotation of a spinning body (since, as mentioned
above, in the case of a quasi-rigid body, $-\vec{\mu}\cdot\vec{B}$
is kinetic energy of rotation) is measurable as well.

\textcolor{black}{In the purely gravitational case, by contrast, the
proper mass is a constant ($m=m_{0}$); the implications for the work
done by the fields on the particle are discussed in Secs.~\ref{sub:Time-components-Static}
and \ref{sub:Summarizing-with-a_simple}.}

\subsection{Center of mass motion\label{sub:Center-of-mass motion}}

Equations~(\ref{tab:Analogy}.1) of Table \ref{tab:Analogy} yield
the \emph{force} on the spinning particle in the electromagnetic and
gravitational case; \emph{not} the acceleration $a^{\alpha}\equiv DU^{\alpha}/d\tau$,
as $P^{\alpha}\ne m_{0}U^{\alpha}$ in general. Setting $m\equiv m_{0}+m'$
in Eq.~(\ref{eq:MomentumMP}), and noting, from decomposition (\ref{eq:FstarDecomp}),
that $\epsilon_{\ \beta\gamma\sigma}^{\alpha}\mu^{\beta}E^{\gamma}U^{\sigma}=\star F^{\beta\alpha}\mu_{\beta}+\mu^{\beta}B_{\beta}U^{\alpha}$,
we can write 
\[
P^{\alpha}=m_{0}U^{\alpha}-\epsilon_{\ \beta\gamma\delta}^{\alpha}S^{\beta}a^{\gamma}U^{\delta}+(m'+\mu^{\alpha}B_{\alpha})U^{\alpha}+\star F_{\beta}^{\ \alpha}\mu^{\beta}.
\]
This is simplified if we consider purely magnetic dipoles ($d^{\alpha}=0$),
and assume $\mu^{\alpha}=\sigma S^{\alpha};$ in that case, cf.~Eq.
\eqref{eq:m}, $m'=-\mu^{\alpha}B_{\alpha}$, and the third term vanishes.
Differentiating, using (\ref{eq:ForceAnalogy}), and noting that,
\emph{if $j^{\alpha}=0$}, $\star F_{\alpha\beta;\tau}U^{\tau}=2B_{[\alpha\beta]}$,
cf. Eq.~(\ref{tab:Analogy}.3a) of Table \ref{tab:Analogy}, we have,
in a region where the charge current density $j^{\alpha}$ is zero
(most of the applications in this paper deal with vacuum), 
\begin{eqnarray}
m_{0}a^{\alpha} & = & qF^{\alpha\beta}U_{\beta}+B^{\alpha\beta}\mu_{\beta}-\mathbb{H}^{\beta\alpha}S_{\beta}-\star F_{\beta}^{\ \alpha}\frac{D\mu^{\beta}}{d\tau}\nonumber \\
 &  & +\epsilon_{\ \beta\gamma\delta}^{\alpha}U^{\delta}\frac{D}{d\tau}(S^{\beta}a^{\gamma}).\label{eq:mavaccum}
\end{eqnarray}
Note the reversed indices in the second term as compared to the expression
for the force (\ref{tab:Analogy}.1a). This leads to a counter-intuitive
dynamical behavior, as we shall exemplify in Sec.~\ref{sub:Radial-Schwa}.

\section{Dynamical manifestations of the symmetries of the magnetic tidal
tensors\label{sub:Symmetries}}

According to Table \ref{tab:Analogy}, both in the case of the electromagnetic
force on a magnetic dipole, and in the case of the gravitational force
on a gyroscope, it is the magnetic/gravitomagnetic tidal tensor, \emph{as
seen by the test particle} of 4-velocity $U^{\alpha}$, that determines
the force exerted upon it. The explicit analogy in Table \ref{tab:Analogy}
is thus ideally suited to compare the two forces, because in this
framework it amounts to comparing $B_{\alpha\beta}$ to $\mathbb{H}_{\alpha\beta}$.
The most important differences between them are: i) $B_{\alpha\beta}$
is linear in the electromagnetic potentials and vector fields, whereas
$\mathbb{H}_{\alpha\beta}$ is not linear in the metric tensor, nor
in the GEM ``vector'' fields (for a detailed discussion of this
aspect, we refer to Secs.~3.5 and 6 of \cite{PaperAnalogies}); ii)
\emph{in vacuum}, $\mathbb{H}_{[\alpha\beta]}=0$ (symmetric tensor),
whereas $B_{\alpha\beta}$ is generically non symmetric, $B_{[\alpha\beta]}\ne0$,
even in vacuum; iii) time components: $\mathbb{H}_{\alpha\beta}$
is spatial with respect to $U^{\alpha}$, whereas $B_{\alpha\beta}$
is not. The two latter differences, which are clear from equations
(\ref{tab:Analogy}.3)-(\ref{tab:Analogy}.4), are the ones in which
we are most interested in the present work. In this section we start
with the physical consequences of the symmetries, and in the next
section we discuss the time projections.

Equation~(\ref{tab:Analogy}.3a) of Table \ref{tab:Analogy} reads
\emph{in vacuum} ($j^{\alpha}=0$) 
\begin{equation}
B_{[\alpha\beta]}=\frac{1}{2}\star\! F_{\alpha\beta;\gamma}U^{\gamma}\ ;\label{eq:VacuumBab}
\end{equation}
this tells us that when the field $F_{\alpha\beta}$ varies along
the worldline of the observer $U^{\alpha}$, that endows $B_{\alpha\beta}$
with an antisymmetric part, implying that $B_{\alpha\beta}$ itself
is non-vanishing. Now, since, in the force (\ref{tab:Analogy}.1a),
$B_{\alpha\beta}$ is the magnetic tidal tensor \emph{as measured
by the particle} (i.e., $U^{\alpha}$ is the test particle's 4-velocity),
this means that whenever the particle moves in a non-homogeneous field,
a force will be exerted on it (except possibly for special orientations
of $\vec{\mu}$). In the inertial frame momentarily comoving with
the particle, this can be interpreted as being due to the time-varying
(in this frame) electric field, which induces, via the law $\nabla\times\vec{B}=\partial\vec{E}/\partial t$,
\emph{a curl} in the magnetic field $\vec{B}$, and implies that the
particle sees a non-vanishing magnetic tidal tensor, cf.~Eqs.~(\ref{eq:Bab-Ba;b})-(\ref{eq:CurlBAccel}).

The gravitomagnetic counterpart $\mathbb{H}_{\alpha\beta}$, by contrast,
is symmetric in vacuum, which means that no analogous induction phenomenon
occurs in gravity. Indeed, even in non-homogeneous fields, there can
be velocity fields for which $\mathbb{H}_{\alpha\beta}=0$, i.e.,
for which \emph{gyroscopes feel no force }(regardless of the direction
of their spin $\vec{S}$). We know that from the curvature invariants,
which we now briefly discuss.

In vacuum the Riemann tensor becomes the Weyl tensor (10 independent
components), which can be irreducibly decomposed (see e.g.~\cite{Maartens:1997fg})
with respect to a unit timelike 4-vector $u^{\alpha}$ into two spatial
tensors, the gravitoelectric $(\mathbb{E}^{u})_{\alpha\beta}\equiv R_{\alpha\gamma\beta\delta}u^{\gamma}u^{\delta}$
and gravitomagnetic $(\mathbb{H}^{u})_{\alpha\beta}\equiv\star R_{\alpha\gamma\beta\delta}u^{\gamma}u^{\delta}$
tidal tensors measured by $u^{\alpha}$: 
\begin{eqnarray}
R_{\alpha\beta}^{\ \ \gamma\delta} & = & 4\left\{ 2u_{[\alpha}u^{[\gamma}+g_{[\alpha}^{\,\,\,\,[\gamma}\right\} (\mathbb{E}^{u})_{\beta]}^{\,\,\,\delta]}\label{eq:Weyldecomp}\\
 &  & +2\left\{ \epsilon_{\alpha\beta\mu\nu}(\mathbb{H}^{u})^{\mu[\delta}u^{\gamma]}u^{\nu}+\epsilon^{\gamma\delta\mu\nu}(\mathbb{H}^{u})_{\mu[\beta}u_{\alpha]}u_{\nu}\right\} .\nonumber 
\end{eqnarray}
The tensors $\mathbb{E}_{\alpha\beta}$ and $\mathbb{H}_{\alpha\beta}$
are both symmetric and traceless (in vacuum), possessing 5 independent
components each, thus encoding the 10 independent components of $R_{\alpha\beta\gamma\delta}$.
Again in vacuum, one can construct the two quadratic scalar invariants
(e.g.~\cite{matte,bel,WyllePRD}), 
\begin{align}
 & \mathbb{E}^{\alpha\gamma}\mathbb{E}_{\alpha\gamma}-\mathbb{H}^{\alpha\gamma}\mathbb{H}_{\alpha\gamma}=\frac{1}{8}R_{\alpha\beta\gamma\delta}R^{\alpha\beta\gamma\delta}\equiv\frac{1}{8}\mathbf{R}\cdot\mathbf{R}\ ,\label{eq:E^2-H^2}\\
 & \mathbb{E}^{\alpha\gamma}\mathbb{H}_{\alpha\gamma}=\frac{1}{16}R_{\alpha\beta\gamma\delta}\star\! R^{\alpha\beta\gamma\delta}\equiv\frac{1}{16}\star\!\mathbf{R}\cdot\mathbf{R}\ .\label{eq:EH}
\end{align}
Note that, in spite of the dependence of $(\mathbb{E}^{u})_{\alpha\beta}$
and $(\mathbb{H}^{u})_{\alpha\beta}$ on the observer 4-velocity $u^{\alpha}$,
the combinations (\ref{eq:E^2-H^2})-(\ref{eq:EH}) are independent
of $u^{\alpha}$ (for this reason we dropped the $u$ superscript
therein).

There is an analogy (a purely \emph{formal} one, cf.~\cite{PaperAnalogies})
with the decomposition of the Maxwell tensor in electric and magnetic
parts \cite{matte,bel,Maartens:1997fg}, and the invariants they form,
which is illuminating for the problem at hand. With respect to a unit
timelike 4-vector $u^{\alpha}$, the Maxwell tensor (6 independent
components) splits irreducibly into the two spatial vectors (3 independent
components each) $(E^{u})^{\alpha}\equiv F^{\alpha\beta}u_{\beta}$
and $(B^{u})^{\alpha}\equiv\star F^{\alpha\beta}u_{\beta}$, as can
be seen from the explicit decomposition (\ref{eq:Fdecomp}), analogous
to (\ref{eq:Weyldecomp}). The fields $(E^{u})^{\alpha}$ and $(B^{u})^{\alpha}$
are covariant definitions for, respectively, the electric and magnetic
fields as measured by an observer of 4-velocity $u^{\alpha}$. In
spite of their $u^{\alpha}$ dependence, combining them one can construct
the two quadratic scalar invariants (e.g.~\cite{matte,bel,Stephani}),
\begin{align}
 & E^{\alpha}E_{\alpha}-B^{\alpha}B_{\alpha}=-\frac{1}{2}F_{\alpha\beta}F^{\alpha\beta}\equiv-\frac{1}{2}\mathbf{F}\cdot\mathbf{F}\ ,\label{eq:E^2 -B^2}\\
 & E^{\alpha}B_{\alpha}=-\frac{1}{4}F_{\alpha\beta}\star F^{\alpha\beta}\equiv-\frac{1}{4}\star\!\mathbf{F}\cdot\mathbf{F}\ ,\label{eq:E.B}
\end{align}
(where again we dropped the $u$ superscripts in $(E^{u})^{\alpha}$
and $(B^{u})^{\alpha}$) formally similar to the quadratic invariants
(\ref{eq:E^2-H^2})-(\ref{eq:EH}). These are actually the only two%
\footnote{This contrasts with the gravitational case, where (\ref{eq:E^2-H^2})-(\ref{eq:EH})
are not the only algebraically independent invariants one can construct
from $R_{\alpha\beta\gamma\delta}$. In vacuum (the simplest case),
they reduce to four, two cubic invariants existing in addition to
the quadratic invariants (\ref{eq:E^2-H^2})-(\ref{eq:EH}), see e.g.~\cite{bel,WyllePRD}.%
} independent scalar invariants one can construct from $F_{\alpha\beta}$.
They have the following interpretation~\cite{Stephani,LandauLifshitz,PaperInvariantes}:
i) if $E^{\alpha}B_{\alpha}\ne0$ then the electric $E^{\alpha}$
and magnetic $B^{\alpha}$ fields are both non-vanishing for all observers;
ii) if $E^{\alpha}E_{\alpha}-B^{\alpha}B_{\alpha}>0$ ($<0$) and
$E^{\alpha}B_{\alpha}=0$, then there are observers for which $B^{\alpha}$
($E^{\alpha}$) is zero.

In the gravitational case, it turns out%
\footnote{We thank L. Wylleman for his input on this issue.%
} (cf.~\cite{McIntosh et al 1994,WyllePRD,PaperInvariantes}) that,
for Petrov type D spacetimes (case of the examples studied below),
and in vacuum, one obtains formally equivalent statements to (i)-(ii)
above, replacing $\mathbf{F}$ by $\mathbf{R}$. That is: i) $\mathbf{\star R\cdot R}\ne0\,\Rightarrow$
$\mathbb{E}_{\alpha\gamma}$ and $\mathbb{H}_{\alpha\gamma}$ are
both non-vanishing for all observers; ii) $\mathbf{\star R\cdot R}=0$,
$\mathbf{R\cdot R}>0$ ($<0$) $\Rightarrow$ there are observers
for which $\mathbb{H}_{\alpha\gamma}$ ($\mathbb{E}_{\alpha\gamma}$)
vanishes. When, \emph{at a given point}, observers exist for which
$\mathbb{H}_{\alpha\gamma}=0$ ($\mathbb{E}_{\alpha\gamma}=0$), the
curvature tensor is dubbed ``purely electric'' (``purely magnetic'')
\emph{at that point}, see e.g.~\cite{WyllePRD,McIntosh et al 1994,BarnesConjecture,VBergh2003}.
Further details and comments on this classification (for general spacetimes),
will be given in~\cite{PaperInvariantes}. The velocity fields for
which $\mathbb{H}_{\alpha\gamma}=0$ will be exemplified below in
gravitational fields --- Schwarzschild and Kerr spacetimes --- with
a clear electromagnetic analogue --- a static point charge and a spinning
charge, respectively --- and we shall see that indeed $\mathbb{H}_{\alpha\gamma}$,
and therefore $F_{{\rm G}}^{\alpha}$, may vanish for moving spinning
particles, which contrasts with the electromagnetic analogue.

\subsection{Radial motion in Schwarzschild spacetime\label{sub:Radial-Schwa}}

\begin{figure}
\includegraphics[width=1\columnwidth]{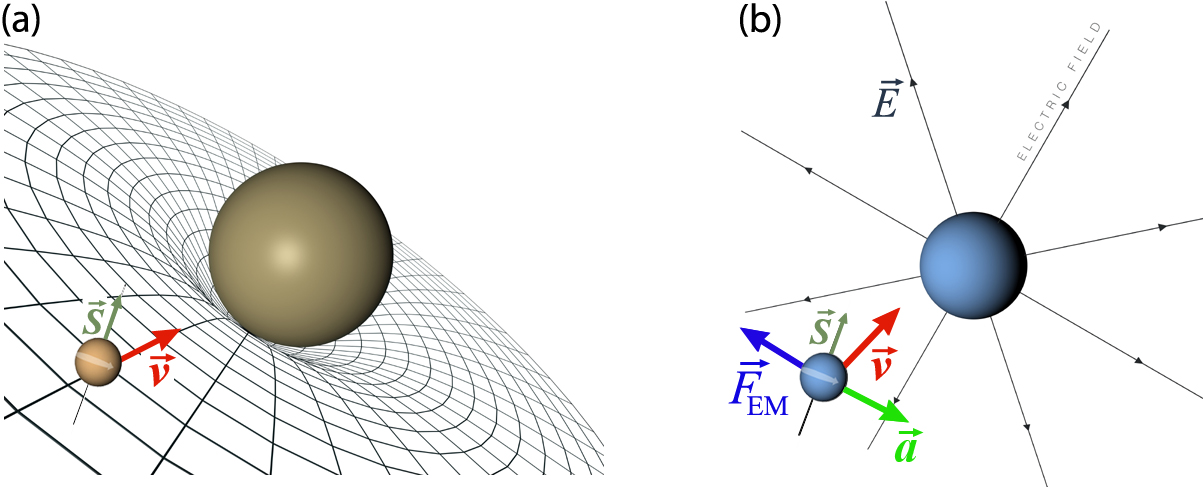}

\caption{\label{fig:Radial}An illustration of the physical consequences of
the different symmetries of the tidal tensors. A gyroscope dropped
from rest in Schwarzschild spacetime will move radially along a geodesic
towards the source, with no force exerted on it. A magnetic dipole
in (initially) radial motion in a Coulomb field, by contrast, feels
a force. Due to the hidden momentum, the force is approximately \emph{opposite}
to the acceleration!}
\end{figure}

The Schwarzschild spacetime is a Petrov type D solution whose quadratic
curvature invariants read 
\begin{equation}
\mathbb{E}^{\alpha\gamma}\mathbb{E}_{\alpha\gamma}-\mathbb{H}^{\alpha\gamma}\mathbb{H}_{\alpha\gamma}=\frac{6M^{2}}{r^{6}}\ ,\qquad\mathbb{E}^{\alpha\gamma}\mathbb{H}_{\alpha\gamma}=0\ .\label{eq:InvariantsSchw}
\end{equation}
In accordance with the classification above, this means that this
is a \emph{purely electric} spacetime, i.e., everywhere there are
observers for which $\mathbb{H}_{\alpha\beta}=0$. Let us find such
observers. The non-zero components of the gravitomagnetic tidal tensor
$\mathbb{H}_{\alpha\beta}\equiv\star R_{\alpha\mu\beta\nu}U^{\mu}U^{\nu}$
seen by an observer of arbitrary 4-velocity $U^{\alpha}=(U^{t},U^{r},U^{\theta},U^{\phi})$,
are, in Schwarzschild coordinates, given\emph{ }by ($\alpha\equiv3M\sin\theta/r$)
\begin{equation}
\begin{array}{c}
\mathbb{H}_{r\theta}=\alpha U^{\phi}U^{t}\ ;\qquad\mathbb{H}_{r\phi}=\alpha U^{t}U^{\theta}\ ;\\
\mathbb{H}_{\theta t}=-\alpha U^{\phi}U^{r}\ ;\qquad\mathbb{H}_{\phi t}=\alpha U^{r}U^{\theta}\ .
\end{array}\label{eq:HabSchwa}
\end{equation}
The condition $\mathbb{H}_{\alpha\beta}=0$ implies $U^{\phi}=U^{\theta}=0$,
whilst leaving $U^{r}$ arbitrary. Thus, observers at rest, or in
radial motion, measure a vanishing $\mathbb{H}_{\alpha\beta}$. Since,
according to Eqs.~(\ref{tab:Analogy}.1) of Table \ref{tab:Analogy},
it is the gravitomagnetic tidal tensor, \emph{as seen by the test
particle}, that determines the force on it, this means that no force
is exerted on a gyroscope at rest or in radial motion: 
\[
F_{{\rm G}}^{\alpha}=-\mathbb{H}^{\beta\alpha}S_{\beta}=0\ ,
\]
i.e., it moves along a geodesic (it is the trivial solution of the
equations of motion with the Mathisson-Pirani condition, see Appendix
\ref{sub:Comparison-of-the-SSC-aplications}), regardless of its spin
$\vec{S}$. For instance, a gyroscope dropped from rest will fall
towards the singularity along a radial geodesic just like a monopole
particle, see Fig. \ref{fig:Radial}(a).

This is not possible in the electromagnetic analogue, due to the symmetries
of $B_{\alpha\beta}$. Consider a magnetic dipole, of 4-velocity $U^{\alpha}$,
in the field of a static point charge $Q$; the force exerted on it
is $F_{{\rm EM}}^{\alpha}=B^{\beta\alpha}\mu_{\beta}$, cf.~Table
\ref{tab:Analogy}, where $B_{\alpha\beta}=\star F_{\alpha\mu;\beta}U^{\mu}$
is the magnetic tidal tensor as seen by the particle. The components
of $B_{\alpha\beta}$, for a generic $U^{\alpha}$, are ($\alpha\equiv3Q\sin\theta/r$)
\begin{equation}
\begin{array}{c}
B_{r\theta}=\alpha U^{\phi};\qquad B_{\theta r}=\alpha2U^{\phi};\qquad B_{r\phi}=-\alpha U^{\theta};\\
B_{\phi r}=-2\alpha U^{\theta};\qquad B_{\theta\phi}=\alpha U^{r};\qquad B_{\phi\theta}=-\alpha U^{r}.
\end{array}\label{eq:BabCoulomb}
\end{equation}
The static observers $U^{i}=0$ are the only ones measuring $B_{\alpha\beta}=0$,
as expected from Eq.~(\ref{eq:VacuumBab}), since the field is inhomogeneous
and therefore not covariantly constant for a moving observer (i.e.,
$\star F_{\alpha\mu;\beta}U^{\mu}\ne0$ if $U^{i}\ne0$). For a radial
velocity $U^{\alpha}=(U^{t},U^{r},0,0)$, the magnetic tidal tensor
reduces to its antisymmetric part, $B_{\alpha\beta}=B_{[\alpha\beta]}$,
with non-vanishing components $B_{\theta\phi}=-B_{\phi\theta}=\alpha U^{r}$.
This means that (except for the special case $\vec{v}\parallel\vec{\mu}$)
a force%
\footnote{The force (\ref{eq:ForceCoulombRadial}) may seem at first sight to
contradict what one might naively expect from the textbook expression
$F_{{\rm EM}}^{i}=-\nabla^{i}(\vec{B}\cdot\vec{\mu})\equiv B^{j;i}\mu_{j}$,
which holds in the particle's momentarily comoving inertial frame,
because the radially moving dipole indeed sees a vanishing magnetic
field \textbf{$\vec{B}$}. However \emph{its curl is non zero} (implying
$B_{i;j}=B_{ij}\ne0$, cf.~Eq. (\ref{eq:Eab-Ea;b})), which, taking
the perspective of such frame, is induced by the time-varying electric
field, by virtue of vacuum equation $\nabla\times\vec{B}=\partial\vec{E}/\partial t$.%
} will be exerted on a magnetic dipole in (initially) radial motion:
\begin{equation}
F_{{\rm EM}}^{0}=0;\ \ F_{{\rm EM}}^{i}=B^{[\alpha i]}\mu_{\alpha}=\ \frac{\gamma Q}{r^{3}}(\vec{v}\times\vec{\mu})^{i}\ ,\label{eq:ForceCoulombRadial}
\end{equation}
where $\vec{v}=\vec{U}/\gamma$ and $\gamma$ is the Lorentz factor.
This force comes \emph{entirely} from the antisymmetric part of $B_{\alpha\beta}$;
it is then natural, given the symmetry of $\mathbb{H}_{\alpha\beta}$
in vacuum, that it has no gravitational counterpart.

It is however important to note that, due the hidden momentum that
the spinning particle possesses, the relation between this force and
the particle's center of mass acceleration is not straightforward.
This is manifest in Eq.~(\ref{eq:mavaccum}); for flat spacetime,
and a particle whose only electromagnetic moment is $\mu^{\alpha}$
($q=0$), it reads: 
\begin{eqnarray*}
m_{0}a^{\alpha} & = & B^{\alpha\beta}\mu_{\beta}+\epsilon_{\ \beta\gamma\delta}^{\alpha}\frac{D}{d\tau}U^{\delta}(S^{\beta}a^{\gamma})-\star F_{\beta}^{\ \alpha}\frac{D\mu^{\beta}}{d\tau}\ .
\end{eqnarray*}
The last term vanishes if one assumes $\mu^{\alpha}=\sigma S^{\alpha}$,
since: $B^{\alpha}(U)=0$ for radial motion; thus, from Eq.~(\ref{eq:EqSpinVector}),
$D\mu_{\mu}/d\tau=\sigma S_{\nu}a^{\nu}U_{\mu}\ $ and $\star F_{\beta}^{\ \alpha}D\mu^{\beta}/d\tau=-\sigma B(U)^{\alpha}S_{\nu}a^{\nu}=0$.
The second term can also be taken to a good approximation as being
zero (which, as explained in Sec.~\ref{sub:Weak-field-regime:},
to an accuracy of order $\mathcal{O}(S^{2})$, amounts to say that
we pick the ``non-helical'' solution allowed by the Mathisson-Pirani
condition). Therefore, since, in this application, $B_{(\alpha\beta)}=0$,
we are led to the conclusion that $m_{0}a^{\alpha}\approx B^{\alpha\beta}\mu_{\beta}=-B^{\beta\alpha}\mu_{\beta}=-F_{{\rm EM}}^{\alpha}$!
{[}See Fig. \ref{fig:Radial}(b).{]} This clearly shows how careful
one must be with the notion of force (understood as $F^{\alpha}=DP^{\alpha}/d\tau$,
with $P^{\alpha}$ defined in the usual way by Eq.~(\ref{eq:Pgeneral})),
because it can significantly differ from $ma^{\alpha}$ when the particle
has hidden momentum.

Finally, it is worth mentioning that the vanishing of $\mathbb{H}_{\alpha\beta}$
for certain velocity fields in the Schwarzschild spacetime is analogous
instead to the vanishing of the magnetic field $B^{\alpha}$ (not
the tidal tensor $B_{\alpha\beta}$) in a Coulomb field. The quadratic
invariants of $F^{\alpha\beta}$ have a structure formally analogous
to the curvature invariants (\ref{eq:InvariantsSchw}): $E^{2}-B^{2}=Q^{2}/r^{4}$,
$E^{\alpha}B_{\alpha}=0$, telling us that there are everywhere observers
for which $B^{\alpha}=0$. For an arbitrary $U^{\alpha}$, the non-vanishing
components of $B^{\alpha}$ are 
\[
B^{\theta}=-\frac{QU^{\phi}}{r^{2}}\sin\theta\ ,\qquad B^{\phi}=\frac{QU^{\theta}}{r^{2}}\csc\theta\ ;
\]
therefore, observers at rest or in purely radial motion measure $B^{\alpha}=0$,
just like with the case of $\mathbb{H}_{\alpha\beta}$ in the Schwarzschild
spacetime. One should however bear in mind that this one is a \emph{purely
formal} analogy, as the parallelism drawn is between gravitational
tidal tensors and electromagnetic fields. The physical effects are
very different: the vanishing of $\mathbb{H}_{\alpha\beta}$ for radial
velocities means that a gyroscope feels no force, whereas the vanishing
of $B^{\alpha}$ does \emph{not} mean that dipoles moving radially
feel no force (which they do, as discussed above), but instead that
they do not undergo Larmor precession ($D\vec{S}/d\tau=0$ in the
comoving frame, cf.~Eq.~(\ref{eq:EqSpinVector})).

\subsection{Equatorial motion in Kerr and Kerr-dS spacetimes\label{sub:EquatorialKerr}}

In this section we compare the motions of gyroscopes in the Kerr and
Kerr-de Sitter spacetimes to magnetic dipoles in the field of a spinning
charge. It is shown that in the equatorial plane there are observers
for which the gravitomagnetic tidal tensor $\mathbb{H}_{\alpha\beta}$
vanishes (i.e., gyroscopes moving with such velocities feel no force),
and that consequently circular geodesics for gyroscopes even exist
in Kerr-dS (independently of the particle's spin). This contrasts
with the electromagnetic system, where observers for which $B_{\alpha\beta}=0$
do not exist at all (consequence of the symmetries of $B_{\alpha\beta}$,
i.e., the laws of electromagnetic induction, as explained above),
and therefore (except for special orientations of $\vec{\mu}$) a
force is always exerted on a magnetic dipole, regardless of its motion.

The vanishing of $\mathbb{H}_{\alpha\beta}$ is instead analogous
to the vanishing of the magnetic \emph{field} $B^{\alpha}$, which
likewise occurs in the equatorial plane, for asymptotically similar
velocity fields. That gives useful insight into the gravitational
problem; for this reason we shall start by the simpler electromagnetic
case.

\subsubsection{A magnetic dipole in the field of a spinning charge.\label{sub:A-magnetic-dipole in spinning charge} }

\begin{figure}
\includegraphics[width=1\columnwidth]{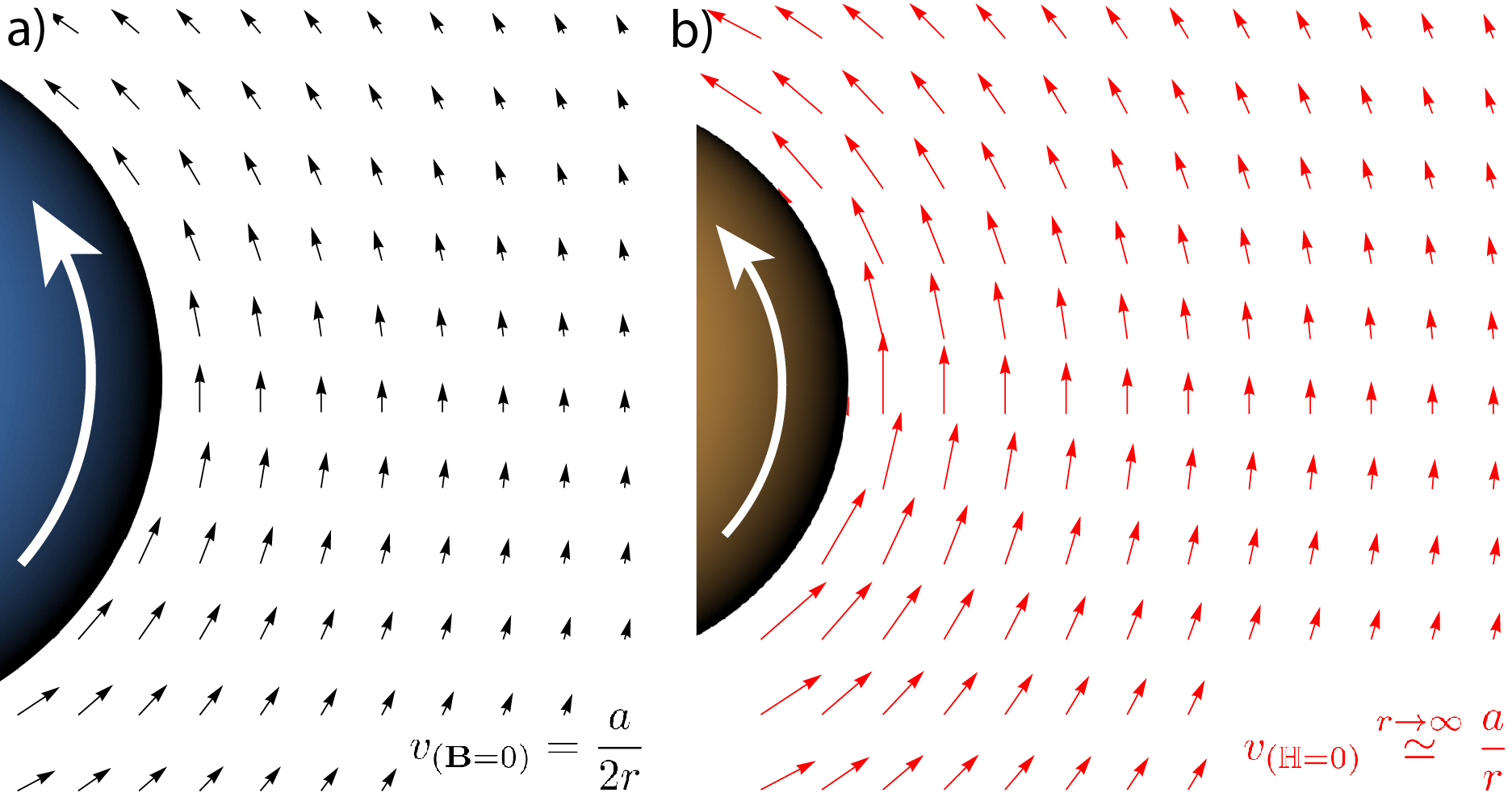}

\caption{\label{fig:EquaKerr}a) Velocity field $\vec{v}_{(\mathbf{B}=0)}$,
which makes the magnetic\emph{ field} $B^{\alpha}$ vanish in the
equatorial plane of a spinning charge; magnetic dipoles with such
velocities do not undergo Larmor \emph{precession}. b) Velocity field
$\vec{v}_{(\mathbb{H}=0)}$ for which the gravitomagnetic\emph{ tidal
tensor} $\mathbb{H}_{\alpha\beta}$ vanishes in the equatorial plane
of Kerr spacetime; gyroscopes moving with such velocities feel \emph{no
force}, $F_{{\rm G}}^{\alpha}=0$. If $\Lambda>0$ (Kerr-dS spacetime),
circular geodesics for gyroscopes even exist (Sec.~\ref{sub:Kerr-dS}).
Asymptotically, $\vec{v}_{(\mathbb{H}=0)}$ and $\vec{v}_{(\mathbf{B}=0)}$
match up to a factor of 2. The velocity $\vec{v}_{(\mathbb{H}=0)}$
however has no \emph{physical} electromagnetic analogue: due to the
laws of electromagnetic induction, for a moving dipole $B_{[\alpha\beta]}\ne0\Rightarrow B_{\alpha\beta}\ne0$
\emph{always}, generically implying $F_{{\rm EM}}^{\alpha}\ne0$.}
\end{figure}

\emph{Velocity field for which} $B^{\alpha}=0$. --- We start by the
electromagnetic system, which will serve as a guide for the gravitational
case. The electromagnetic field produced by a spinning charge (magnetic
moment $\vec{\mu}_{{\rm s}}$) is described by the 4-potential $A^{\alpha}=(\phi,\vec{A})$:
\begin{equation}
\phi=\frac{Q}{r}\ ,\ \ \ \vec{A}=\frac{\vec{\mu}_{{\rm s}}\times\vec{r}}{r^{3}}=\frac{\mu_{{\rm s}}}{r^{3}}\vec{e}_{\phi}\ .\label{EMsphere}
\end{equation}
The invariant structure for this electromagnetic field is 
\begin{equation}
\left\{ \begin{array}{l}
{\displaystyle \vec{E}^{2}-\vec{B}^{2}=\frac{Q^{2}}{r^{4}}-\frac{\mu_{{\rm s}}^{2}(5+3\cos2\theta)}{2r^{6}}>0\ ,}\\
\\
\vec{E}\cdot\vec{B}={\displaystyle \frac{2\mu_{{\rm s}}Q\cos\theta}{r^{5}}\,(=0\mbox{ in the equatorial plane)}\ ,}
\end{array}\right.\label{eq:InvSphere}
\end{equation}
the first inequality always holding assuming the classical gyromagnetic
ratio $\mu_{{\rm s}}/J=Q/2M$ (corresponding to a source in which
the charge and mass are identically distributed). Expressions (\ref{eq:InvSphere})
tell us that in the equatorial plane $\theta=\pi/2$ there are observers
that measure $B^{\alpha}$ to be zero (since $\vec{E}\cdot\vec{B}=0$
and $\vec{E}^{2}-\vec{B}^{2}>0$ therein). It is straightforward to
obtain the 4-velocity of such observers. The magnetic field $B^{\alpha}=\star F^{\alpha\beta}U_{\beta}$
seen by an arbitrary observer of 4-velocity $U^{\alpha}=(U^{t},U^{r},U^{\theta},U^{\phi})$
is given by: 
\begin{eqnarray*}
B^{r} & = & \frac{2\mu_{{\rm s}}\cos\theta}{r^{3}}U^{t}\ ,\qquad B^{\theta}=\left(\frac{\mu_{{\rm s}}U^{t}}{r^{4}}-\frac{U^{\phi}Q}{r^{2}}\right)\sin\theta\ ,\\
B^{\phi} & = & \frac{QU^{\theta}}{r^{2}\sin\theta}\ ,\qquad B^{t}=\frac{\mu_{{\rm s}}}{r^{3}}\left(2U^{r}\cos\theta+rU^{\theta}\sin\theta\right)\ .
\end{eqnarray*}
Thus, the condition $B^{r}=0$ implies $\theta=\pi/2$ (i.e., equatorial
plane, as expected); in the equatorial plane, $B^{t}=0$ implies $U^{\theta}=0$,
and $B^{\theta}=0$ implies 
\begin{equation}
\frac{d\phi}{dt}=\frac{U^{\phi}}{U^{t}}=\frac{\mu_{{\rm s}}}{Qr^{2}}=\frac{J}{2Mr^{2}}\equiv\omega_{(\mathbf{B}=0)}\ \label{eq:EMvplot}
\end{equation}
where \emph{in the third} equality again we assumed $\mu_{{\rm s}}/J=Q/2M$.
Therefore, observers with angular velocity (\ref{eq:EMvplot}) measure
a vanishing magnetic field in the equatorial plane. No restriction
is imposed on the radial component of the velocity, apart from the
normalization condition $U^{\alpha}U_{\alpha}=-1$. The velocity field
corresponding to the case $U^{r}=0$ is plotted in Fig.~\ref{fig:EquaKerr}a).
The vanishing of $B^{\alpha}$ for these observers comes from an exact
cancellation between the magnetic field generated by the relative
translation of the source and the field produced by its rotation.
It means that a magnetic dipole possessing a velocity of the form
(\ref{eq:EMvplot}) does not undergo Larmor precession, since the
second term of Eq.~(\ref{eq:EqSpinVector}) vanishes.

In \cite{EPAPS} we investigate the corresponding gravitational problem,
i.e., if there are boost velocities for which gyroscopes in the Kerr
spacetime do not ``precess''.

$B_{\alpha\beta}$ \emph{never vanishes.}--- The force (\ref{tab:Analogy}.1a)
exerted on the dipole, however, does \emph{not} vanish, as it is only
the magnetic field $B^{\alpha}$, not the tidal tensor $B_{\alpha\beta}$,
that vanishes for the velocity fields of the type (\ref{eq:EMvplot}).
As measured by a generic observer $U^{\alpha}$, \textbf{$B_{\alpha\beta}$}
has the following components in the equatorial plane 
\begin{align}
 & B_{r\theta}=\alpha(r^{2}QU^{\phi}-3\mu_{{\rm s}}U^{t})\ ;\quad B_{\theta r}=\alpha(2r^{2}QU^{\phi}-3\mu_{{\rm s}}U^{t})\ ;\nonumber \\
 & B_{r\phi}=-\alpha Qr^{2}U^{\theta}\ ;\qquad B_{\phi r}=-2\alpha Qr^{2}U^{\theta}\ ;\nonumber \\
 & B_{r\phi}=-\alpha Qr^{2}U^{\theta}\ ;\qquad B_{\phi r}=-2\alpha Qr^{2}U^{\theta}\ ;\label{eq:BabScharge}\\
 & B_{tr}=3\alpha\mu_{{\rm s}}U^{\theta}\ ;\qquad B_{t\theta}=3\alpha\mu_{{\rm s}}U^{r}\ ,\nonumber 
\end{align}
with $\alpha\equiv1/r^{3}$. Thus we see that in order to make $B_{(\alpha\beta)}$
vanish, we must have $U^{\theta}=U^{r}=0$ and 
\begin{equation}
\frac{d\phi}{dt}=\frac{U^{\phi}}{U^{t}}=2\frac{\mu_{{\rm s}}}{Qr^{2}}=\frac{J}{Mr^{2}}\equiv\omega_{(B_{(\alpha\beta)}=0)}\label{eq:vnoBabSym}
\end{equation}
(differs from a factor of 2 from the angular velocity (\ref{eq:EMvplot})
which makes $B^{\alpha}$ vanish; the second equality again assumes
$\mu_{{\rm s}}/J=Q/2M$). However, $B_{[\alpha\beta]}$ only vanishes
if $\vec{v}=0$; hence it is not possible to find any observer for
which $B_{\alpha\beta}=B_{(\alpha\beta)}+B_{[\alpha\beta]}=0$. Again,
the fact that $B_{[\alpha\beta]}$ cannot vanish for a moving observer
is a direct consequence of Maxwell's equations, or the laws of electromagnetic
induction: a dipole moving relative to the spinning charge always
sees a varying electromagnetic field; that endows $B_{\alpha\beta}$
with an antisymmetric part, by virtue (from the point of view of a
momentarily comoving inertial frame) of the vacuum equation $\nabla\times\vec{B}=\partial E/\partial t$,
or, covariantly, by Eq.~(\ref{tab:Analogy}.3a). Note that this is
true even if one considers a dipole in a circular equatorial trajectory
around the central source: $D\!\star\! F_{\alpha\beta}/d\tau=2B_{[\alpha\beta]}\ne0$
along such worldline, which is due to the variation of the electric
field along the curve (it is constant in magnitude, but varying in
direction). 

\subsubsection{A gyroscope in Kerr spacetime\label{sub:A-gyroscope-in-Kerr}}

\emph{Velocity for which} $\mathbb{H}_{\alpha\beta}=0$.--- From what
we learned in the electromagnetic problem, we expect the existence
of observers for which $\mathbb{H}_{\alpha\beta}$ vanishes, based
on two observations. First, we have seen that in the equatorial plane
of the electromagnetic system there are velocities (\ref{eq:vnoBabSym})
for which the magnetic tidal tensor reduces to its antisymmetric part
$B_{\alpha\beta}=B_{[\alpha\beta]}$; since the gravitomagnetic tidal
tensor is symmetric in vacuum: $\mathbb{H}_{\alpha\beta}=\mathbb{H}_{(\alpha\beta)}$,
it is natural to expect, in the spirit of the analogy, that $\mathbb{H}_{\alpha\beta}=0$
in the corresponding gravitational setup. Secondly, there is a close
analogy between the invariants of the two systems. The Kerr spacetime
is of Petrov type D, hence a classification for the curvature tensor
based on quadratic invariants formally analogous to the one for $F_{\alpha\beta}$
applies, as discussed in Sec. \ref{sub:Symmetries}. The invariants
(\ref{eq:E^2-H^2})-(\ref{eq:EH}) read (e.g.~\cite{cherubini:02})
\begin{align}
\mathbf{R}\cdot\mathbf{R} & =\frac{48M^{2}}{\Sigma^{6}}(a^{4}\cos^{4}\theta-14a^{2}r^{2}\cos^{2}\theta+r^{4})\ \cdot\nonumber \\
 & \ \ \ (r^{2}-a^{2}\cos^{2}\theta)\nonumber \\
\star\mathbf{R}\cdot\mathbf{R} & =\frac{96M^{2}ra}{\Sigma^{6}}(a^{2}\cos^{2}\theta-3r^{2})(3a^{2}\cos^{2}\theta-r^{2})\cos\theta\ ,\label{eq:*R.R_Kerr}
\end{align}
where $\Sigma\equiv r^{2}+a^{2}\cos^{2}\theta$. For large $r$ we
have the structure:\textcolor{black}{{} 
\[
\left\{ \begin{array}{l}
{\displaystyle \mathbb{E}^{\alpha\gamma}\mathbb{E}_{\alpha\gamma}-\mathbb{H}^{\alpha\gamma}\mathbb{H}_{\alpha\gamma}\stackrel{r\rightarrow\infty}{\simeq}\frac{6M^{2}}{r^{6}}>0}\ ,\\
\\
\mathbb{E}^{\alpha\gamma}\mathbb{H}_{\alpha\gamma}\stackrel{r\rightarrow\infty}{\simeq}{\displaystyle \frac{18JM\cos\theta}{r^{7}}\ (=0\mbox{ \ in the equat. plane})\ ,}
\end{array}\right.
\]
}\textcolor{black}{\emph{formally}} \textcolor{black}{analogous to
its electromagnetic counterpart (\ref{eq:InvSphere}). Note in particular
that the result $\mathbb{E}^{\alpha\gamma}\mathbb{H}_{\alpha\gamma}=0$
for the equatorial plane ($\theta=\pi/2$) is} \textcolor{black}{\emph{exact}}\textcolor{black}{,
cf.~Eq.~(\ref{eq:*R.R_Kerr}). Since $\mathbf{R}\cdot\mathbf{R}>0$
therein, this means that} \textcolor{black}{\emph{in the equatorial
plane there are observers for which $\mathbb{H}_{\alpha\beta}$ vanishes,}}\textcolor{black}{{}
in analogy with the vanishing of $B^{\alpha}$ in the equatorial plane
of the field of a spinning charge.} It is straightforward to determine
the 4-velocity of such observers. In the equatorial plane, the non-zero
components of the gravitomagnetic tidal tensor $\mathbb{H}_{\alpha\beta}\equiv\star R_{\alpha\mu\beta\nu}U^{\mu}U^{\nu}$
seen by an arbitrary observer of 4-velocity $U^{\alpha}=(U^{t},U^{r},U^{\theta},U^{\phi})$,
are given (exactly) by 
\begin{eqnarray}
\mathbb{H}_{r\theta} & = & \alpha\left[\left(2a^{2}+r^{2}\right)U^{\phi}U^{t}-a\left(a^{2}+r^{2}\right)\left(U^{\phi}\right)^{2}-a\left(U^{t}\right)^{2}\right]\nonumber \\
\mathbb{H}_{r\phi} & = & \alpha\left(a^{2}+r^{2}\right)\left(aU^{\phi}-U^{t}\right)U^{\theta}\ ,\nonumber \\
\mathbb{H}_{rt} & = & \alpha a\left(aU^{\phi}-U^{t}\right)U^{\theta}\ ,\label{eq:HabKerr}\\
\mathbb{H}_{\theta\phi} & = & \alpha a\left[\left(a^{2}+r^{2}\right)U^{\phi}-aU^{t}\right]U^{r}\ ,\nonumber \\
\mathbb{H}_{\theta t} & = & -\alpha\left[\left(a^{2}+r^{2}\right)U^{\phi}-aU^{t}\right]U^{r}\ ,\nonumber \\
\mathbb{H}_{\phi\phi} & = & -2\alpha aU^{r}U^{\theta}\ =\ \mathbb{H}_{tt}\ ,\quad\mathbb{H}_{\phi t}=\alpha\left(2a^{2}+r^{2}\right)U^{r}U^{\theta}\ ,\nonumber 
\end{eqnarray}
where $\alpha\equiv3M/r^{3}$. It is easily seen that in order to
make all components vanish we must have $U^{\theta}=0$ (i.e.~the
observer must move in the equatorial plane, as expected and in analogy
with the electromagnetic case above) and 
\begin{equation}
\frac{d\phi}{dt}=\frac{U^{\phi}}{U^{t}}=\frac{a}{a^{2}+r^{2}}\equiv\omega_{(\mathbb{H}=0)}\ .\label{velocityKerr}
\end{equation}
Thus, observers with angular velocity $\omega=\omega_{(\mathbb{H}=0)}$
measure a vanishing gravitomagnetic tidal tensor in the equatorial
plane. Again, no restriction is imposed on $U^{r}$, apart from the
normalization condition $U^{\alpha}U_{\alpha}=-1$. The velocity field
corresponding to the case $U^{r}=0$ is plotted in Fig.~\ref{fig:EquaKerr}b).
It is interesting to note that, asymptotically, $\omega_{(\mathbb{H}=0)}$
matches the angular velocity (\ref{eq:vnoBabSym}) for which the \emph{symmetric
part} of the magnetic tidal tensor $B_{\alpha\beta}$ vanishes in
the electromagnetic analogue (and, up to a factor of 2, the angular
velocity (\ref{eq:EMvplot}) for which $B^{\alpha}$ vanishes).

As discussed above, $\omega_{(\mathbb{H}=0)}$ has no electromagnetic
counterpart; the magnetic tidal tensor $B_{\alpha\beta}$ can never
vanish for a moving observer, due to Eq.~(\ref{eq:VacuumBab}), i.e.,
the laws of electromagnetic induction. We have thus here another illustration
of the physical consequences of the different symmetries of $\mathbb{H}_{\alpha\beta}$
as compared to $B_{\alpha\beta}$, signaling the absence of electromagnetic-like
induction effects in the physical gravitational forces. Note that
these differences are manifest even in the weak field and slow motion
regime, since taking the field to be weak (either by going far away
from the source, or by taking $a$ to be small) only amounts to making
the velocity for which $F_{{\rm G}}^{\alpha}$ vanishes smaller, since
$|v|\approx a/r$. That illustrates how misleading the usual treatments
in the literature on ``gravitoelectromagnetism'' in the framework
of the linearized theory (e.g.~\cite{Ruggiero:2002hz,Gravitation and Inertia})
can be, naively casting the force on a gyroscope as an expression
of the type $\vec{F}_{{\rm G}}=K\nabla(\vec{S}\cdot\vec{H})$ (similar
to the electromagnetic force on a magnetic dipole). This regime is
studied in detail in Sec.~\ref{sub:Weak-field-regime:}.

Finally, it is interesting to note that the angular velocity (\ref{velocityKerr})
appeared before in apparently unrelated contexts; it coincides%
\footnote{We thank O. Semerák for pointing this out to us.%
} with the angular velocity of the ``Carter canonical observers''
(e.g.~\cite{SemerakStationaryFrames}), which are observers that
measure the photons of the principal null congruences (see p. 902
of \cite{Gravitation}) to be in purely radial motion. It also appeared
in a recent paper \cite{VelocityKerr}, Eq.~(30) therein, where it
is shown that the Kerr metric can be obtained by a rescaling of an
orthonormal tetrad field in Minkowski space, constructed from spheroidal
coordinates in differential rotation, each spheroidal shell $r=constant$
rotating rigidly with an angular velocity that is precisely $\omega_{(\mathbb{H}=0)}$.

\emph{No circular geodesics for spinning material particles in Kerr
spacetime }--- the vanishing of $F_{{\rm G}}^{\alpha}$ for gyroscopes
moving with angular velocity (\ref{velocityKerr}) makes one wonder
if a spinning particle can move along circular geodesics around a
Kerr black hole, which we shall now check. Equation~(\ref{velocityKerr})
corresponds to \emph{prograde motion}; the angular velocity of prograde
circular geodesics reads (e.g.~\cite{RindlerPerlick}): 
\begin{equation}
\omega_{{\rm geo}}\equiv\frac{U_{{\rm geo}}^{\phi}}{U_{{\rm geo}}^{t}}=\frac{1}{a+\sqrt{\frac{r^{3}}{M}}}\,.\label{eq:CircularKerr}
\end{equation}
Equating this expression to (\ref{velocityKerr}), we obtain $r=a^{2}/M$;
this solution, however, lies inside the horizon: since $r_{+}=M+\sqrt{M^{2}-a^{2}}$,
the condition $r\ge r_{+}$ implies 
\[
\frac{a^{2}}{M}\ge M+M\sqrt{1-a^{2}/M^{2}}\ \Leftrightarrow\ 1-A^{2}\le-\sqrt{1-A^{2}}
\]
where we defined the dimensionless parameter $A\equiv a/M$. Note
that $A=1$ is the extreme Kerr case, and $A>1$ corresponds to a
naked singularity; therefore (excluding the naked singularity scenario)
the circular orbit would exist only in the extreme case, it would
be precisely at the horizon, and thus it would be a null geodesic.
Otherwise, no circular geodesics exist with angular velocity (\ref{velocityKerr}),
and so $\mathbb{H}_{\alpha\beta}\ne0$ along any time-like circular
geodesic.

The only possibility of having $F_{{\rm G}}^{\alpha}=\mathbb{H}^{\beta\alpha}S_{\beta}=0$
would then be if $S^{\alpha}$ was an eigenvector of $\mathbb{H}_{\beta}^{\ \alpha}$
corresponding to a zero eigenvalue; that does not lead to circular
geodesics however, because $S^{\alpha}$ cannot remain an eigenvector.
For $U^{\alpha}=(U^{t},0,0,U^{\phi})$, the only eigenvectors of $\mathbb{H}_{\beta}^{\ \alpha}$
with zero eigenvalue are $U^{\alpha}$ and $\mathbf{e}_{\phi}$; $S^{\alpha}$
(orthogonal to $U^{\alpha}$) cannot remain in the eigenspace spanned
by $U^{\alpha}$ and $\mathbf{e}_{\phi}$, by virtue of the transport
law (\ref{eq:EqSpinVector}), which can be seen as follows. Consider
a frame rigidly rotating with an angular velocity $\omega_{{\rm geo}}$
corresponding to a geodesic at some value of $r$ (the associated
coordinates are obtained from the Boyer-Lindquist coordinates by the
transformation $t'=t$, $r'=r$, $\theta'=\theta$, $\phi'=\phi-\omega_{{\rm geo}}t$),
and the orthonormal basis $\mathbf{e}_{\hat{\alpha}'}$ tied to it,
such that $\mathbf{e}_{\hat{t}'}=\mathbf{U}$, and $\mathbf{e}_{\hat{r}'}$,
$\mathbf{e}_{\hat{\theta}'}$ $\mathbf{e}_{\hat{\phi}'}$ follow from
normalizing $\mathbf{e}_{r}$, $\mathbf{e}_{\theta}$ and $(\mathbf{h}^{U})\cdot\mathbf{e}_{\phi}$,
respectively. (Here $(\mathbf{h}^{U})$ is the projector orthogonal
to $U^{\alpha}$, cf. Sec. \ref{sub:Notation-and-conventions}). In
such frame the gyroscope's CM is at rest, therefore Eq. (\ref{eq:EqSpinVector})
applies, $dS^{\hat{i}}/d\tau=(\vec{S}\times\vec{\Omega})^{\hat{i}}$;
moreover the gravitomagnetic field $\vec{H}=2\vec{\Omega}$ takes
the very simple form $\vec{H}=-2\sqrt{M/r^{3}}\vec{e}_{\hat{\theta}'}$,
cf. Eq. (41) of \cite{RindlerPerlick}. Hence, for an initial $\vec{S}=S^{\hat{\phi'}}\vec{e}_{\hat{\phi}'}$,
we have $d\vec{S}/d\tau=S\Omega\vec{e}_{\hat{r}}$ and therefore $\vec{S}$
cannot remain parallel to $\mathbf{e}_{\hat{\phi}'}$ (thus $S^{\alpha}$
does not remain in the eigenspace of $U^{\alpha}$ and $\mathbf{e}_{\phi}$).
We then conclude that no circular geodesics for spinning classical
particles are possible in the Kerr spacetime.

\subsubsection{Circular geodesics in Kerr-dS spacetimes\label{sub:Kerr-dS}}

The failure to obtain circular geodesics for spinning material particles
in the previous section was due to the fact that the angular velocity
of circular geodesics in the Kerr spacetime dies off as $r^{-2/3}$,
whereas the angular velocity for which $\mathbb{H}_{\alpha\beta}=0$
dies off as $r^{-2}$; in other words, geodesics are ``too fast''.
But they should be possible in other spacetimes; in this spirit, Kerr-de
Sitter comes as natural candidate, since a repulsive $\Lambda$ should
``slow down'' the circular geodesics. This is indeed the case. In
Boyer-Lindquist coordinates, the metric takes the form (e.g.~\cite{StuchlikKovar})
\begin{eqnarray}
ds^{2} & = & -\frac{\Delta_{r}}{\chi^{2}\Sigma}\left(dt-a\sin^{2}\theta d\phi\right)^{2}+\frac{\Sigma}{\Delta_{r}}dr^{2}+\frac{\Sigma}{\Delta_{\theta}}d\theta^{2}\nonumber \\
 &  & +\frac{\Delta_{\theta}\sin^{2}\theta}{\chi^{2}\Sigma}\left[adt-(a^{2}+r^{2})d\phi\right]^{2}\,,\label{eq:kerrds}
\end{eqnarray}
where 
\[
\begin{array}{c}
{\displaystyle \Delta_{r}\equiv r^{2}-2Mr+a^{2}-\frac{\Lambda}{3}r^{2}(r^{2}+a^{2})}\ ;\quad{\displaystyle \chi\equiv1+\frac{\Lambda}{3}a^{2}}\ ;\\
{\displaystyle \Delta_{\theta}=1+\frac{\Lambda}{3}a^{2}\cos^{2}\theta}\ ;\qquad\Sigma\equiv r^{2}+a^{2}\cos^{2}\theta\ .
\end{array}
\]
Since $\Lambda\ne0\Rightarrow R_{\mu\nu}=\Lambda g_{\mu\nu}$, the
vacuum classification based on scalar invariants used in the previous
section does not apply herein to the Riemann tensor. However, a similar
classification holds for the Weyl tensor $C_{\alpha\beta\gamma\delta}$
(again, since it is of Petrov type D), see e.g.~\cite{WyllePRD}.
The relationship between $\mathbb{H}_{\alpha\beta}$ and the magnetic
part of the Weyl tensor, $\mathcal{H}_{\alpha\beta}\equiv\star C_{\alpha\mu\beta\nu}U^{\mu}U^{\nu}$,
can be obtained from the expression of $R_{\alpha\beta\gamma\delta}$
in terms of $C_{\alpha\beta\gamma\delta}$, e.g. Eq. (2) of \cite{cherubini:02};
it reads 
\[
\mathbb{H}_{\alpha\beta}=\mathcal{H}_{\alpha\beta}+\frac{1}{2}\epsilon_{\alpha\beta\sigma\gamma}U^{\gamma}R^{\sigma\lambda}U_{\lambda}\ .
\]
This tells us that, for this spacetime, $\mathbb{H}_{\alpha\beta}=\mathcal{H}_{\alpha\beta}$.
Therefore, solving for $\mathbb{H}_{\alpha\beta}=0$ amounts to solving
for $\mathcal{H}_{\alpha\beta}=0$, which reduces to the same procedure
of the previous section, but this time using the invariants of the
Weyl tensor. The invariants have a similar structure, similarly leading
to the conclusion that in the equatorial plane there are observers
for which $\mathcal{H}_{\alpha\beta}=\mathbb{H}_{\alpha\beta}=0$.
Actually, the gravitomagnetic tidal tensor for the metric~(\ref{eq:kerrds})
is obtained by simply multiplying expressions (\ref{eq:HabKerr})
by $9/(3+a^{2}\Lambda)^{2}$: 
\[
(\mathbb{H}_{{\rm Kerr-dS}})_{\alpha\beta}=\frac{9}{(3+a^{2}\Lambda)^{2}}(\mathbb{H}_{{\rm Kerr}})_{\alpha\beta}\ .
\]
Thus, the angular velocity of the observers for which $\mathbb{H}_{\alpha\beta}=0$
is given by the same Eq.~(\ref{velocityKerr}). Now we need to check
if this velocity field can correspond to circular geodesics. We can
easily derive the geodesic equations from the Euler-Lagrange equations
\begin{equation}
\frac{d}{d\tau}\left(\frac{\partial\mathcal{L}}{\partial U^{\mu}}\right)-\frac{\partial\mathcal{L}}{\partial x^{\mu}}=0\label{eq:EL}
\end{equation}
with Lagrangian $\mathcal{L}=g_{\mu\nu}U^{\mu}U^{\nu}/2$. To compute
the circular geodesics we only need the $r$-equation, $d(g_{rr}U^{r})/d\tau=g_{\mu\nu,r}U^{\mu}U^{\nu}/2$,
which for circular equatorial orbits yields 
\begin{equation}
(\omega_{{\rm geo}})_{\pm}=\frac{-Ma+\frac{\Lambda}{3}ar^{3}\pm\sqrt{Mr^{3}-\frac{\Lambda}{3}r^{6}}}{r^{3}-a^{2}M+\frac{\Lambda}{3}a^{2}r^{3}}\,,\label{eq:ang-velocity}
\end{equation}
which reduces to the Kerr case, Eq.~\eqref{eq:CircularKerr}, when
$\Lambda=0$.

There are two things we need to check: first, that the geodesics lie
outside the black hole event horizon (and inside the cosmological
horizon), and second, that the geodesics are time-like. The horizons
are located at the real roots of $\Delta_{r}=0$, which gives the
equation 
\begin{equation}
r^{2}-2Mr+a^{2}-\frac{\Lambda}{3}r^{2}(r^{2}+a^{2})=0\,.\label{eq:horizons}
\end{equation}
To find spinning particles that follow circular geodesics, we have
to equate the \emph{prograde} solutions of Eq.~\eqref{eq:ang-velocity}
to \eqref{velocityKerr}, 
\begin{equation}
\frac{a}{a^{2}+r^{2}}=\frac{-Ma+\frac{\Lambda}{3}ar^{3}+\sqrt{Mr^{3}-\frac{\Lambda}{3}r^{6}}}{r^{3}-a^{2}M+\frac{\Lambda}{3}a^{2}r^{3}}\,.\label{eq:circ-geo-eq}
\end{equation}
We cannot analytically solve this equation for $r$ in general, but
for our purposes it suffices to numerically show that such an $r$
exists for some particular cases of $a$ and $\Lambda$. Consider
for example the case $a/M=0.8$, $\Lambda M^{2}=0.001$. Solving equation~\eqref{eq:circ-geo-eq}
for $r$, we find, as the only acceptable solution, $r\simeq14.2025M$
(the other roots are either complex or fall within the horizon). This
geodesic is time-like and lies outside the event horizon, as one can
see from Eq. \eqref{eq:horizons}. Obviously, several other solutions
of~\eqref{eq:circ-geo-eq} for different values of $a$ and $\Lambda$
are possible. We generically find that, for fixed $a/M$, decreasing
values of $\Lambda M^{2}$ correspond to solutions of~\eqref{eq:circ-geo-eq}
with increasing values of $r$.

\begin{framed}%
\emph{Sec.~\ref{sub:Symmetries} in brief} --- the physical consequences
of the different symmetries of $B_{\alpha\beta}$ and $\mathbb{H}_{\alpha\beta}$ 
\begin{itemize}
\item In electromagnetism, due to vacuum equation $B_{[\alpha\beta]}=\star F_{\alpha\beta;\gamma}U^{\gamma}/2$,
a force $F_{{\rm EM}}^{\alpha}=B^{\beta\alpha}\mu_{\beta}$ is exerted
on the dipole \emph{whenever it moves} in an inhomogeneous field (except
for very special orientations of $\vec{\mu}$). 
\item In gravity, $\mathbb{H}_{[\alpha\beta]}=0$, and there are velocity
fields for which $\mathbb{H}_{\alpha\beta}=0$, i.e., for which \emph{gyroscopes
feel no force};

\begin{itemize}
\item in the examples studied, they correspond to the situations where,
in the electromagnetic analogue, $B_{\alpha\beta}=B_{[\alpha\beta]}$; 
\item there are even geodesic motions for spinning particles: radial geodesics
in Schwarzschild, circular geodesics in Kerr-dS spacetimes. 
\end{itemize}
\item Formal analogy between the quadratic scalar invariants of $R_{\alpha\beta\gamma\delta}$
and $F_{\alpha\beta}$ is useful to obtain velocities for which $\mathbb{H}_{\alpha\beta}=0$. \end{itemize}
\end{framed}

\section{Manifestations of the time projections of the tidal tensors --- the
work done on the test particle\label{sub:Time-Components}}

A fundamental difference between the gravitational and electromagnetic
interactions concerns the time projections of the forces $F_{{\rm G}}^{\alpha}$
and $F_{{\rm EM}}^{\alpha}$ in the different frames, which we shall
explore in this section%
\footnote{A (very) preliminary version of some of the results herein was presented
in \cite{ProcEre2009}.%
}. We start by explaining the meaning of the time projection of a force
in a given frame, and its relation with the work done by it and the
particle's energy.

Consider a congruence of observers $\mathcal{O}(u)$ with 4-velocity
$u^{\alpha}$, and let $U^{\alpha}$ denote the 4-velocity of a test
particle. The following relation generically holds \cite{The many faces}:
\begin{equation}
U^{\alpha}=\gamma(u^{\alpha}+v^{\alpha});\quad\gamma\equiv-u^{\alpha}U_{\alpha}=\frac{1}{\sqrt{1-v^{\alpha}v_{\alpha}}}\ ,\label{eq:U_u}
\end{equation}
where $v^{\alpha}=U^{\alpha}/\gamma-u^{\alpha}$ is the velocity of
the test particle relative%
\footnote{Let $(t,x^{i})$ be the coordinate system of a locally \emph{inertial}
frame momentarily comoving with the observer; in such frame $u^{i}=0$
and $v^{i}=dx^{i}/dt$ is the ordinary 3-velocity of the test particle.%
} to $\mathcal{O}(u)$. The energy of the particle relative to $\mathcal{O}(u)$
is $E\equiv-P^{\alpha}u_{\alpha}$, and its rate of change per unit
proper time (the ``power equation'') is 
\begin{equation}
\frac{dE}{d\tau}=-F^{\alpha}u_{\alpha}-P^{\alpha}u_{\alpha;\beta}U^{\beta}\,,\label{eq:DE/dt}
\end{equation}
where $F^{\alpha}\equiv DP^{\alpha}/d\tau$ denotes the 4-force. Thus
we see that the variation of the particle's energy relative to $\mathcal{O}(u)$
consists of two terms: the time projection of $F^{\alpha}$ along
$u^{\alpha}$, plus a term depending on the variation of $u^{\alpha}$
along the particle's worldline. The first term is interpreted as the
rate of work, \emph{as measured by} $\mathcal{O}(u)$, done by the
force on the test particle (per unit proper time $\tau$). In order
to better understand it, it is useful to split $F^{\alpha}$ into
its components parallel and orthogonal to the particle's CM worldline,
\[
F^{\alpha}=F_{\parallel}^{\alpha}+F_{\perp}^{\alpha};\qquad F_{\parallel}^{\alpha}\equiv-F^{\beta}U_{\beta}U^{\alpha};\quad\ F_{\perp}^{\alpha}\equiv(h^{U})_{\ \beta}^{\alpha}F^{\beta}\ ;
\]
the first term of \eqref{eq:DE/dt} then reads, using \eqref{eq:U_u},
\begin{equation}
-F^{\alpha}u_{\alpha}=-\gamma F^{\beta}U_{\beta}+F_{\perp}^{\alpha}v_{\alpha}\,.\label{eq:WorkForce}
\end{equation}
Forces orthogonal to $U^{\alpha}$ ($F^{\alpha}=F_{\perp}^{\alpha}$)
are the more familiar ones; it is the case of the forces on point
particles with no internal structure (monopole particles). Let us
start by this simplest case. Such particles have a momentum parallel
to the 4-velocity, $P^{\alpha}=mU^{\alpha}$, and constant mass $m=m_{0}$;
the force is thus parallel to the acceleration, $F^{\alpha}\equiv DP^{\alpha}/d\tau=m_{0}a^{\alpha}$,
which implies $F_{\parallel}^{\alpha}=-F^{\beta}U_{\beta}U^{\alpha}=0$
(due to the condition $U^{\beta}U_{\beta}=-1$). That leads to $-F^{\alpha}u_{\alpha}=F^{\alpha}v_{\alpha}$,
telling us that the time-projection (in the frame $\mathcal{O}(u)$)
of $F^{\alpha}$ is the familiar power $\vec{F}\cdot\vec{v}$ (see
e.g.~\cite{MathissonNeueMechanik,MollerBook}). If we take an inertial
frame, so that the second term of (\ref{eq:DE/dt}) vanishes, then
$-F^{\alpha}u_{\alpha}=dE/d\tau=m_{0}d\gamma/d\tau$, i.e., $F^{\alpha}v_{\alpha}=m_{0}d\gamma/d\tau$
is the rate of variation of the particle's kinetic energy of translation.
It is clear in particular that $dE/d\tau=F^{\alpha}v_{\alpha}=0$
in a frame comoving with the particle. An example of such a force
is the Lorentz force, $F^{\alpha}=qF^{\alpha\beta}U_{\beta}=qE^{\alpha}$,
for which $F^{\alpha}U_{\alpha}=0$, and whose projection along $u^{\alpha}$
reads $-u_{\alpha}F^{\alpha}=qv_{\alpha}E^{\alpha}$, yielding the
power transferred by the electric force to the moving particle (relative
to $\mathcal{O}(u)$).

However, if the particle has internal structure then its internal
degrees of freedom may store energy (e.g., kinetic energy of rotation
about the center of mass), and so the particle's proper mass $m=-P^{\alpha}U_{\alpha}$
no longer has to be a constant (cf. Sec.~\ref{sub:Mass-of-the}).
Moreover, the momentum will not be parallel to $U^{\alpha}$, as the
particle in general possesses hidden momentum, cf.~Sec.~\ref{sub:Momentum-of-the-Particle}.
These, together (as we shall see next), endow $F^{\alpha}$ with a
non-vanishing component $F_{\parallel}^{\alpha}$ along $U^{\alpha}$,
which is the new ingredient. $F_{\parallel}^{\alpha}$ is the rate
of work done by the force as measured in the frame comoving with the
particle.

Let us turn our attention now to the second term of Eq.~(\ref{eq:DE/dt}).
Decomposing (e.g.~\cite{The many faces,PaperAnalogies,Maartens:1997fg})
\begin{equation}
u_{\alpha;\beta}=-a(u)_{\alpha}u_{\beta}+\omega_{\alpha\beta}+\theta_{\alpha\beta}\label{eq:uKinDecomp}
\end{equation}
where $a(u)^{\alpha}=u_{\ ;\beta}^{\alpha}u^{\beta}$ is the observers'
acceleration (\emph{not} the particle's!), $\omega_{\alpha\beta}\equiv(h^{u})_{\alpha}^{\lambda}(h^{u})_{\beta}^{\nu}u_{[\lambda;\nu]}$
is the vorticity, and $\theta_{\alpha\beta}\equiv(h^{u})_{\alpha}^{\lambda}(h^{u})_{\beta}^{\nu}u_{(\lambda;\nu)}$
is the shear/expansion tensor of the observer congruence ($\theta_{\alpha\beta}\equiv\sigma_{\alpha\beta}+\theta(h^{u})_{\alpha\beta}/3$,
where $\sigma_{\alpha\beta}$ is the traceless shear and $\theta$
the expansion scalar). Let us denote by $G(u)^{\alpha}=-a(u)^{\alpha}$
the ``gravitoelectric field'' \cite{PaperAnalogies,The many faces}
measured by the observers. Decomposing $P^{\alpha}=mU^{\alpha}+P_{{\rm hid}}^{\alpha}$,
cf.~Eq.~(\ref{eq:HiddenMomentum}), and using (\ref{eq:U_u}) and
\eqref{eq:uKinDecomp}, the second term of Eq.~(\ref{eq:DE/dt})
becomes: 
\begin{align}
-P^{\alpha}u_{\alpha;\beta}U^{\beta} & =m\gamma^{2}[G(u)_{\alpha}-\theta_{\alpha\beta}v^{\beta}]v^{\alpha}\nonumber \\
 & \quad+\gamma P_{{\rm hid}}^{\alpha}\left[G(u)_{\alpha}-\left(\omega_{\alpha\beta}+\theta_{\alpha\beta}\right)v^{\beta}\right]\,.\label{eq:WorkGEM}
\end{align}
This part of $dE/d\tau$ depends only on the kinematical quantities
of the observer congruence, not on the physical force $F^{\alpha}$.
In other words, it is an artifact of the reference frame, which vanishes
if it is locally inertial. Its importance (in a non-local sense) cannot
however be overlooked. To understand this, consider a simple example,
a monopole particle in Kerr spacetime, from the point of view of the
congruence of \emph{static observers} (\textcolor{black}{cf.~Sec.~\ref{sub:Notation-and-conventions},
point \ref{enu:Static-observers})}. Since the congruence is rigid,
$\theta_{\alpha\beta}=0$; also, for a monopole particle, $P_{{\rm hid}}^{\alpha}=0$,
and, in a gravitational field, $F^{\alpha}=0$ (the particle moves
along a geodesic). Therefore, the energy variation reduces to $dE/d\tau=-P^{\alpha}u_{\alpha;\beta}U^{\beta}=m\gamma^{2}G(u)_{\alpha}v^{\alpha}$,
which is the rate of ``work'' done by the gravitoelectric ``force''
\cite{Natario,PaperAnalogies,The many faces} $m\gamma^{2}G(u)^{\alpha}$.
(In the Newtonian limit, it reduces to the work of the Newtonian force
$m\vec{G}$). Hence we see that (\ref{eq:WorkGEM}) is the part of
(\ref{eq:DE/dt}) that encodes the change in translational kinetic
energy of a particle (relative to the static observers) which occurs
due to the gravitational field without the action of any (physical,
covariant) force, and that is non-zero for particles in geodesic motion.

Substituting Eqs.~(\ref{eq:WorkForce}) and (\ref{eq:WorkGEM}) into
(\ref{eq:DE/dt}), we obtain a generalization, for the case of test
particles with varying $m$ and hidden momentum, of the ``power equation''
(6.12) of \cite{The many faces} (the latter applying to monopole
particles only).

\subsection{\label{sub:Time-components-in-CM-frame }Time components in test
particle's frame }

\begin{figure}
\includegraphics[width=1\columnwidth]{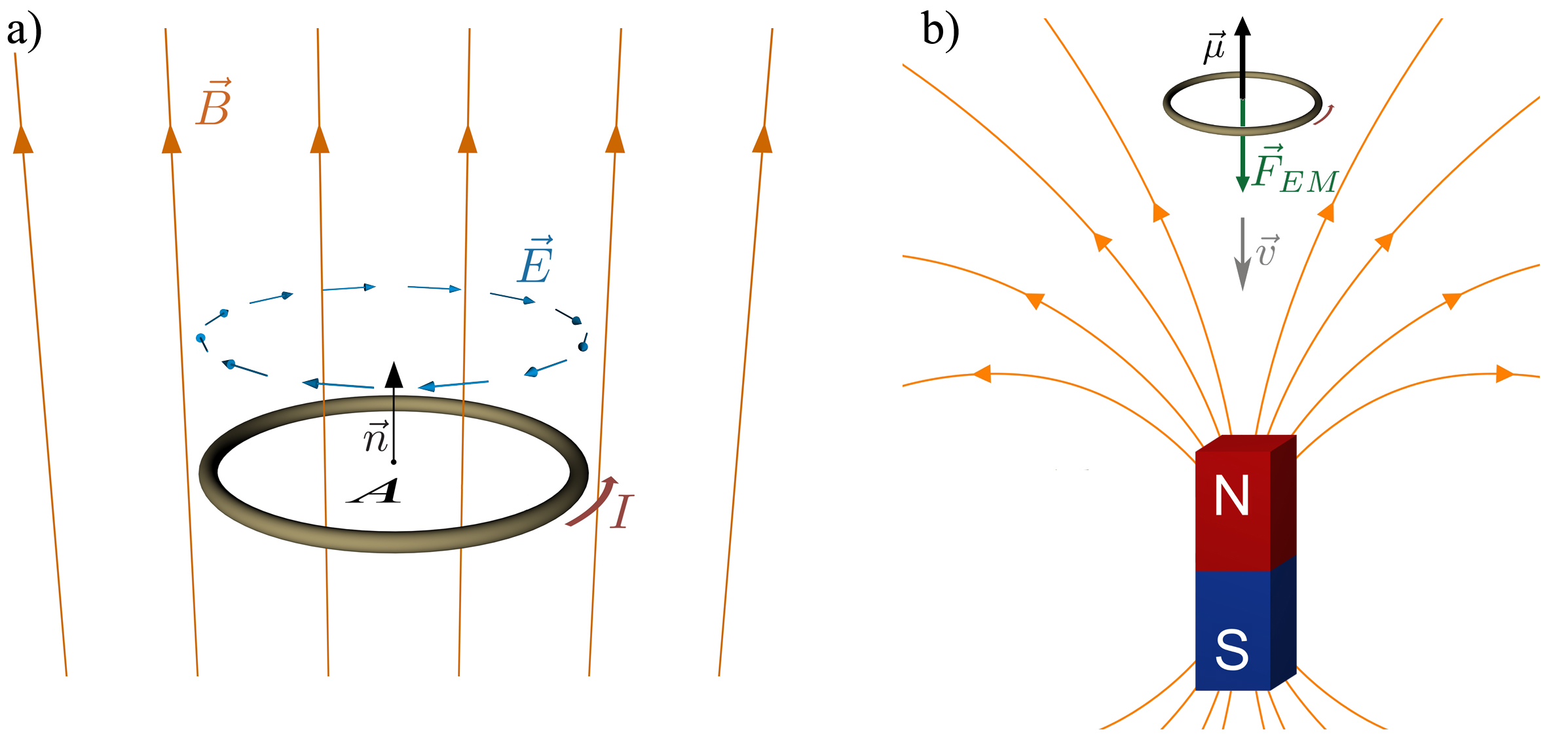}

\caption{\label{fig:DipoleMagnet}A magnetic dipole (depicted as a current
loop) falling in the inhomogeneous magnetic field of a strong magnet,
from the point of view of two different frames: a) the particle's
rest frame; b) the rest frame of the strong magnet (static observers).
Here $\vec{\mu}=IA\vec{n}$; $A\equiv$ area of the loop; $I\equiv$
current through the loop; $\vec{n}\equiv$ unit vector normal to the
loop; $\vec{E}\equiv$ induced electric field. In the dipole's frame
non-vanishing work is done on it by $\vec{E}$, at a rate $\mathcal{P}_{{\rm ind}}=-F_{{\rm EM}}^{\alpha}U_{\alpha}$,
which is reflected in a variation of proper mass $m$. From the point
of view of static observers $u^{\alpha}$, the work is zero ($-F_{{\rm EM}}^{\alpha}u_{\alpha}=\mathcal{P}_{{\rm ind}}+\mathcal{P}_{{\rm trans}}=0$),
manifesting that a stationary magnetic field does no work. That may
be regarded as an exact cancellation between $\mathcal{P}_{{\rm ind}}$
and the rate of variation of the particle's translational kinetic
energy, $\mathcal{P}_{{\rm trans}}$.}
\end{figure}

A fundamental difference between the tensorial structure of $\mathbb{H}_{\alpha\beta}$
and $B_{\alpha\beta}$ is that whereas the former is spatial, in \emph{both}
indices, with respect to the observer $U^{\alpha}$ measuring it,
$(\mathbb{H}^{U})_{\alpha\beta}U^{\beta}=(\mathbb{H}^{U})_{\alpha\beta}U^{\alpha}=0$
(this follows from the symmetries of the Riemann tensor), the latter
is not: $(B^{U})_{\alpha\beta}U^{\alpha}=0$, but $(B^{U})_{\alpha\beta}U^{\beta}=\star F_{\alpha\gamma;\beta}U^{\gamma}U^{\beta}\ne0$
in general. This means that whereas $F_{{\rm G}}^{\alpha}$ is orthogonal
to the particle's worldline, $F_{{\rm EM}}^{\alpha}$ has a nonvanishing
projection along it (i.e., a time projection in the particle's CM
frame), $F_{{\rm EM}}^{\alpha}U_{\alpha}\ne0$. Let us see its physical
meaning. First note, from Eq.~(\ref{tab:Analogy}.4a), that 
\begin{equation}
F_{{\rm EM}}^{\alpha}U_{\alpha}=B^{\beta\alpha}U_{\alpha}\mu_{\beta}=\epsilon_{\beta\delta\mu\nu}U^{\delta}E^{[\mu\nu]}\mu^{\beta}\ ,\label{eq:Induction1}
\end{equation}
showing that it consists of a coupling between $\mu^{\alpha}$ and
the space projection of the antisymmetric part of the electric tidal
tensor $E_{\alpha\beta}$ measured in the particle's CM frame, which,
as discussed in Sec.~\ref{sub:analogy based on tidal tensors}, encodes
Faraday's law of induction. Indeed, if one chooses the CM frame to
be locally non-shearing and non-rotating (as one can always do), we
may replace $E_{\alpha\beta}$ by the covariant derivative of the
electric field $E_{\alpha;\beta}$, cf.~Eq.~(\ref{eq:Eab-Ea;b}),
and Eq.~(\ref{eq:Induction1}) becomes therein, in vector notation,
$F_{{\rm EM}}^{\alpha}U_{\alpha}=-(\nabla\times\vec{E})\cdot\vec{\mu}$.
Its significance becomes clear if one thinks about the magnetic dipole
as a small current loop of area $A$ and magnetic moment $\vec{\mu}=\vec{n}AI$,
see Fig.~\ref{fig:DipoleMagnet}a. It then follows: 
\begin{equation}
-F_{{\rm EM}}^{\alpha}U_{\alpha}=(\nabla\times\vec{E})\cdot\vec{n}AI=I\oint_{{\rm loop}}\vec{E}\equiv\mathcal{P}_{{\rm ind}}\ ,\label{eq:Induction}
\end{equation}
where in the second equality we first used the fact that the loop
is (by definition) infinitesimal, so $(\nabla\times\vec{E})\cdot\vec{n}A=\int_{\Sigma^{(2)}}(\nabla\times\vec{E})\cdot d\vec{\Sigma}$
for a 2-surface $\Sigma^{(2)}$ enclosed by the loop, and then applied
the Stokes theorem in the 3-D local rest space of the dipole. Here
$\vec{E}$ is the \emph{induced electric field}, coming from the induction
law%
\footnote{\label{fn:DB/dt} This generalization of the Maxwell-Faraday law for
accelerated frames is needed if one is to deal with the electric and
magnetic fields measured in the test particle's frame, which in general
accelerates. One could instead base the analysis in the inertial frame
\emph{momentarily} comoving with it, as done in Sec.~V of \cite{CHPRD},
where $\partial\vec{B}/\partial\tau=-\nabla\times\vec{E}$ holds;
the two treatments are equivalent.%
} (\ref{eq:MaxFardayGenVector}). 

Thus $-F_{{\rm EM}}^{\alpha}U_{\alpha}\equiv\mathcal{P}_{{\rm ind}}$
is the rate of work transferred to the dipole by Faraday's law of
induction. Using Eqs.~(\ref{eq:dm/dt0}) and (\ref{eq:Phid-Contrib-m}),
we see that it consists of the variation of the proper mass $m$,
\emph{minus} the projection along $U^{\alpha}$ of the derivative
of the hidden momentum (to which only the ``electromagnetic'' hidden
momentum contributes): 
\begin{equation}
\mathcal{P}_{{\rm ind}}=\frac{dm}{d\tau}-\frac{DP_{{\rm hid}}^{\alpha}}{d\tau}U_{\alpha}=\frac{dm}{d\tau}-\frac{DP_{{\rm hidEM}}^{\alpha}}{d\tau}U_{\alpha}\,.\label{eq:PindGen}
\end{equation}
Note, from Eq.~(\ref{eq:DE/dt}), that $\mathcal{P}_{{\rm ind}}$
is the variation of the dipole's energy $E=-P_{0}$ \emph{as measured
in a momentarily comoving inertial frame}.

The induction phenomenon in Eq.~(\ref{eq:Induction}) has no counterpart
in gravity. Since $\mathbb{H}_{\alpha\beta}$ is spatial relative
to $U^{\alpha}$, we \textit{always have} 
\begin{equation}
F_{{\rm G}}^{\alpha}U_{\alpha}=0\ ,\label{eq:FgProj}
\end{equation}
and the proper mass $m$ is a constant (since also $P^{\alpha}a_{\alpha}=-U_{\alpha}DP_{{\rm hid}}^{\alpha}/d\tau=0$,
cf.~Eqs.~\eqref{eq:dm/dt0}-\eqref{eq:Phid-Contrib-m}). That is,
the \emph{energy} of the gyroscope, \emph{as measured in its CM rest
frame,} is constant. We see thus that the spatial character of the
gravitational tidal tensors \emph{precludes} induction effects analogous
to the electromagnetic ones.

\subsection{Time components as measured by static observers\label{sub:Time-components-Static}}

\subsubsection{Electromagnetism\label{sub:TimeProj_Static_EM}}

With respect to an arbitrary congruence of observers of 4-velocity
$u^{\alpha}$, the time projection of the force exerted on a magnetic
dipole is, cf.~Eq.~(\ref{eq:WorkForce}), 
\begin{equation}
-F_{{\rm EM}}^{\alpha}u_{\alpha}=-\gamma F_{{\rm EM}}^{\alpha}U_{\alpha}+F_{{\rm EM}\perp}^{\alpha}v_{\alpha}=\gamma\mathcal{P}_{{\rm ind}}+F_{{\rm EM}\perp}^{\alpha}v_{\alpha}\label{eq:Pmech+Pind}
\end{equation}
where, in accordance with the discussion above, we identify $\mathcal{P}_{{\rm ind}}=-F_{{\rm EM}}^{\alpha}U_{\alpha}$
as the power transferred to the dipole by Faraday's induction, and
$F_{\perp{\rm EM}}^{\alpha}v_{\alpha}$ is the power transferred by
the component of $F_{{\rm EM}}^{\alpha}$ orthogonal to the particle's
worldline. Consider now a congruence of observers along whose worldlines
the fields are covariantly constant, $F_{\ \ ;\gamma}^{\alpha\beta}u^{\gamma}=0$;
the time projection of the force with respect to them vanishes: 
\begin{equation}
-F_{{\rm EM}}^{\alpha}u_{\alpha}\equiv-\frac{DP_{\alpha}}{d\tau}u^{\alpha}=-\star F_{\gamma\beta;\alpha}U^{\beta}\mu^{\gamma}u^{\alpha}=0.\label{eq:FEMstatic}
\end{equation}
This tells us that the total work done on the dipole, \emph{as measured
by such observers}, is zero. Take these observers to form, moreover,
an inertial frame; these will be dubbed in this context ``\emph{static}''
or ``laboratory''%
\footnote{\label{fn:Static observers}The reason for such denominations is that,
in the electromagnetic setups herein (the magnet in Fig.~\ref{fig:DipoleMagnet}b,
the spinning/non-spinning charges of Secs.~\ref{sub:Symmetries}
and \ref{sub:Weak-field-regime:}), only the observers \emph{at rest}
relative to the sources obey the condition $F_{\ \ ;\gamma}^{\alpha\beta}u^{\gamma}=0$.
Note that even for e.g. observers $u'^{\alpha}$ in circular motion
around a Coulomb charge we have $F_{\ \ ;\gamma}^{\alpha\beta}u'^{\gamma}\ne0$
(as can be seen replacing $U^{\alpha}\rightarrow u'^{\alpha}$ in
Eqs. (\ref{eq:BabCoulomb}), which imply $\star F_{\ \ ;\gamma}^{\alpha\beta}u'^{\gamma}=2(B^{u'})_{[\alpha\beta]}\ne0$
when $u'^{i}\ne0$), even though $u'^{\alpha}$ is in that case a
symmetry of $F_{\alpha\beta}$, $\mathcal{L}_{u'}F_{\alpha\beta}=0$,
and $F_{\alpha\beta}$ is time-independent in the co-rotating frame.%
} observers. In this case $u_{\alpha;\beta}=0$ and the second term
of Eq. (\ref{eq:DE/dt}) vanishes; therefore, the energy of the particle,
$E=-P_{\alpha}u^{\alpha}$, is a conserved quantity in such frame.
Using Eq.~(\ref{eq:HiddenMomentum}), we can write it in the form
\begin{equation}
E=m+T+E_{{\rm hid}}={\rm constant},\label{eq:P0staticDecomp}
\end{equation}
where we dub $E_{{\rm hid}}\equiv-P_{{\rm hid}}^{\alpha}u_{\alpha}$
the ``hidden energy'' (i.e., the time component of the hidden momentum),
and $T\equiv(\gamma-1)m$ is the kinetic energy of translation of
the center of mass, as measured in this frame (in the Newtonian regime,
$T\approx mv^{2}/2$). In the (very scarce, to the authors' knowledge)
literature addressing this problem, a cancellation between the variations
of $T$ and $m$ is suggested in \cite{Wald et al}, or, for the case
of a spherical spinning charged body, of $T$ and kinetic energy of
rotation about the CM \cite{Young,YoungQuestion66,Deissler} (which
agrees with the former assertion, since for such a body, $dm/d\tau$
is essentially a variation of kinetic energy of rotation, as we shall
see in Sec.~\ref{sub:Electromagnetic Torque-and-force on Spherical}).
Equation (\ref{eq:P0staticDecomp}) shows however that, in the general
case when $P_{{\rm hid}}^{\alpha}\ne0$, the energy exchange occurs
between three components, with $E_{{\rm hid}}$ also playing a role.
A suggestive example are the bobbings of a particle with magnetic
dipole moment orbiting a cylindrical charge considered in Sec.~III.B.1
of \cite{Wald et al 2010} (and illustrated in Fig.~1 of \cite{ProcERE2011}).

In this work we are especially interested in the case $P_{{\rm hid}}^{\alpha}=0$
($\Rightarrow E_{{\rm hid}}=0$), so that $m+T={\rm constant}$; i.e.,
the energy exchange, due to the action of the force $F_{{\rm EM}}^{\alpha}$,
occurs only between proper mass and translational kinetic energy.
It follows also that $\mathcal{P}_{{\rm ind}}=dm/d\tau$. Therefore,
from (\ref{eq:DE/dt}) and (\ref{eq:Pmech+Pind}) (and since $u_{\alpha;\beta}=0$),
\begin{equation}
\frac{dE}{d\tau}=-F_{{\rm EM}}^{\alpha}u_{\alpha}=\mathcal{P}_{{\rm ind}}+\mathcal{P}_{{\rm trans}}=0\ ,\label{eq:Ptrans+Pind}
\end{equation}
where 
\begin{equation}
\mathcal{P}_{{\rm trans}}\equiv\frac{dT}{d\tau}=F_{{\rm EM}\perp}^{\alpha}v_{\alpha}+\left(\gamma-1\right)\frac{dm}{d\tau}\label{eq:Ptrans}
\end{equation}
is the rate of variation of \emph{translational kinetic energy}, and
we noted that $F_{{\rm EM}\perp}^{\alpha}v_{\alpha}=md\gamma/d\tau$.
An example is the problem depicted in Fig.~\ref{fig:DipoleMagnet}b):
a magnetic dipole falling along the symmetry axis of the field generated
by a strong magnet. (We have $P_{{\rm hid}}^{\alpha}=0$ for this
configuration%
\footnote{That this is a solution of the equations of motion supplemented with
Mathisson-Pirani condition can be seen by arguments analogous to the
ones given in Appendix \ref{sub:Comparison-of-the-SSC-aplications}
for the gravitational counterpart.%
}). From the point of view of the static observers, $\vec{E}(u)=0$
and only magnetic field $\vec{B}(u)$ is present; we know that the
latter can do no work, because if we think about the dipole as a current
loop (cf.~Fig.~\ref{fig:DipoleMagnet}) and consider the force exerted
in each of its individual moving charges, we see that the magnetic
force $\vec{F}=q(\vec{v}\times\vec{B})$ is always orthogonal to the
velocity $\vec{v}$ of the charges, so that no work can be done. It
is thus quite natural that $F_{{\rm EM}}^{\alpha}u_{\alpha}=0$. According
to Eq.~(\ref{eq:Ptrans+Pind}), this arises from an exact cancellation
between $\mathcal{P}_{{\rm trans}}$ and $\mathcal{P}_{{\rm ind}}$:
on the one hand there is an attractive spatial force $\vec{F}_{{\rm EM}}$
causing the dipole to gain translational kinetic energy; on the other
hand there is a variation of its internal energy (proper mass $m$)
by induction, which allows for the total work to vanish (in agreement
with the reasoning in~\cite{Wald et al}, p. 21). Further remarks
on this issue are given in Secs.~\ref{sub:Summarizing-with-a_simple}
and Appendix \ref{sub:Conserved-quantities,-proper}.

\textcolor{black}{It is worth mentioning that this cancellation solves
an apparent paradox that has for long been discussed in the literature~\cite{Coombes,Young,Deissler,Wald et al}
--- that on the one hand a force is exerted on a magnetic dipole placed
in a non-homogeneous magnetic field, causing it to move, whilst on
the other hand }$\vec{B}$ can do no work in any charge/current distribution\textcolor{black}{.
The analysis above generalizes and reformulates, in a relativistic
covariant framework, the arguments in~\cite{Coombes,Young,YoungQuestion66,Deissler},
and supports the claim in \cite{Wald et al} that the solution of
the apparent paradox lies on the variation of $m$. It is also useful,
to make these points more clear, to compare with the cases of a monopole
charged particle, and of an electric dipole subject to an electromagnetic
field: there is also a force on the particle, which is set into motion
gaining kinetic energy; but, in these cases,} \textcolor{black}{\emph{the
electric field is}} \textcolor{black}{\emph{doing work}}\textcolor{black}{,
there is a potential energy involved, and the gain in translational
kinetic energy is} \textcolor{black}{\emph{not}}\textcolor{black}{~canceled
out by} \textcolor{black}{a variation of the particle's proper mass
$(m$ is constant for a monopole particle, and also for an electric
dipole if one assumes that the dipole vector is parallel transported).
These cases are discussed in detail in Appendix \ref{sub:Conserved-quantities,-proper}.}

\subsubsection{Gravity\label{sub:TimeProj_Static_Grav}}

In gravity, where $F_{{\rm G}}^{\alpha}U_{\alpha}=0$ (i.e., the induction
effects are absent), we have, for arbitrary observers $u^{\alpha}$,
\begin{equation}
-F_{{\rm G}}^{\alpha}u_{\alpha}=F_{{\rm G}}^{\alpha}v_{\alpha}.\label{eq:TimeProjFg}
\end{equation}
This implies that a cancellation similar to the one in Eq.~\eqref{eq:Ptrans+Pind}
does not occur. Except when $v^{\alpha}\perp F_{{\rm G}}^{\alpha}$,
$F_{{\rm G}}^{\alpha}$ does work whenever the particle moves relative
to the reference frame; in particular it is so from the point of view
of static observers in a stationary spacetime (i.e., a stationary
gravitomagnetic tidal field does work on mass currents), \emph{by
contrast with its electromagnetic counterpart}. And there is a potential
energy associated with such work, as we shall now show.

A conserved quantity for a spinning particle in a stationary spacetime
is (e.g.~\cite{Dixon1970I,Wald et al 2010,Semerak I,Hartl}) 
\begin{equation}
E_{{\rm tot}}=-P^{\alpha}\xi_{\alpha}+\frac{1}{2}\xi_{\alpha;\beta}S^{\alpha\beta}={\rm constant}\ ,\label{eq:Etot}
\end{equation}
where \textcolor{black}{$\bm{\xi}\equiv\partial/\partial t$} is the
time \emph{Killing vector field}. Consider the congruence of static
observers%
\footnote{See point \ref{enu:Static-observers} of Sec.~\ref{sub:Notation-and-conventions}.
In stationary asymptotically flat spacetimes, such as the Kerr metric
studied below, these are observers rigidly fixed to the asymptotic
inertial rest frame of the source
. They are thus the closest analogue of the flat spacetime notion
of observers at rest relative to the source in the electromagnetic
systems above.%
}, of 4-velocity parallel to $\xi^{\alpha}$: $u^{\alpha}=\xi^{\alpha}/\xi$,
where \textcolor{black}{$\xi\equiv\sqrt{-\xi^{\alpha}\xi_{\alpha}}$
is their lapse, or redshift factor (see e.g. \cite{The many faces,GEM User Manual}).}
The first term of (\ref{eq:Etot}), $-P^{\alpha}\xi_{\alpha}=E\xi$,
is the ``Killing energy'', a conserved quantity for the case of
a non-spinning particle ($S^{\alpha\beta}=0$) in geodesic motion,
which yields its energy with respect to the\emph{ }static observers\emph{
at infinity}%
\footnote{If the particle is in a bounded orbit, one can imagine this measurement
process as follows: let $E_{{\rm tot}}(\tau_{1})$ be the total energy
of the particle at $\tau_{1}$; if, at that instant, the particle
was by some process converted into light and sent to infinity, the
resulting radiation would reach infinity with an energy $E=-u^{\alpha}P_{\alpha}=E_{{\rm tot}}(\tau_{1})$. %
}. It \textcolor{black}{can be interpreted as its ``total energy''
(rest mass + kinetic + ``Newtonian potential energy'') in a gravitational
field (e.g.~\cite{General Relativity}). The energy $E_{{\rm tot}}$
can likewise be interpreted as the energy at infinity for the case}
of \textcolor{black}{a spinning particle. To see the interpretation
of the second term in} (\ref{eq:Etot}), 
\begin{equation}
V\equiv\frac{1}{2}\xi_{\alpha;\beta}S^{\alpha\beta}\ ,\label{eq:VGrav}
\end{equation}
consider the case when $P_{{\rm hid}}^{\alpha}=0$. We have 
\begin{eqnarray}
0=\frac{dE_{{\rm tot}}}{d\tau} & = & -F_{{\rm G}}^{\alpha}\xi_{\alpha}-mU^{\alpha}U^{\beta}\xi_{\alpha;\beta}+\frac{dV}{d\tau}\label{eq:TotPower}\\
\Leftrightarrow\xi F_{{\rm G}}^{\alpha}u_{\alpha} & = & \frac{dV}{d\tau},\label{eq:dV/dt}
\end{eqnarray}
where we used the Killing equation $\xi_{(\alpha;\beta)}=0$. The
quantity $-\xi F_{{\rm G}}^{\alpha}u_{\alpha}=\xi F_{{\rm G}}^{\alpha}v_{\alpha}$
is the rate work (per unit of particle's proper time $\tau$) of $F_{{\rm G}}^{\alpha}$,
as measured by the static observers\emph{ at infinity,} and thus $V$
is the spin-curvature \emph{potential energy} associated with that
work%
\footnote{\textcolor{black}{One may check explicitly that $dV/d\tau=\xi_{\alpha;\beta\gamma}S^{\alpha\beta}U^{\gamma}=\xi F_{G}^{\alpha}v_{\alpha}$,
noting that $DS^{\alpha\beta}/d\tau=0$ if $P_{{\rm hid}}^{\alpha}=0$,
and using the general relation for a Killing vector $\xi_{\mu;\nu\lambda}=R_{\lambda\sigma\mu\nu}\xi^{\sigma}$.}%
}.

In order to compare with the electromagnetic equation~(\ref{eq:P0staticDecomp}),
note that $d\xi/d\tau=-\gamma G(u)_{\alpha}v^{\alpha}$, and that
for $P_{{\rm hid}}^{\alpha}=0$ we have $E=\gamma m=m+T$. Thus we
can write $dE_{{\rm tot}}/d\tau=d(\xi E+V)/d\tau$ in the form 
\begin{equation}
\xi\frac{dT}{d\tau}-\xi m\gamma^{2}G_{\alpha}v^{\alpha}+\frac{dV}{d\tau}=0.\label{eq:EnergyBalanceGrav}
\end{equation}
The second term accounts for the ``power'' of the gravitoelectric
``force'' $m\gamma^{2}\vec{G}(u)$ (which is \emph{not} a physical
force, arising, as explained above, from the observers' acceleration);
it reduces to the variation of Newtonian potential energy in the weak
field slow motion limit. \textcolor{black}{Equation~(\ref{eq:EnergyBalanceGrav})
tells us that the variation of translational kinetic energy $T$ comes
from the spin-curvature potential energy $V$, and from the power
transferred by $m\gamma^{2}\vec{G}(u)$ ($m$ being constant); this
contrasts with the case of the magnetic dipole discussed above, where
(again for $P_{{\rm hid}}^{\alpha}=0$) the variation of kinetic energy
comes from the variation of proper mass $m$, with no potential energy
being involved. In terms of the work done on the particle, $F_{{\rm G}}^{\alpha}$
is thus more similar to the electromagnetic forces exerted on a monopole
charge or on an electric dipole (for $Dd^{\alpha}/d\tau=0$), where
the proper mass is likewise constant and the energy exchange is between
$T$ and potential energy (see Appendix \ref{sub:Conserved-quantities,-proper}
for more details).}

\begin{figure}
\includegraphics[width=0.5\textwidth]{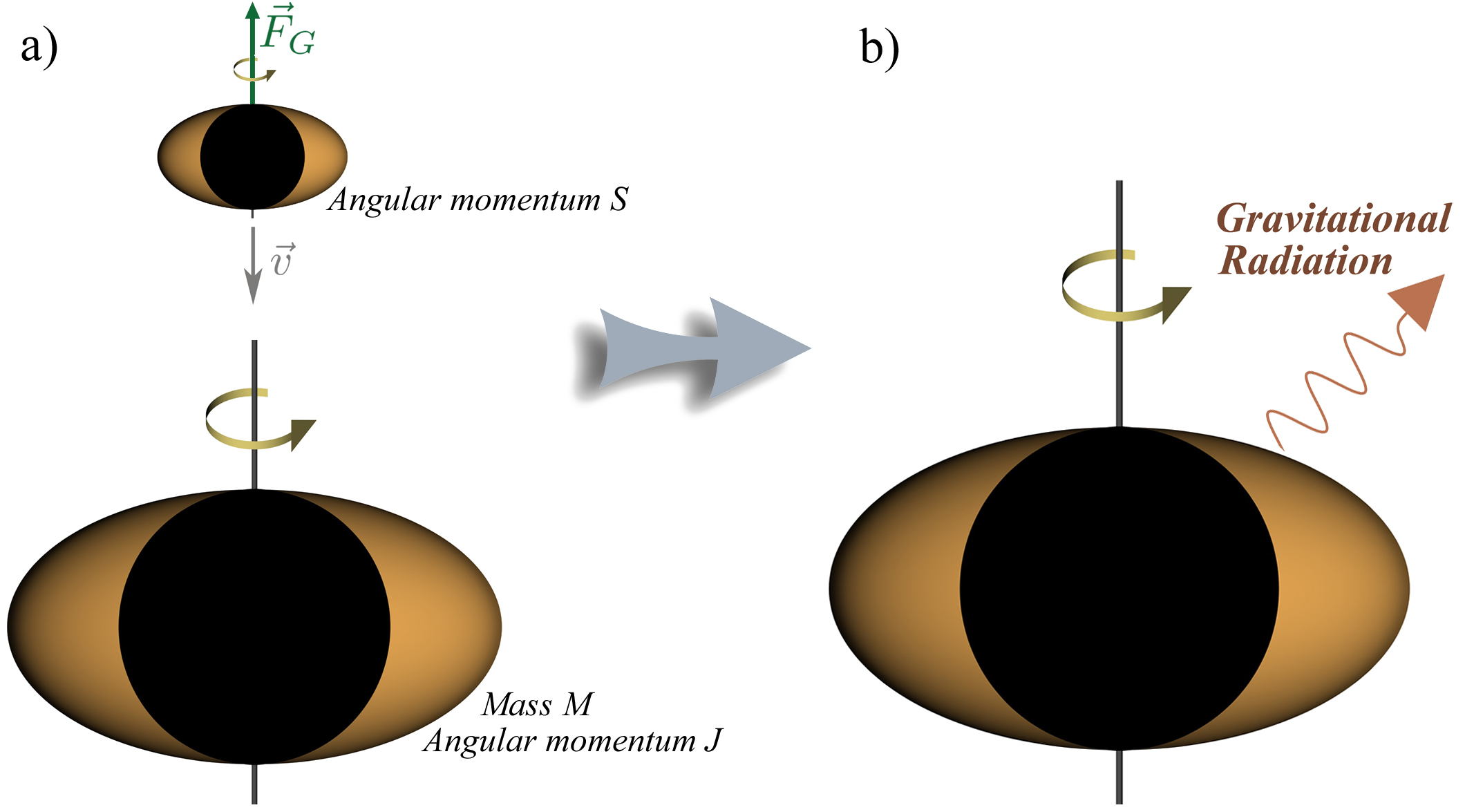}

\caption{{\small{{{\label{fig:Hawking}a) Gyroscope (small Kerr black hole)
in the field of a large Kerr black hole; b) black hole merger. Evidence
that, unlike its electromagnetic counterpart, the gravitomagnetic
tidal field }}}}\emph{\small{{{does}}}}{\small{{{ work: the
spin-dependent part of the energy released is the work (as measured
by the static observers at infinity) of $F_{{\rm G}}^{\alpha}$.}}}}}
\end{figure}

There is a known consequence of the fact that $F_{{\rm G}}^{\alpha}$
does work (and of the interaction energy $V)$: the spin dependence
of the upper bounds for the energy released by gravitational radiation
when two black holes collide (Fig.~\ref{fig:Hawking}b), obtained
by Hawking \cite{Hawking} from the area law. 

In order to see this, consider the apparatus in Fig.~\ref{fig:Hawking}:
two Kerr black holes with their spins aligned, a large one (mass $M$,
spin $J=aM$), which is the source, and a small one (4-velocity $U^{\alpha}$,
spin $S\equiv\sqrt{S^{\alpha}S_{\alpha}}$), which we take to be the
test particle, falling into the former along the symmetry axis (how
this is set up with the Mathisson-Pirani spin condition is discussed
in Appendix \ref{sub:Comparison-of-the-SSC-aplications}). For this
setup $\mathbf{U}=U^{0}\mathbf{e}_{0}+U^{r}\mathbf{e}_{r}$, $\mathbf{S}=S^{0}\mathbf{e}_{0}+S^{r}\mathbf{e}_{r}$,
where $\mathbf{e}_{\alpha}\equiv\partial/\partial_{\alpha}$ are Boyer-Lindquist
coordinate basis vectors; and $P_{{\rm hid}}^{\alpha}=0$. Moreover,
$V$ becomes a \emph{pure spin-spin} potential energy, since, for
radial motions, $F_{{\rm G}}^{\alpha}=0$ if $J=0$, cf.~Eqs.~\eqref{eq:HabSchwa}.
Using $S_{\alpha\beta}=\epsilon_{\alpha\beta\gamma\delta}U^{\delta}S^{\gamma}$
(as follows from the condition $S^{\alpha\beta}U_{\beta}=0$, cf.
Sec.~\ref{sub:Center-of-mass}), and noting that $S^{r}U^{0}-S^{0}U^{r}=S$
(as follows from $S^{\alpha}U_{\alpha}=0$, and, along the axis, $g_{00}=-1/g_{rr}$),
one obtains, for Eq.~\eqref{eq:VGrav}, 
\[
V(r)=\pm\frac{2aMSr}{(a^{2}+r^{2})^{2}}\ =\int_{\infty}^{\tau(r)}\xi F_{{\rm G}}^{\alpha}u_{\alpha}d\tau\,,
\]
the $+/-$ sign applying to the case when $\vec{S}$ and $\vec{J}$
are parallel/anti-parallel. The second equality follows from Eq.~(\ref{eq:dV/dt});
this result can be checked noting that, in Boyer-Lindquist coordinates,
$\xi F_{{\rm G}}^{\alpha}u_{\alpha}=(F_{{\rm G}})_{0}$, and computing
explicitly the time component $(F_{{\rm G}})_{0}$ for axial fall,
Eq.~(37) of \cite{Wald}. Thus we see that $V(r)$ is \emph{minus}
the work done by $F_{{\rm G}}^{\alpha}$ as the particle goes from
infinity to $r$. Let us comment on the presence of the lapse factor
$\xi$ in the integral above. Computing the work of $F_{{\rm G}}^{\alpha}$
does not amount to integrating the power measured by the local static
observers, $-F_{{\rm G}}^{\alpha}u_{\alpha}=F_{{\rm G}}^{\alpha}v_{\alpha}$
(i.e., summing up the work elements $dW\equiv F_{{\rm G}}^{\alpha}v_{\alpha}d\tau$),
as that would mean adding \textcolor{black}{energies measured by different
observers; instead, we should integrate quantity} $\xi F_{{\rm G}}^{\alpha}v_{\alpha}$,
which can be thought as summing up work elements \textcolor{black}{measured
by the static observer} \textcolor{black}{\emph{at infinity.}}

Let us now analyze the problem of the black hole merger. The increase
of translational kinetic energy of the small black hole during the
fall is given by Eq.~(\ref{eq:EnergyBalanceGrav}). The term $m\gamma^{2}G_{\alpha}v^{\alpha}$
is the gain in kinetic energy due to the ``Newtonian'' attraction,
and exists regardless of $S^{\alpha}$; the term involving the spin-spin
potential energy $V$, however, will cause the test particle's kinetic
energy and, therefore, the energy available to be released by gravitational
radiation in the collision, to depend on $S$. Upper bounds for this
energy, which are, accordingly, spin dependent, were obtained in \cite{Hawking}
by a totally independent method. From such limits, and for the setup
in Fig.~\ref{fig:Hawking}, Wald \cite{Wald} obtained an expression
(Eq.~(35) therein) for the amount of energy $\Delta E_{s}$ by which
the upper bound is increased/reduced when $\vec{S}$ is parallel/anti-parallel
to $\vec{J}$, comparing with the case $S=0$ (fall along a geodesic).
It turns out that this energy is precisely \emph{minus} the value
of $V(r)$ at the horizon $r_{+}$, $\Delta E_{s}=-V(r_{+})$; that
is, it is \emph{the work done by} $F_{{\rm G}}^{\alpha}$ on the small
black hole as it comes from infinity to the horizon: $\Delta E_{s}=\int_{\infty}^{\tau(r_{+})}(-\xi F_{{\rm G}}^{\alpha}u_{\alpha})d\tau$.

We close this section with some remarks on the meaning of the work
done by the gravitomagnetic tidal field. One can associate to the
static observers in the Kerr spacetime a gravitomagnetic ``vector''
field $\vec{H}$ (see Sec.~\ref{sub:Spin-``precession''--}, and
e.g.~\cite{PaperAnalogies,The many faces,Natario,Gravitation and Inertia};
in the weak field regime this field is well known to be very similar
to its electromagnetic counterpart, e.g.~\cite{Ciufolini Lageos,Gravitation and Inertia,Ruggiero:2002hz}),
causing inertial (i.e., fictitious) ``forces'' on test particles
of the type $\vec{v}\times\vec{H}$, formally similar to the magnetic
force $q\vec{v}\times\vec{B}$. Namely, the force is orthogonal to
the velocity; hence this analogy might lead one to believe that, similarly
to its magnetic counterpart, the gravitomagnetic field cannot do work
on test particles. One must bear in mind, however, that $\vec{H}$
(by contrast with $\vec{B}$) has no local existence, as it is a mere
artifact of the reference frame; hence it would never be involved
in a covariant quantity like the 4-force $DP^{\alpha}/d\tau$, or
the work done by it. Moreover, both in electromagnetism and in gravity,
\emph{it is the tidal fields that yield the force}; that is manifest
in force Eqs.~(\ref{tab:Analogy}.1) of Table \ref{tab:Analogy}.
The electromagnetic tidal tensors herein are essentially derivatives
of the fields; for this reason we were able to argue in terms of the
fields in the applications depicted in Fig.~\ref{fig:DipoleMagnet}
(even though it is their derivatives that show up in the equations).
But the gravitational tidal tensors cannot be cast as derivatives
of the GEM fields, even in the weak field regime, except under very
special conditions (see Sec.~3.5 of \cite{PaperAnalogies}); the
force $F_{{\rm G}}^{\alpha}$ is thus in general very different from
its electromagnetic counterpart. Namely, it is so whenever the test
particle moves relative to the source --- so that the work of $F_{{\rm G}}^{\alpha}$
can dramatically differ from that of $F_{{\rm EM}}^{\alpha}$, which
is well exemplified by the contrast herein: as measured in the test
particle's frame, we have $F_{{\rm EM}}^{\alpha}U_{\alpha}\ne0$,
$F_{{\rm G}}^{\alpha}U_{\alpha}=0$; as measured by the static observers
$u^{\alpha}$, we have precisely the \emph{opposite} situation: $F_{{\rm EM}}^{\alpha}u_{\alpha}=0$,
$F_{{\rm G}}^{\alpha}u_{\alpha}\ne0$.%
\begin{framed}%
\emph{Sec.~\ref{sub:Time-Components} in brief} --- the work done
on the particle (magnetic dipole vs gyroscope) 
\begin{itemize}
\item The time projection of the force, $-F^{\alpha}u_{\alpha}$, is the
rate at which it does work on the particle, as measured by an observer
of 4-velocity $u^{\alpha}$. 
\end{itemize}
Time projections along the particle's worldline ($U^{\alpha}$) 
\begin{itemize}
\item Electromagnetic is non-vanishing, $F_{{\rm EM}}^{\alpha}U_{\alpha}\ne0$;
it is the rate of work done by Faraday's induction law, arising from
$E_{[\alpha\beta]}$ (or equivalently, from $B_{\alpha\beta}U^{\beta}$);
reflected in a variation of $m$. 
\item Gravitational is zero, $F_{{\rm G}}^{\alpha}U_{\alpha}=0$; the gyroscope's
proper mass $m$ is constant; no analogous induction effect (as $\mathbb{H}_{\alpha\beta}U^{\beta}=0$). 
\end{itemize}
Time projections relative to \emph{static observers} ($u^{\alpha}$) 
\begin{itemize}
\item Electromagnetic is zero, $F_{{\rm EM}}^{\alpha}u_{\alpha}=0$ $\Rightarrow$
a stationary electromagnetic field does no work on magnetic dipoles. 
\item Gravitational is non-zero, $F_{{\rm G}}^{\alpha}u_{\alpha}\ne0$ $\Rightarrow$
gravitomagnetic (tidal) field does work --- there is a spin-curvature
potential energy; embodies Hawking-Wald spin-spin interaction energy. \end{itemize}
\end{framed}

\section{Weak field regime and gravitational spin-spin force\label{sub:Weak-field-regime:}}

In the previous two sections we discussed the crucial differences
between the gravitational and electromagnetic forces on a spinning
particle that are manifest in the symmetries and time projections
of the tidal tensors. However, in the literature (e.g.~\cite{Ruggiero:2002hz,Gravitation and Inertia,Harris1991,Wald et al 2010})
concerning the weak field, slow motion regime --- where the non-linearities
of the gravitational field are negligible, and one might indeed expect
a similarity between the gravitational and electromagnetic interactions
--- they are usually portrayed as being very similar. In this section
we will study this regime, and dissect the impact of the aforementioned
differences. We shall consider the basic example of analogous physical
systems: a magnetic dipole in the electromagnetic field of a spinning
charge (charge $Q$, magnetic moment $\mu_{{\rm s}}$), and a gyroscope
in the gravitational field of a spinning mass (mass $M$, angular
momentum $J$), asymptotically described by the Kerr solution.

We start by briefly describing the approximations that we will use.
The electromagnetic potentials are, exactly, $\phi\equiv Q/r$ and
$\vec{A}\equiv\vec{\mu}_{{\rm s}}\times\vec{r}/r^{3}$; for the gravitational
field we take the linearized Kerr metric 
\begin{equation}
ds^{2}=-\left(1+2\Phi\right)dt^{2}+2\mathcal{A}_{j}dtdx^{j}+\left(1-2\Phi\right)\delta_{ij}dx^{i}dx^{j}\ ,\label{eq:Linearmetric}
\end{equation}
with the gravitational ``potentials'' $\Phi\equiv-M/r$, $\vec{\mathcal{A}}\equiv-2\vec{J}\times\vec{r}/r^{3}$.
The gravitational tidal tensors, are, consistently, linearized in
the potentials. The electromagnetic tidal tensors are linear in the
potentials, hence no weak field assumption is made in the forces (\ref{eq:spin-spin}),
(\ref{eq:FEMLinMoving})-(\ref{eq:FEMLinMoving0}). The expression
for the acceleration (\ref{eq:mavaccum}), however, involves a term
of second order in the electromagnetic fields, which is to be neglected
in a coherent comparison with linearized gravity. In the computation
of the electromagnetic and gravitational tidal tensors involved in
the forces (\ref{eq:FEMLinMoving})-(\ref{eq:FEMLinMoving0}), (\ref{eq:FGLin})-(\ref{eq:FGLin0}),
exerted on slowly moving test particles (velocity $v$), only terms
up to first order in $v$ are kept (as usual in slow motion approximations,
e.g.~\cite{Wald}). The relationship with the post-Newtonian approximations
in e.g. \cite{WillPoissonBook,WillBook,Faye2006,KidderWill,TagoshiOhashiOwen,DSX,Kaplan}
is established in \cite{EPAPS}.

Let us first consider stationary setups, where the test particle is
at rest relative to the central source (or singularity, for the case
of a black hole); i.e., at rest with respect to the static observers
$u^{\alpha}$ (cf.~point \ref{enu:Static-observers} of Sec.~\ref{sub:Notation-and-conventions}).
For these observers, the \emph{linearized} gravitational tidal tensors
match the electromagnetic ones, identifying the appropriate parameters:
\begin{equation}
(\mathbb{E}^{u})_{ij}\simeq\frac{M}{r^{3}}\delta_{ij}-\frac{3Mr_{i}r_{j}}{r^{5}}\stackrel{M\leftrightarrow Q}{=}(E^{u})_{ij}\,,\label{eq:EijLinStatic}
\end{equation}

\begin{align}
(\mathbb{H}^{u})_{ij} & \simeq3\left[\frac{(\vec{r}\cdot\vec{J})}{r^{5}}\delta_{ij}+2\frac{r_{(i}J_{j)}}{r^{5}}-5\frac{(\vec{r}\cdot\vec{J})r_{i}r_{j}}{r^{7}}\right]\label{eq:HijBijLinStatic}\\
 & {}\stackrel{J\leftrightarrow\mu_{{\rm s}}}{=}(B^{u})_{ij}\nonumber 
\end{align}
(all the time components vanish identically for these observers).
Therefore, the force exerted on a gyroscope \emph{whose center of
mass is at rest} relative to the central mass is similar (apart for
a minus sign) to its electromagnetic counterpart, identifying $\mu_{{\rm s}}\leftrightarrow J$
and $\mu\leftrightarrow S$, 
\begin{equation}
F_{{\rm G}}^{i}=-\mathbb{H}^{ji}S_{j}\stackrel{\ \{J,S\}\leftrightarrow\{\mu_{{\rm s}},\mu\}}{\simeq}-F_{{\rm EM}}^{i}\,.\label{eq:spin-spin}
\end{equation}
In other words, there is, for stationary setups, a gravitational spin-spin
force similar to its electromagnetic counterpart; this result is due
to Wald \cite{Wald}.

\emph{Manifestation of the different symmetries} --- in the general
case, where the dipole/gyroscope is allowed to move, however, Table
\ref{tab:Analogy} makes clear that the two forces differ, because
$\mathbb{H}_{\alpha\beta}$ remains symmetric, whereas $B_{\alpha\beta}$
acquires an antisymmetric part. This leads to key differences in the
dynamics (already exemplified in Sec.~\ref{sub:Symmetries}), which
are non-negligible in the weak field\emph{ }and slow motion approximation,
as we shall now see. Consider the test particles to be moving with
3-velocity $\vec{v}$ relative to the central sources. The magnetic
tidal tensor as seen by the moving dipole, $B_{\alpha\beta}$, can
be obtained in terms of the tidal tensors $(E^{u})_{\alpha\beta},\ (B^{u})_{\alpha\beta}$
measured by the static observers, using the decomposition 
\begin{equation}
\star F_{\alpha\beta;\gamma}=2u_{[\alpha}(B^{u})_{\beta]\gamma}-\epsilon_{\alpha\beta\lambda\sigma}u^{\sigma}(E^{u})_{\ \gamma}^{\lambda}\,.\label{eq:FcovStarDecomp}
\end{equation}
The force (\ref{tab:Analogy}.2a) exerted on the magnetic dipole reads,
\emph{to first order} in $v$, 
\begin{align}
 & F_{{\rm EM}}^{i}\simeq B^{ji}\mu_{j}\ \simeq\ (B^{u})^{ji}\mu_{j}-(E^{u})^{li}\epsilon_{\ kl}^{j}v^{k}\mu_{j}\,,\label{eq:FEMLinMoving}\\
 & F_{{\rm EM}}^{0}=B^{i0}\mu_{i}=0\,.\label{eq:FEMLinMoving0}
\end{align}
The gravitomagnetic tidal tensor as seen by the moving gyroscope,
$\mathbb{H}_{\alpha\beta}$, can analogously be obtained in terms
of the tidal tensors $(\mathbb{E}^{u})_{\alpha\beta},\ (\mathbb{H}^{u})_{\alpha\beta}$
measured by the static observers, using the dual of decomposition
(\ref{eq:Weyldecomp}), 
\begin{eqnarray*}
\star R_{\alpha\beta}^{\ \ \ \gamma\delta} & = & 4\epsilon_{\ \ \alpha\beta}^{\lambda\tau}u_{\lambda}u^{[\gamma}(\mathbb{E}^{u})_{\tau}^{\,\,\,\delta]}-2\epsilon_{\ \alpha\beta}^{\tau\ \ [\gamma}(\mathbb{E}^{u})_{\tau}^{\,\,\,\delta]}\\
 &  & +4(\mathbb{H}^{u})_{[\beta}^{\ [\delta}u^{\gamma]}u_{\alpha]}+\epsilon_{\ \ \alpha\beta}^{\lambda\tau}\epsilon^{\gamma\delta\mu\nu}(\mathbb{H}^{u})_{\mu\tau}u_{\lambda}u_{\nu}\,.
\end{eqnarray*}
The force exerted on the gyroscope reads, to linear order in the fields,
and to first order in $v$, 
\begin{align}
F_{{\rm G}}^{i} & \simeq-\mathbb{H}^{ji}S_{j}\ \simeq\ -(\mathbb{H}^{u})^{ji}S_{j}+2(\mathbb{E}^{u})_{\ }^{l(i}\epsilon_{\ \ kl}^{j)}v^{k}S_{j}\,,\label{eq:FGLin}\\
F_{{\rm G}}^{0} & \simeq-\mathbb{H}^{i0}S_{i}\ \simeq-\ (\mathbb{H}^{u})^{ji}v_{j}S_{i}\,.\label{eq:FGLin0}
\end{align}
We note that, to this accuracy, the spatial part of the forces, apart
from global signs and a factor of two in the second term of (\ref{eq:FGLin})
as compared to (\ref{eq:FEMLinMoving}), differ essentially in the
fact that the former expression is \emph{symmetrized} in $\{i,j\}$,
whereas the latter is \emph{not}. Thus the differences in the symmetries
of the tidal tensors, discussed in Sec.~\ref{sub:Symmetries}, \emph{are
manifest to leading order}. (Explicit expressions for $\vec{F}_{{\rm G}}$
and $\vec{F}_{{\rm EM}}$ are given in \cite{EPAPS}).

Also the differences in the time components, studied in Sec \ref{sub:Time-Components},
are manifest in Eqs.~\eqref{eq:FGLin0}, \eqref{eq:FEMLinMoving0}
herein: $F_{{\rm G}}^{0}\simeq-F_{{\rm G}}^{\alpha}u_{\alpha}\ne0$,
telling us that, from the point of view of the static observers $\mathcal{O}(u)$,
non-negligible work is done on the gyroscope; but $F_{{\rm EM}}^{0}=-F_{{\rm EM}}^{\alpha}u_{\alpha}=0$
(an exact result, cf.~Eq.~(\ref{eq:FEMstatic})), telling us that
no work is done on the dipole. One may also check that whereas $F_{{\rm G}}^{\alpha}U_{\alpha}=0$,
$F_{{\rm EM}}^{\alpha}$ has a non-vanishing time projection in the
particle's frame, which, to first order in $v$, reads $F_{{\rm EM}}^{\alpha}U_{\alpha}\simeq(B^{u})^{ji}\mu_{j}v_{i}$.

It should however be noted that the forces above do not translate
in a trivial fashion into accelerations; $F^{\alpha}$ is in general
not even parallel to $a^{\alpha}$, as the test particles possess
hidden momentum. Assuming $\mu^{\alpha}=\sigma S^{\alpha}$, from
Eq.~(\ref{eq:mavaccum}) we have, in the electromagnetic case, 
\begin{align}
m_{0}a^{i} & \simeq F_{{\rm EM}}^{i}+2B^{[ij]}\mu_{j}\simeq B^{ij}\mu_{j}\nonumber \\
 & \simeq(B^{u})^{ij}\mu_{j}-(E^{u})^{lj}\epsilon_{\ kl}^{i}v^{k}\mu_{j}\,.\label{eq:maEMLin}
\end{align}
The last two terms of (\ref{eq:mavaccum}) are herein neglected. As
for the term $\star F_{\beta}^{\ i}D\mu^{\beta}/d\tau$, it follows
from Eq.~(\ref{eq:EqSpinVector}), for $\mu^{\alpha}=\sigma S^{\alpha}$,
that it is of second order in the fields, thus to be neglected in
a coherent comparison with linearized gravity. The term $DP_{{\rm hidI}}^{\alpha}/d\tau=\epsilon_{\ \beta\gamma\delta}^{\alpha}U^{\delta}D(S^{\gamma}a^{\beta})/d\tau$
is also negligible in this approximation if, among the many possible
solutions \cite{Helical} allowed by the condition $S^{\alpha\beta}U_{\beta}=0$,
we choose the ``non-helical'' one; actually, imposing $P_{{\rm hidI}}^{\alpha}\approx0$
amounts, \emph{in this application} (not in general), to picking such
a solution, as we explain in detail in \cite{EPAPS}. The explicit
result, substituting $(E^{u})_{ij}$ and $(B^{u})_{ij}$ from Eqs.~(\ref{eq:EijLinStatic})-(\ref{eq:HijBijLinStatic}),
reads 
\begin{align}
m_{0}a^{i}\simeq & \frac{3}{r^{5}}\left[(\vec{r}\cdot\vec{\mu}_{{\rm s}})\mu^{i}+2r^{(j}\mu_{{\rm s}}^{i)}\mu_{j}-5\frac{(\vec{r}\cdot\vec{\mu}_{{\rm s}})(\vec{r}\cdot\vec{\mu})r^{i}}{r^{2}}\right]\nonumber \\
 & +\frac{Q}{r^{3}}\left[2\vec{v}\times\vec{\mu}+\frac{3\vec{r}[(\vec{v}\times\vec{r})\cdot\vec{\mu}]}{r^{2}}+3\frac{(\vec{v}\cdot\vec{r})\vec{\mu}\times\vec{r}}{r^{2}}\right]^{i}.\label{eq:Explicit_maLinEM}
\end{align}

In the gravitational system we have, from Eq.~(\ref{eq:mavaccum}),
\begin{equation}
m_{0}a^{i}\simeq F_{{\rm G}}^{i}\simeq-(\mathbb{H}^{u})^{ji}S_{j}+2(\mathbb{E}^{u})_{\ }^{l(i}\epsilon_{\ \ kl}^{j)}v^{k}S_{j}\ .\label{eq:maGLin}
\end{equation}
Again the last term of (\ref{eq:mavaccum}) is negligible for the
non-helical representation (in the purely gravitational case, to this
accuracy, taking $P_{{\rm hidI}}^{\alpha}\approx0\Rightarrow P^{\alpha}\approx mU^{\alpha}$
works generically as a means of picking such a representation \cite{PlyatskoCQG2011}),
as explained in detail in \cite{EPAPS}. The explicit result reads
\begin{align}
m_{0}a^{i}\simeq & -\frac{3}{r^{5}}\left[(\vec{r}\cdot\vec{J})S^{i}+2r^{(j}J^{i)}S_{j}-5\frac{(\vec{r}\cdot\vec{J})(\vec{r}\cdot\vec{S})r^{i}}{r^{2}}\right]\nonumber \\
 & -\frac{3M}{r^{3}}\left[\vec{v}\times\vec{S}+\frac{2\vec{r}[(\vec{v}\times\vec{r})\cdot\vec{S}]}{r^{2}}+\frac{(\vec{v}\cdot\vec{r})\vec{S}\times\vec{r}}{r^{2}}\right]^{i}.\label{eq:Explicit_maLinG}
\end{align}
Comparing with (\ref{eq:Explicit_maLinEM}) we note that all the terms
in the gravitational equation have an electromagnetic counterpart.
However, the spin-orbit interaction terms (second lines of Eqs.~(\ref{eq:Explicit_maLinEM})
and (\ref{eq:Explicit_maLinG})) all have differing factors; these
factors reflect, in this regime, the consequences of the different
symmetries of the tidal tensors, and account for the contrasting effects
studied in Sec.~\ref{sub:Symmetries}. One may check, for instance,
why (\ref{eq:Explicit_maLinG}), but not (\ref{eq:Explicit_maLinEM}),
allows for radial motion in the field of \emph{static} \emph{source}s
($\vec{\mu}_{{\rm s}}=\vec{J}=0$): if $\vec{v}$ is radial, $\vec{v}\times\vec{r}=0$
and $\vec{v}\times\vec{S}=-(\vec{v}\cdot\vec{r})\vec{S}\times\vec{r}/r^{2}$,
so the first and third terms of the second line of Eq. (\ref{eq:Explicit_maLinG})
cancel out, yielding $m_{0}a^{i}=0$. But such cancellation does not
occur in the electromagnetic equation (\ref{eq:Explicit_maLinEM}),
which yields $m_{0}a^{i}\ne0$.

To conclude: from Eqs.~(\ref{tab:Analogy}.3) of Table \ref{tab:Analogy}
we expected that if the fields do not vary along the test particle's
worldline (so that $F_{\alpha\beta;\gamma}U^{\gamma}=0$) then $F_{{\rm EM}}^{\alpha}$
and $F_{{\rm G}}^{\alpha}$ should be similar in the weak field approximation,
since $B_{\alpha\beta}$ and $\mathbb{H}_{\alpha\beta}$ have the
same symmetries and the non-linearities of the later are negligible;
and that otherwise, when $F_{\alpha\beta;\gamma}U^{\gamma}\ne0$,
differences should arise, due to the differing symmetries of the tidal
tensors. In the application herein, this translates into the following:
when the test particles are at rest with respect to the sources the
two forces indeed are similar; however, in the general dynamical case
where the particles move, the two forces differ significantly even
to first order in the velocity (and in the fields), cf.~Eqs.~(\ref{eq:Explicit_maLinEM}),
(\ref{eq:Explicit_maLinG}). Thus the tidal tensor formalism makes
transparent an aspect that can be rephrased as in \cite{Wald}: the
spin-spin interactions in gravity and electromagnetism are very similar
(in this regime), but the ``spin-orbit'' interactions are substantially
different.

In the literature concerning the weak field gravito-electromagnetic
analogy (e.g.~\cite{Gravitation and Inertia,Ruggiero:2002hz,Harris1991,Black Holes}),
the gravitational force acting on a gyroscope is commonly cast in
the form $\vec{F}_{{\rm G}}=K\nabla(\vec{H}\cdot\vec{S})/2$ (where
$K$ is some constant depending on the convention, and $H^{i}\equiv\epsilon^{ijk}g_{0k,j}/K$
is the gravitomagnetic field), similar to its electromagnetic counterpart
$\vec{F}_{{\rm EM}}=\nabla(\vec{B}\cdot\vec{\mu})$, seemingly implying
a similarity between the two interactions. We emphasize that such
expressions are not suited to describe dynamics, as they hold only
if the gyroscope's center of mass is at \emph{rest} in a \emph{stationary}
field (this is usually overlooked in the literature, despite the assertion
in~\cite{Wald}, where this analogy was originally presented, that
it was derived under these conditions). A detailed discussion of these
issues and comparison with the results in the literature is given
in \cite{EPAPS}.

\begin{framed}%
\emph{Sec.~\ref{sub:Weak-field-regime:} in brief} 
\begin{itemize}
\item In the \emph{stationary,} weak field regime, when the particles are
at rest with respect to the sources, the gravitational and electromagnetic
interactions are very similar, having a similar spin-spin force. 
\item When the test particles move, the differences (made clear in the symmetries
of the tidal tensors) \emph{are of leading order}, thus non-negligible
in any slow motion approximation. \end{itemize}
\end{framed}

\section{Beyond pole-dipole --- the torque on the spinning particle\label{sec:Beyond-pole-dipole}}

In the pole-dipole approximation, as we have seen in Sec.~\ref{sub:Spin-``precession''--},
it follows from Eq.~(\ref{eq:EqSpinVector}) that purely magnetic
dipoles with $\vec{\mu}=\sigma\vec{S}$ have $S^{2}$ as a constant
of the motion. This might be somewhat surprising. If one imagines
the magnetic dipole as a spinning charged body, one expects, in a
time-varying magnetic field, the induced electric field to exert in
general (due to its curl) a net torque on it, which will accelerate%
\footnote{Unlike the dipole torque $\vec{\tau}=\vec{\mu}\times\vec{B}$, the
torque due to the induced electric field will not in general be orthogonal
to $\vec{S}$, and hence will change its magnitude. For instance,
in the example in Fig.~\ref{fig:Torques}a) below, $\vec{E}_{{\rm ind}}$
has circular lines around $\vec{S}$, so that $\vec{\tau}_{{\rm ind}}\parallel\vec{S}$.%
} its rotation. Indeed, we have seen in Sec.~\ref{sub:Time-Components}
that the induced electric field does work on the spinning body, causing
a variation $dm/d\tau=-\vec{\mu}\cdot D\vec{B}/d\tau$ of its proper
mass $m$. Such variation is known, from the non-relativistic treatments
in~\cite{Young,Deissler}, where a rigid spherical body is considered,
to be a variation of rotational kinetic energy%
\footnote{It is not cast therein as a variation of proper mass $m$ (as those
are non-relativistic treatments), but of the Hamiltonian term $-\vec{\mu}\cdot\vec{B}$.%
}. Thus we expect it to be associated to a variation of the spinning
angular velocity, and hence of $S^{2}$. However, the dipole torque
in Eqs.~(\ref{eq:EqSpinVector})-(\ref{eq:Spin3+1}) consists only
of the term $\vec{\mu}\times\vec{B}$ (which is there even if the
field is constant, and conserves $S^{2}$); there is no term coupling
to the \emph{derivatives} of the electromagnetic fields, i.e., no
sign of induction phenomena.

As we shall see below, this apparent inconsistency is an artifact
inherent to the pole-dipole approximation, where terms $\mathcal{O}(a^{2})$
($a\equiv$ size of the particle), which are of quadrupole order,
are neglected. Indeed, whereas the contribution of the work done by
the induced electric field to the body's energy is of the type $\vec{\mu}\cdot\vec{B}$,
i.e., of dipole order, the associated torque involves \emph{the trace
of} the quadrupole moment of the charge distribution. Moreover, there
is no analogous torque in the gravitational case, confirming the absence
of an analogous gravitational induction effect.

For clarity, we will treat the two interactions (electromagnetic and
gravitational) separately.

\subsection{Electromagnetic torque\label{sub:Electromagnetic-torque}}

We start by the electromagnetic case in flat spacetime. The equation
for the spin evolution of an extended spinning charged body subject
to an electromagnetic field is, up to quadrupole order (e.g.~Eq.~(8.5)
of \cite{Dixon1967}), 
\begin{eqnarray}
\frac{DS_{{\rm can}}^{\alpha\beta}}{d\tau} & = & 2P_{{\rm Dix}}^{[\alpha}U^{\beta]}+2Q^{\theta[\beta}F_{\ \ \theta}^{\alpha]}\nonumber \\
 &  & +2m_{\ \ \rho\mu}^{[\alpha}F^{\beta]\mu;\rho}\ ,\label{eq:DSQuadrupole}
\end{eqnarray}
where $P_{{\rm Dix}}^{\alpha}$ and $S_{{\rm can}}^{\alpha\beta}$
are defined by Eqs.~(\ref{eq:Pcan}) and (\ref{eq:Scan}), and consist
on the sum of the \emph{physical} momenta $P^{\alpha},\ S^{\alpha\beta}$,
Eqs.~\eqref{eq:Pgeneral}-\eqref{eq:Sab}, plus \emph{electromagnetic
terms}, see Appendix \ref{sec:DixonEqs}. It is shown in \cite{BaileyIsrael}
that $S_{{\rm can}}^{\alpha\beta}$ and $P_{{\rm Dix}}^{\alpha}+qA^{\alpha}\equiv P_{{\rm can}}^{\alpha}$
are the canonical momenta associated to the Lagrangian of the system.
In the equation above $Q^{\alpha\beta}$ is the electromagnetic dipole
moment as defined in (\ref{eq:Qab}), and $m^{\alpha\beta\gamma}$
is an electromagnetic quadrupole moment, defined as 
\begin{equation}
m^{\alpha\beta\gamma}\equiv\frac{4}{3}Q^{(\alpha\beta)\gamma};\qquad Q^{\alpha\beta\gamma}\equiv\mathcal{J}^{\alpha[\beta\gamma]}+\frac{1}{2}q^{\alpha[\beta}U^{\gamma]}\ ,\label{eq:Qabc}
\end{equation}
where $\mathcal{J}^{\alpha\beta\gamma}$ and $q^{\alpha\beta}$ are,
respectively, the current and charge ``quadrupole moments''%
\footnote{\label{fn:QuadrupoleDenomination}Following the convention in e.g.
\cite{Wald et al 2010,Dixon1970II,Dixon1974III}, we dub integrals
of the type (\ref{eq:chargequadrupole})-(\ref{eq:currentquadrupole})
``quadrupole moments''. However, frequently in the literature the
term ``charge quadrupole moment'' refers to the traceless part of
$q^{\alpha\beta}$. Note that $q^{\alpha\beta}\ne0$ for a uniform
spherical body, contrary with its traceless part, which measures a
type of deviation from spherical symmetry (more consistent with the
actual picture of a quadrupole of charges). Sometimes (e.g.~\cite{Gravitation},
p.~977) $q^{\alpha\beta}$ is called the ``second moment of the
charge''.%
}, see Eqs.~(3.8)-(3.9) of \cite{Dixon1967}:%
\footnote{\label{fn:w_n}In Eq.~(3.8) of \cite{Dixon1967}, $w^{\gamma}d\Sigma_{\gamma}$,
instead of $d\Sigma$, appears; \textcolor{black}{however, $w^{\hat{\gamma}}=n^{\hat{\gamma}}+\mathcal{O}(x^{2})$,
cf.~Eq.~(\ref{eq:w}), yielding a correction to the integrand of
order $\mathcal{O}(x^{4})$, negligible to quadrupole order (where
only terms up to $\mathcal{O}(x^{2})$ are to be kept \cite{Madore:1969})}.
Hence we can take therein $w^{\gamma}d\Sigma_{\gamma}\simeq n^{\gamma}d\Sigma_{\gamma}=d\Sigma$,
cf.~Eq.~\eqref{eq:dSigma}.%
} 
\begin{eqnarray}
q^{\hat{\alpha}\hat{\beta}} & \equiv & \int_{\Sigma(\tau,U)}x^{\hat{\alpha}}x^{\hat{\beta}}j^{\hat{\gamma}}d\Sigma_{\hat{\gamma}}\ ;\label{eq:chargequadrupole}\\
\mathcal{J}^{\hat{\alpha}\hat{\beta}\hat{\nu}} & \equiv & \int_{\Sigma(\tau,U)}x^{\hat{\alpha}}x^{\hat{\beta}}j^{\hat{\nu}}d\Sigma\ .\label{eq:currentquadrupole}
\end{eqnarray}
In flat spacetime, the normal coordinates $\{x^{\hat{\alpha}}\}$
are just a rectangular coordinate system originating at $z^{\alpha}(\tau)$.
Decomposing $\mathcal{J}^{\alpha\beta\nu}$ into its projections parallel
and orthogonal to $U^{\nu}$, we obtain 
\begin{equation}
\mathcal{J}^{\alpha\beta\nu}=q^{\alpha\beta}U^{\nu}+\mathcal{J}^{\alpha\beta\gamma}(h^{U})_{\gamma}^{\nu}\ ,\label{eq:JabcDecomp}
\end{equation}
where we noted that, in flat spacetime, $\Sigma(\tau,U)$ is a \emph{hyperplane}
orthogonal to $n^{\alpha}=U^{\alpha}$, thus $-j^{\nu}U_{\nu}d\Sigma=j^{\gamma}d\Sigma_{\gamma}$.

Using Eq.~(\ref{eq:P'_S'}ii), we may re-write Eq.~(\ref{eq:DSQuadrupole})
explicitly in terms of the physical angular momentum $S^{\alpha\beta}$:
\begin{equation}
\frac{DS^{\alpha\beta}}{d\tau}=\frac{DS_{{\rm can}}^{\alpha\beta}}{d\tau}-\frac{DS^{'\alpha\beta}}{d\tau}\ ;\quad S'^{\alpha\beta}=F_{\ \ \sigma}^{[\alpha}q^{\beta]\sigma}\ .\label{eq:DSQuadII}
\end{equation}
Note that $DS'^{\alpha\beta}/d\tau$ is a quadrupole type contribution.

We are interested in the torque $\tau^{\alpha}$, \textcolor{black}{i.e.,
the vector that measures the rate of deviation of the spin vector
from Fermi-Walker transport, Eq. (\ref{eq:FermiTransport}):} 
\begin{equation}
\tau^{\alpha}\equiv\frac{D_{F}S^{\alpha}}{d\tau}\Rightarrow\tau^{\sigma}=\frac{1}{2}\epsilon_{\alpha\beta}^{\ \ \sigma\delta}U_{\delta}\frac{DS^{\alpha\beta}}{d\tau}\ .\label{eq:FermiTorque}
\end{equation}
Using Eqs.~(\ref{eq:DSQuadrupole}), (\ref{eq:DSQuadII}), it follows
that: 
\begin{eqnarray}
\frac{D_{F}S^{\alpha}}{d\tau} & = & \tau_{{\rm DEM}}^{\alpha}+\tau_{{\rm QEM}}^{\alpha}\ ;\label{eq:TorqueEMtotal}\\
\tau_{{\rm DEM}}^{\sigma} & \equiv & \epsilon_{\ \alpha\beta\nu}^{\sigma}U^{\nu}(d^{\alpha}E^{\beta}+\mu^{\alpha}B^{\beta})\ ;\label{eq:Tdem}\\
\tau_{{\rm QEM}}^{\sigma} & \equiv & \tau_{{\rm QEMcan}}^{\sigma}-\tau'^{\sigma}\ ;\label{eq:Tqemtot}\\
\tau_{{\rm QEMcan}}^{\sigma} & \equiv & \epsilon_{\ \alpha\beta\nu}^{\sigma}U^{\nu}m_{\ \ \rho\mu}^{[\alpha}F^{\beta]\mu;\rho}\ ;\label{eq:TqemCan}\\
\tau'^{\sigma} & = & \frac{1}{2}\epsilon_{\ \alpha\beta\nu}^{\lambda}U^{\nu}E^{[\alpha\beta]}\left(q_{\ \gamma}^{\gamma}\delta_{\lambda}^{\sigma}-q_{\ \lambda}^{\sigma}\right)\nonumber \\
 &  & +\frac{1}{2}\epsilon_{\ \alpha\beta\nu}^{\sigma}U^{\nu}F_{\ \gamma}^{\alpha}\frac{Dq^{\beta\gamma}}{d\tau}\ .\label{eq:Tau'}
\end{eqnarray}
Here $\tau_{{\rm DEM}}^{\alpha}$ is the dipole torque already present
in Eq.~(\ref{eq:SpinPirani}), i.e., just a covariant form for $\vec{\tau}=\vec{\mu}\times\vec{B}+\vec{d}\times\vec{E}$.
We split the quadrupole torque $\tau_{{\rm QEM}}^{\sigma}$ into two
parts. The first part, $\tau_{{\rm QEMcan}}^{\sigma}$, which we may
dub the ``canonical electromagnetic quadrupole torque'', is the
torque%
\footnote{In the literature concerning Dixon's multipole scheme, $\tau_{{\rm QEMcan}}^{\sigma}$
is commonly portrayed as the quadrupole torque, see e.g.~\cite{Wald et al 2010}.
However, it is clear from Eq.~(\ref{eq:TorqueEMtotal}) that it is
not the \emph{total} quadrupole torque $\tau_{{\rm QEM}}^{\sigma}$,
and the results below show how crucial this distinction is.%
} coming from the third term of (\ref{eq:DSQuadrupole}) (i.e.~the
quadrupole contribution to $DS_{{\rm can}}^{\alpha\beta}/d\tau$).
The second part, $\tau'^{\sigma}\equiv\frac{1}{2}\epsilon_{\alpha\beta}^{\ \ \sigma\delta}U_{\delta}DS^{'\alpha\beta}/d\tau$,
plays a crucial role in this discussion, since the first term of (\ref{eq:Tau'})
is minus the torque due to the \emph{electric field induced} in the
CM frame by the Maxwell-Faraday law (\ref{eq:MaxFardayGenVector}).
This is what we shall now see.

\subsubsection{The induction torque}

\textcolor{black}{Consider the rectangular coordinates $\{x^{\hat{\alpha}}\}$
to be comoving with the particle's CM, }$\partial_{\hat{0}}=\mathbf{U}$\textcolor{black}{.
In such a frame, the torque (about the CM) due to the induced electric
field is $\vec{\tau}_{{\rm ind}}=\int\rho_{{\rm c}}\vec{x}\times\vec{E}_{{\rm ind}}d^{3}x$,
where $\rho_{c}\equiv-j^{\alpha}U_{\alpha}$ is the charge density
in the CM frame.} Let us expand $\vec{E}$ in a Taylor series around
the CM: $E^{\hat{i}}=E_{{\rm CM}}^{\hat{i}}+E_{{\rm CM}}^{\hat{i},\hat{j}}x_{\hat{j}}+...$
(for the integral above, to quadrupole order, only terms up to linear
order in $\vec{x}$ are to be kept in this expansion), which, splitting
$E_{{\rm CM}}^{\hat{i},\hat{j}}=E_{{\rm CM}}^{[\hat{i},\hat{j}]}+E_{{\rm CM}}^{(\hat{i},\hat{j})}$,
we may write as 
\[
E^{\hat{i}}=E_{{\rm CM}}^{\hat{i}}-\frac{1}{2}[\vec{x}\times(\nabla\times\vec{E})_{{\rm CM}}]^{\hat{i}}+E_{{\rm CM}}^{(\hat{i},\hat{j})}x_{\hat{j}}\ .
\]
The second term is the part of $\vec{E}$ that has a curl, that is,
the induced electric field: $\vec{E}_{{\rm ind}}(x)\approx-\vec{x}\times(\nabla\times\vec{E})_{{\rm CM}}/2$.
(The third term may be cast as a gradient of some scalar function,
thus not related with induction). Therefore, recalling the definition
of $q^{\alpha\beta}$, Eq.~\eqref{eq:chargequadrupole} above, we
have 
\begin{eqnarray}
\tau_{{\rm ind}}^{\hat{i}} & = & -\frac{1}{2}(\nabla\times\vec{E}_{{\rm CM}})^{\hat{j}}\int\rho_{{\rm c}}\left[x^{\hat{i}}x_{\hat{j}}-\delta_{\ \hat{j}}^{\hat{i}}x^{2}\right]d^{3}x\nonumber \\
 & = & -\frac{1}{2}(\nabla\times\vec{E}_{{\rm CM}})^{\hat{j}}\left[q_{\ \hat{j}}^{\hat{i}}-\delta_{\ \hat{j}}^{\hat{i}}q_{\ \gamma}^{\gamma}\right]\ ,\label{eq:tau_ind_i}
\end{eqnarray}
which, by relations (\ref{eq:Eab-Ea;b}), is a non-covariant form
for 
\begin{align}
\tau_{{\rm ind}}^{\alpha} & =\frac{1}{2}\epsilon_{\ \mu\nu\lambda}^{\sigma}U^{\lambda}E^{[\mu\nu]}\left[q_{\ \sigma}^{\alpha}-\delta_{\ \sigma}^{\alpha}q_{\ \gamma}^{\gamma}\right]\label{eq:tauind}\\
 & =\frac{1}{2}B_{\ \beta}^{\sigma}U^{\beta}\left[q_{\ \sigma}^{\alpha}-\delta_{\ \sigma}^{\alpha}q_{\ \gamma}^{\gamma}\right]\ ,\label{eq:tauindBab}
\end{align}
i.e., the first term of (\ref{eq:Tau'}). In the second equality we
used Eqs.~(\ref{tab:Analogy}.4a) of Table \ref{tab:Analogy}.

\subsubsection{Rigid spinning charged body\label{sub:Rigid-spinning-charged}}

Consider the case when the test particle is a charged, ``quasi-rigid''
body \cite{Dixon1970I,EulerTop,EhlersRudoplh}, rotating with an angular
velocity $\Omega^{\alpha}$, defined as follows. If $A^{\alpha}$
is a spatial vector with origin at the CM, orthogonal to the CM 4-velocity
$U^{\alpha}$, and \emph{fixed} to the body, then 
\begin{equation}
\frac{D_{F}A^{\alpha}}{d\tau}=\Omega_{\ \beta}^{\alpha}A^{\beta}\ ;\quad\Omega_{\alpha\beta}=\epsilon_{\beta\alpha\mu\nu}\Omega^{\mu}U^{\nu}\ .\label{eq:Omega}
\end{equation}
Let $U_{{\rm p}}^{\alpha}$ be the 4-velocity field of the points
in the body; we may write $U_{{\rm p}}^{\alpha}=\gamma_{{\rm p}}(U^{\alpha}+v_{{\rm p}}^{\alpha})$;
$\gamma_{{\rm p}}\equiv-U_{{\rm p}}^{\alpha}U_{\alpha}$, cf.~Eqs.~\eqref{eq:U_u},
where $v_{{\rm p}}^{\alpha}=\Omega_{\ \beta}^{\alpha}x^{\beta}$ is
the velocity of a point in the body relative to the CM frame. It follows
that the charge 4-current density is 
\[
j^{\alpha}=\rho_{{\rm c}}(U^{\alpha}+v_{{\rm p}}^{\alpha})=\rho_{{\rm c}}(U^{\alpha}+\Omega_{\ \beta}^{\alpha}x^{\beta})\ ,
\]
whose space part reads, in the CM frame, $j(\vec{r})=\rho_{{\rm c}}\vec{\Omega}\times\vec{x}$;
here $\rho_{{\rm c}}=-j^{\alpha}U_{\alpha}$ is the charge density
as measured in the CM frame. The magnetic dipole moment, Eq.~(\ref{eq:mu_a}),
then becomes 
\begin{align}
\mu^{\hat{\alpha}} & =\frac{\Omega^{\hat{\beta}}}{2}\left[\delta_{\ \hat{\beta}}^{\hat{\alpha}}q_{\ \gamma}^{\gamma}-q_{\ \hat{\beta}}^{\hat{\alpha}}\right]\ ,\label{eq:mu_rigid}
\end{align}
where we used \eqref{eq:chargequadrupole}, and noted that $\rho_{{\rm c}}d\Sigma=j^{\gamma}d\Sigma_{\gamma}$.
Thus the rate of work done on this body by the induction torque $\tau_{{\rm ind}}^{\alpha}$,
$\mathcal{P}=\tau_{{\rm ind}}^{\alpha}\Omega_{\alpha}$, is, from
Eqs.~(\ref{eq:tauind})-(\ref{eq:tauindBab}), 
\begin{equation}
\tau_{{\rm ind}}^{\alpha}\Omega_{\alpha}=-\epsilon_{\beta\mu\nu\lambda}U^{\lambda}E^{[\mu\nu]}\mu^{\beta}=-B_{\ \beta}^{\alpha}U^{\beta}\mu_{\alpha}=-F_{{\rm EM}}^{\alpha}U_{\alpha}\ .\label{eq:WorkTauind}
\end{equation}
That is, we obtain precisely the work $\mathcal{P}_{{\rm ind}}=-F_{{\rm EM}}^{\alpha}U_{\alpha}$
of Sec.~\ref{sub:Time-components-in-CM-frame }, Eqs.~(\ref{eq:Induction1})-(\ref{eq:Induction}).
This is the result we seek: we have just proved that the work transferred
to the body by Faraday's law of induction, which, to pole-dipole order,
is manifest in the projection along $U^{\alpha}$ of the dipole force
$F_{{\rm EM}}^{\alpha}$ (and in the variation of the proper mass
$dm/d\tau$), is indeed associated to an induction torque, which causes
$S^{2}$ to vary as expected (since $\tau_{{\rm ind}}^{\alpha}$ is
not orthogonal to $S^{\alpha}$ in general). This torque was known
to exist from some non-relativistic treatments \cite{Young,YoungQuestion66,Deissler}
dealing with the special case of spinning\emph{ spherical} charged
bodies. It just happens that it is \emph{not} manifest to pole-dipole
order, as it involves the second moment of the charge $q_{\alpha\beta}$,
which is of quadrupole order%
\footnote{\label{fn:MomentInertia}Note also that in order to assign a moment
of inertia $I_{\alpha\beta}$ and an angular velocity to a spinning
particle one must go beyond dipole order, as $I_{\alpha\beta}=(h^{U})_{\alpha\beta}(m_{{\rm Q}})_{\ \tau}^{\tau}-(m_{{\rm Q}})_{\alpha\beta}$
(cf. e.g.~\cite{Gravitation}), where $(m_{{\rm Q}})_{\alpha\beta}$
is the mass ``quadrupole'', Eq.~(\ref{eq:MassQuadrupole}).%
}. But the rate of work that this torque does, $\tau_{{\rm ind}}^{\alpha}\Omega_{\alpha}$,
is manifest to dipole order, since, for a rigid body, $q_{\alpha\beta}$
and $\Omega^{\alpha}$ combine into the magnetic dipole moment $\mu^{\alpha}$,
by virtue of Eq.~(\ref{eq:mu_rigid}).

\subsubsection{Torque on spherical charged body\label{sub:Electromagnetic Torque-and-force on Spherical}}

In this context, and in view of a comparison with the gravitational
problem, it is interesting to consider the case of a uniform, spherical
charged body, whose quadrupole moments of $j^{\alpha}$ reduce to
the trace of $q_{\alpha\beta}$, so that we expect the total quadrupole
torque on the particle $\tau^{\alpha}$ to come essentially from $\tau_{{\rm ind}}^{\alpha}$.

First let us explicitly compute the quadrupole moments for this type
of body. It is clear that the charge quadrupole, Eq.~(\ref{eq:chargequadrupole}),
is such that, in rectangular coordinates $\{x^{\hat{\alpha}}\}$ originating
at the center of mass and comoving with it ($\partial_{\hat{0}}=\mathbf{U}$),
its time components are zero, $q^{\hat{0}\hat{0}}=q^{\hat{0}\hat{i}}=0$,
and its spatial part reduces to its trace, $q^{\hat{i}\hat{j}}=\delta^{\hat{i}\hat{j}}q_{\ \hat{k}}^{\hat{k}}/3$.

Such tensor is covariantly written as 
\begin{equation}
q^{\alpha\beta}=\frac{1}{3}q_{\ \tau}^{\tau}(h^{U})^{\alpha\beta}\ .\label{eq:qabsphere}
\end{equation}
As for the tensor $\mathcal{J}^{\alpha\beta\gamma}$, due to the axisymmetry
and the reflection symmetry with respect to the equatorial plane,
all of its spatial components $\mathcal{J}^{\hat{i}\hat{j}\hat{k}}$
in the CM frame vanish. The only non-vanishing components are $\mathcal{J}^{\hat{i}\hat{j}\hat{0}}=\int_{x^{\hat{0}}=0}x^{\hat{i}}x^{\hat{j}}j^{\hat{0}}d^{3}x=q^{\hat{i}\hat{j}}$.
Hence $\mathcal{J}^{\alpha\beta\gamma}(h^{U})_{\gamma}^{\nu}=0$;
thus, by virtue of Eq.~(\ref{eq:JabcDecomp}), $\mathcal{J}^{\alpha\beta\nu}=q^{\alpha\beta}U^{\nu}$,
and therefore 
\begin{eqnarray}
\mathcal{J}^{\alpha\beta\nu} & = & \frac{1}{3}q_{\ \sigma}^{\sigma}U^{\nu}(h^{U})^{\alpha\beta}\ .\label{eq:JabcSphere}
\end{eqnarray}
Substituting (\ref{eq:qabsphere}) and (\ref{eq:JabcSphere}) into
(\ref{eq:Qabc}), we have 
\begin{equation}
Q^{\alpha\beta\gamma}=\frac{1}{2}q_{\ \tau}^{\tau}(h^{U})^{\alpha[\beta}U^{\gamma]}=\frac{1}{2}q_{\ \tau}^{\tau}g^{\alpha[\beta}U^{\gamma]}\ .\label{eq:QabcSphere}
\end{equation}

Let us now compute the quadrupole torque exerted on the body, Eq.
\eqref{eq:Tqemtot}. Substituting (\ref{eq:QabcSphere}), (\ref{eq:Qabc})
into Eq.~(\ref{eq:TqemCan}), we obtain 
\begin{equation}
\tau_{{\rm QEMcan}}^{\sigma}=\frac{1}{3}q_{\ \gamma\ }^{\gamma}U^{[\alpha}F_{\ \ \ ;\lambda}^{\beta]\lambda}\epsilon_{\alpha\beta}^{\ \ \sigma\delta}U_{\delta}=0\,,\label{eq:torqueQEMcanShere}
\end{equation}
the second equality holding in vacuum (which is the problem at hand)
by virtue of Maxwell's equations $F_{\ \ ;\beta}^{\alpha\beta}=4\pi j^{\alpha}$.
This means that $\tau_{{\rm QEM}}^{\sigma}=-\tau'^{\sigma}$. In order
to compute $\tau'^{\sigma}$, Eq.~(\ref{eq:Tau'}), we must give
a law of evolution for $q_{\alpha\beta}$. Eq.~(\ref{eq:qabsphere})
guarantees that the body is spherical; we also demand $dq_{\alpha}^{\alpha}/d\tau=0$,
so that it has constant size (in a comoving frame). Together, these
relations imply that $q_{\alpha\beta}$ is Fermi-Walker transported,
$D_{F}q_{\alpha\beta}/d\tau=0$, i.e., it has constant components
in an orthonormal tetrad comoving with the body's CM, as expected.
It then follows from Eqs.~(\ref{eq:Tqemtot}), (\ref{eq:Tau'}) and
(\ref{eq:tauind})-(\ref{eq:tauindBab}) that the quadrupole torque
reduces to 
\begin{align}
 & \tau_{{\rm QEM}}^{\sigma}=-\tau'^{\sigma}=\tau_{{\rm ind}}^{\sigma}+\frac{1}{6}\epsilon_{\ \alpha\beta\lambda}^{\sigma}U^{\lambda}a^{\alpha}E^{\beta}q_{\ \gamma}^{\gamma}\,;\label{eq:TqemSpherical}\\
 & \tau_{{\rm ind}}^{\sigma}=-\frac{q_{\ \gamma}^{\gamma}}{3}\epsilon_{\ \alpha\beta\lambda}^{\sigma}U^{\lambda}E^{[\alpha\beta]}=-\frac{q_{\ \gamma}^{\gamma}}{3}B_{\ \beta}^{\sigma}U^{\beta}\ .\label{eq:TindSpherical}
\end{align}
In other words, up to an acceleration dependent term (arising from
the Fermi-Walker transport of $q_{\alpha\beta}$), $\tau_{{\rm QEM}}^{\sigma}$
is \emph{the torque due to the induced electric field}.

To compare with the results known in the literature, consider a body
with uniform charge and mass densities. For such a body we may write
$2\sigma I_{\ \beta}^{\alpha}=(q_{\gamma}^{\ \gamma}(h^{U})_{\beta}^{\alpha}-q_{\ \beta}^{\alpha})$,
where $\sigma\equiv q/2m$ is the classical gyromagnetic ratio and
$I_{\alpha\beta}$ the moment of inertia tensor (cf. Footnote \ref{fn:MomentInertia});
in the spherical case we have $q_{\sigma}^{\ \sigma}/3=\sigma I_{\sigma}^{\ \sigma}/3=\sigma I$,
where $I=I_{zz}=I_{xx}=I_{yy}$ denotes the moment of inertia of the
sphere with respect to any axis of rotation passing through its center.
Thus $\tau_{{\rm ind}}^{\alpha}=-\sigma I\epsilon_{\ \mu\nu\lambda}^{\alpha}U^{\lambda}E^{[\mu\nu]}=-\sigma IB_{\ \beta}^{\alpha}U^{\beta}$.
In the CM frame, and in vector notation, the total torque (\ref{eq:FermiTorque})
on such body reads 
\begin{equation}
\vec{\tau}\equiv\frac{D\vec{S}}{d\tau}=\vec{\tau}_{{\rm DEM}}+\vec{\tau}_{{\rm QEM}}=\vec{\mu}\times\vec{B}-\sigma I\frac{D\vec{B}}{d\tau}-\frac{\sigma I}{2}\vec{a}\times\vec{E}\ ,\label{eq:DfermiSSphere}
\end{equation}
which is the relativistic generalization of Eq.~(1) of~\cite{Young},
or Eq.~(6) of~\cite{YoungQuestion66} (those non-relativistic results
follow from Eq.~(\ref{eq:DfermiSSphere}) above by replacing $\tau\rightarrow t$,
and neglecting the acceleration dependent term).

\emph{Work done on the particle and rotational kinetic energy.---}
Let us now compute the work, $\tau^{\sigma}\Omega_{\sigma}$, done
by the total torque $\tau^{\alpha}=\tau_{{\rm DEM}}^{\alpha}+\tau_{{\rm QEM}}^{\alpha}$
on the particle. First note that, for a ``quasi-rigid'' body, the
relation $S^{\alpha}=I^{\alpha\beta}\Omega_{\beta}$ holds~\cite{Dixon1970I};
which, for a uniform spherical body, becomes 
\begin{equation}
S^{\alpha}=I\Omega^{\alpha}\ .\label{eq:S_Omega}
\end{equation}
Hence, assuming the proportionality $\mu^{\alpha}=\sigma S^{\alpha}$,
it follows from Eq.~(\ref{eq:Tdem}) (with $d^{\alpha}=0$, which
is the problem at hand) that the work of the dipole torque is zero,
$\tau_{{\rm DEM}}^{\alpha}\Omega_{\alpha}=0$. Thus, $\tau^{\sigma}\Omega_{\sigma}=\tau_{{\rm QEM}}^{\alpha}\Omega_{\alpha}$.
From Eqs.~(\ref{eq:mu_rigid}) and (\ref{eq:qabsphere}), we have
\begin{equation}
\mu^{\alpha}=\frac{1}{3}\Omega^{\alpha}q_{\ \gamma}^{\gamma}\ ,\label{eq:muSpherical}
\end{equation}
and therefore, from Eqs.~(\ref{eq:TqemSpherical}) and (\ref{eq:PhidDecompMP}),
\[
\tau^{\sigma}\Omega_{\sigma}=\tau_{{\rm ind}}^{\sigma}\Omega_{\sigma}+\frac{1}{2}\epsilon_{\ \alpha\beta\lambda}^{\sigma}U^{\lambda}a^{\alpha}E^{\beta}\mu_{\sigma}=\tau_{{\rm ind}}^{\sigma}\Omega_{\sigma}-\frac{1}{2}P_{{\rm hidEM}}^{\alpha}a_{\alpha}.
\]
Now consider the case when there is no electromagnetic hidden momentum
($P_{{\rm hidEM}}^{\alpha}=0$), as is the case of the setup in Fig.
\ref{fig:Torques}a); then $\tau^{\sigma}\Omega_{\sigma}=\tau_{{\rm ind}}^{\sigma}\Omega_{\sigma}$.
On the other hand, from Eqs.~(\ref{eq:FermiTorque}) and (\ref{eq:S_Omega}),
we have that $\tau^{\sigma}=ID_{F}\Omega^{\sigma}/d\tau$ and $2\tau^{\sigma}\Omega_{\sigma}=Id(\Omega^{2})/d\tau$.
Therefore, using (\ref{eq:WorkTauind}), we obtain 
\begin{equation}
\frac{1}{2}I\frac{d(\Omega^{2})}{d\tau}=\tau_{{\rm ind}}^{\sigma}\Omega_{\sigma}=-F_{{\rm EM}}^{\alpha}U_{\alpha}\ .\label{eq:KinRotTimeProj}
\end{equation}
Note that $I\Omega^{2}/2$ is the body's kinetic energy of rotation
about its CM, see e.g.~\cite{Dixon1970I,EulerTop}; hence Eq. (\ref{eq:KinRotTimeProj})
tells us that, for this setup, the rate of variation of the body's
kinetic energy of rotation equals the rate of work, \emph{as measured
in the CM frame}, done by the dipole force $F_{{\rm EM}}^{\alpha}$
on the particle (that is, its projection $-F_{{\rm EM}}^{\alpha}U_{\alpha}$).

Observing, from Eqs.~ (\ref{tab:Analogy}.1a) of Table \ref{tab:Analogy},
(\ref{eq:FstarDecomp}), and (\ref{eq:PhidDecompMP}), that $P_{{\rm hidEM}}^{\alpha}=0$
implies $F_{{\rm EM}}^{\alpha}U_{\alpha}=\mu_{\alpha}DB^{\alpha}/d\tau$,
and using $\mu^{\alpha}=\sigma S^{\alpha}$, together with Eqs.~(\ref{eq:TorqueEMtotal}),
(\ref{eq:Tdem}), (\ref{eq:TqemSpherical}) and (\ref{eq:TindSpherical}),
we can rewrite Eq.~(\ref{eq:KinRotTimeProj}) as 
\begin{align}
\frac{I}{2}\frac{d(\Omega^{2})}{d\tau}= & -\frac{d(B^{\alpha}\mu_{\alpha})}{d\tau}+B^{\alpha}\frac{D\mu_{\alpha}}{d\tau}\label{eq:KinRotYoung}\\
= & -\frac{d(B^{\alpha}\mu_{\alpha})}{d\tau}-\frac{\sigma q_{\gamma}^{\gamma}}{6}\left[\frac{d(B^{2})}{d\tau}+\epsilon_{\ \alpha\beta\delta}^{\sigma}U^{\delta}a^{\alpha}E^{\beta}B_{\sigma}\right],\nonumber 
\end{align}
which is the relativistic generalization of Eq.~(10) of \cite{YoungQuestion66}
(the acceleration dependent term is absent therein). From this equation
we see that, for this setup, the varying part of the mass, $-B^{\alpha}\mu_{\alpha}$,
present in the dipole approximation, Eq.~(\ref{eq:m}), is \emph{kinetic
energy of rotation} (\emph{not} potential energy, as claimed in some
literature, e.g.~\textcolor{black}{\cite{Dixon1970I,WeyssenhoffRaabe,Dixon1965}}).
This establishes, in a relativistic covariant formulation, and in
the context of Dixon's multipole approach, the claims in \textcolor{black}{\cite{Coombes,Young,YoungQuestion66,Deissler}}.
The second terms in the right members of Eqs.~(\ref{eq:KinRotYoung}),
of quadrupole order, are not manifest in the \emph{dipole order} mass
equation (\ref{eq:dm2}), since to that accuracy $B^{\alpha}D\mu_{\alpha}/d\tau=0$,
by virtue of Eq.~(\ref{eq:EqSpinVector}).

\subsection{Gravitational torque\label{sub:Gravitational-torque}}

The equation for the spin evolution of an extended body in a gravitational
field is, up to quadrupole order~\cite{Dixon1974III,Wald et al 2010}
\begin{equation}
\frac{DS^{\kappa\lambda}}{d\tau}=2P^{[\kappa}U^{\lambda]}+\ \frac{4}{3}J^{\mu\nu\rho[\kappa}R_{\ \rho\mu\nu}^{\lambda]}\ ,\label{eq:DSQuadGrav}
\end{equation}
leading to the torque (cf.~Eq.~\eqref{eq:FermiTorque}) 
\begin{equation}
\frac{D_{F}S^{\sigma}}{d\tau}=\tau_{{\rm QG}}^{\sigma}\ ,\qquad\tau_{{\rm QG}}^{\sigma}\equiv\frac{4}{6}J^{\mu\nu\rho[\kappa}R_{\ \rho\mu\nu}^{\lambda]}\epsilon_{\kappa\lambda}^{\ \ \sigma\delta}U_{\delta}\ .\label{eq:TauQG}
\end{equation}
Here (cf.~Eqs.~(9.12) of \cite{Dixon1974III} or (5.29) of \cite{Dixon1967})

\begin{equation}
J^{\alpha\beta\gamma\delta}=\frac{1}{2}\left(t^{\gamma[\alpha\beta]\delta}-t^{\delta[\alpha\beta]\gamma}\right)-U^{[\alpha}p^{\beta][\gamma\delta]}-U^{[\gamma}p^{\delta][\alpha\beta]}\ ,\label{eq:Jabcd}
\end{equation}
where the moments $t^{\alpha\beta\gamma\delta}$ and $p^{\alpha\beta\gamma}$
can be written, in Riemann normal coordinates $\{x^{\hat{\alpha}}\}$,
as 
\begin{eqnarray}
t^{\hat{\alpha}\hat{\beta}\hat{\gamma}\hat{\delta}} & \equiv & \int_{\Sigma(\tau,U)}x^{\hat{\alpha}}x^{\hat{\beta}}T^{\hat{\gamma}\hat{\delta}}d\Sigma\ ;\label{eq:tabcd}\\
p^{\hat{\alpha}\hat{\beta}\hat{\gamma}} & \equiv & \int_{\Sigma(\tau,U)}x^{\hat{\alpha}}x^{\hat{\beta}}J^{\hat{\gamma}}d\Sigma\ ,\label{eq:pabc}
\end{eqnarray}
\textcolor{black}{where $J^{\gamma}\equiv-T^{\gamma\delta}n_{\delta}$
is the mass/energy current as measured by the observers orthogonal
to $\Sigma(\tau,U)$ (so that $T^{\gamma\delta}d\Sigma_{\delta}=J^{\gamma}d\Sigma$,
cf.~Eq.~\eqref{eq:dSigma}). Expressions \eqref{eq:tabcd}-\eqref{eq:pabc}
correspond}%
\footnote{\textcolor{black}{Noting that $w^{\hat{\sigma}}d\Sigma_{\hat{\sigma}}=d\Sigma+\mathcal{O}(x^{2})$,
cf.~Footnote \ref{fn:w_n}, and that, in the system $\{x^{\hat{\alpha}}\}$,
the bitensors in }\cite{Dixon1974III}\textcolor{black}{{} read $\sigma^{\hat{\alpha}}=-x^{\hat{\alpha}}$,
$\sigma_{\ \hat{\beta}}^{\hat{\alpha}}=\delta_{\ \hat{\beta}}^{\hat{\alpha}}+\mathcal{O}(x^{2})$,
$\Theta^{\hat{\kappa}\hat{\lambda}\hat{\mu}\hat{\nu}}=\delta^{\hat{\kappa}(\hat{\mu}}\delta^{\hat{\nu})\hat{\lambda}}+\mathcal{O}(x^{2})$,
$H_{\hat{\alpha}\hat{\beta}}=\delta_{\hat{\alpha}\hat{\beta}}+\mathcal{O}(x^{2})$;
so the corrections due to them in \eqref{eq:tabcd}-\eqref{eq:pabc}
are integrands of order $\mathcal{O}(x^{4})$, negligible to quadrupole
order (where only terms up to $\mathcal{O}(x^{2})$ are to be kept).}%
}\textcolor{black}{, in flat spacetime, to Eqs.~(5.2)-(5.3) of }\cite{Dixon1967}\textcolor{black}{,
and, in curved spacetime, to Eqs.~(9.4) and (9.11) of }\cite{Dixon1974III}\textcolor{black}{.
They are }\textcolor{black}{\emph{tensors}}\textcolor{black}{{} (similarly
to the expressions in }\cite{Dixon1974III}\textcolor{black}{), since
the use of Riemann normal coordinates $\{x^{\hat{\alpha}}\}$ amounts
to defining the moments in terms of the exponential map (}see \cite{Madore:1969,CostaNatario2014}).

\textcolor{black}{The tensor $p^{\alpha\beta\gamma}$ has the interpretation
of the quadrupole moment of the mass current, analogous to the quadrupole
moment of the charge current} $\mathcal{J}^{\alpha\beta\gamma}$\textcolor{black}{,
Eq.~(\ref{eq:currentquadrupole}). Note moreover that $-t^{\alpha\beta\gamma\delta}U_{\delta}=p^{\alpha\beta\gamma}$,
since $n_{\hat{\alpha}}=U_{\hat{\alpha}}+\mathcal{O}(x^{2})$, cf.
Eq. (\ref{eq:n-U}), and therefore, to quadrupole order, we may take
$J^{\gamma}\simeq-T^{\gamma\delta}U_{\delta}$ in \eqref{eq:pabc}.
}We may thus decompose $t^{\alpha\beta\gamma\delta}$ as 
\begin{equation}
t^{\alpha\beta\gamma\delta}=p^{\alpha\beta\gamma}U^{\delta}+p^{\alpha\beta\sigma}(h^{U})_{\ \sigma}^{\delta}U^{\gamma}+t^{\alpha\beta\lambda\sigma}(h^{U})_{\ \lambda}^{\gamma}(h^{U})_{\ \sigma}^{\delta}\ .\label{eq:tabcdDecomp}
\end{equation}
Similarly, $p^{\alpha\beta\gamma}$ may also be decomposed as 
\begin{equation}
p^{\alpha\beta\gamma}=(m_{{\rm Q}})^{\alpha\beta}U^{\gamma}+p^{\alpha\beta\lambda}(h^{U})_{\lambda}^{\gamma}\ ,\label{eq:p_abc_decomp}
\end{equation}
analogous to \eqref{eq:JabcDecomp}, where 
\begin{equation}
(m_{{\rm Q}})^{\hat{\alpha}\hat{\beta}}=\int_{\Sigma(\tau,U)}x^{\hat{\alpha}}x^{\hat{\beta}}J^{\hat{\gamma}}d\Sigma_{\hat{\gamma}}\label{eq:MassQuadrupole}
\end{equation}
is the mass ``quadrupole'' (or ``second moment of the mass'';
see \cite{ThorneHartle85,Gravitation,Wald et al 2010} and Footnote
\ref{fn:QuadrupoleDenomination}), analogous to the charge quadrupole
(\ref{eq:chargequadrupole}).

\subsubsection{Torque on ``spherical'' body\label{sub:Garv. Torque-on-spinning sphere}}

Our goal in this section is to consider the gravitational analogue
of the problem in Sec.~\ref{sub:Electromagnetic Torque-and-force on Spherical}.
Therein we considered a spherical charged body in flat spacetime,
whose charge quadrupole moment was shown to reduce to its trace, $q_{\ \beta}^{\alpha}=q_{\ \tau}^{\tau}(h^{U})_{\ \beta}^{\alpha}/3$,
and whose current quadrupole was $\mathcal{J}^{\alpha\beta\nu}=q_{\ \sigma}^{\sigma}U^{\nu}(h^{U})^{\alpha\beta}/3$.
We prescribe the analogous test body for the gravitational problem
by demanding it to have \emph{an analogous multipole structure} (i.e.,
an analogous ``gravitational skeleton''~\cite{MathissonNeueMechanik}),
rather than demanding its shape to be ``spherical'', which in a
general curved spacetime is not a well defined notion. (A body with
such multipole structure will of course be a sphere in the case of
flat spacetime; and otherwise may be thought of as one if the field
is not too strong). As shown above, the quadrupole moment $p^{\alpha\beta\gamma}$,
Eq.~(\ref{eq:pabc}), has an analogous definition to $\mathcal{J}^{\alpha\beta\gamma}$,
Eq.~(\ref{eq:currentquadrupole}), only with $J^{\alpha}$ in the
place of $j^{\alpha}$; hence its structure must be (analogously to
Eq.~(\ref{eq:JabcSphere})): 
\begin{equation}
p^{\alpha\beta\gamma}=\frac{1}{3}(m_{{\rm Q}})_{\ \tau}^{\tau}(h^{U})^{\alpha\beta}U^{\gamma}\ .
\end{equation}
The last term of (\ref{eq:tabcdDecomp}) is the quadrupole moment
of the space part of $T^{\gamma\delta}$, $(h^{U})_{\ \lambda}^{\gamma}(h^{U})_{\ \sigma}^{\delta}T^{\lambda\sigma}\equiv T_{\perp}^{\gamma\delta}$,
which has no electromagnetic analogue. For a ``quasi-rigid'' spinning
body, we have (e.g.~\cite{Frehland1971,EhlersRudoplh}) $T^{\alpha\beta}({\rm p})=\rho U_{{\rm {\rm p}}}^{\alpha}U_{{\rm p}}^{\beta}+s^{\alpha\beta}$,
where $s^{\alpha\beta}$ are the stresses, $U_{{\rm p}}^{\alpha}=\gamma_{{\rm p}}(U^{\alpha}+v_{{\rm p}}^{\alpha})$
is the 4-velocity of the (rotating) mass element at the point ${\rm p}$
of the body, $v_{{\rm p}}^{\alpha}$ is the spatial velocity of ${\rm p}$
relative to the center of mass frame, and $\gamma_{{\rm p}}=-U_{{\rm p}}^{\alpha}U_{\alpha}$,
see decomposition \eqref{eq:U_u}. Hence $T_{\perp}^{\alpha\beta}=\rho\gamma_{{\rm p}}^{2}v_{{\rm {\rm p}}}^{\alpha}v_{{\rm p}}^{\beta}+s_{\perp}^{\alpha\beta}$,
its two terms being of the same order of magnitude $\sim\rho v_{{\rm p}}^{2}$
(e.g.~\cite{Frehland1971}). For non-relativistic rotation speeds
$v_{{\rm p}}\ll1$, we have $\|T_{\perp}^{\alpha\beta}\|\ll\rho$,
and therefore the last term of (\ref{eq:tabcdDecomp}) is negligible
compared to the others. It then follows: 
\begin{eqnarray}
J^{\alpha\beta\gamma\delta} & \approx & -(m_{{\rm Q}})_{\ \tau}^{\tau}U^{[\alpha}g^{\beta][\gamma}U^{\delta]}\ ,\label{eq:JabcdSphere}
\end{eqnarray}
(in agreement with Eq.~(7.31) of \cite{Dixon1970I}). Substituting
in Eq.~(\ref{eq:TauQG}), we obtain the gravitational torque: 
\[
\tau_{{\rm QG}}^{\sigma}=-\frac{1}{3}(m_{{\rm Q}})_{\ \tau}^{\tau}U^{[\alpha}R_{\ \ \mu}^{\beta]}U^{\mu}\epsilon_{\alpha\beta}^{\ \ \sigma\delta}U_{\delta}=0\ ,
\]
the second equality holding for vacuum ($R^{\mu\nu}=0)$, which (as
in the electromagnetic case) is the problem at hand. Thus, no gravitational
torque is exerted, up to quadrupole order, on a spinning spherical
body%
\footnote{This is consistent with the results from the post-Newtonian treatment
in e.g. \cite{ThorneHartle85}, where the approximate vacuum expression
$\tau_{{\rm QG}}^{i}\approx\epsilon_{\ jk}^{i}[\mathbb{E}_{\ l}^{j}\mathscr{J}{}^{lk}+4\mathbb{H}_{\ l}^{j}\mathscr{S}{}^{lk}/3]$
(Eq. (1.9c) therein) is derived. In our notation, $\mathscr{S}^{jk}=\epsilon_{\ \ lm}^{(k}p^{j)lm}$,
$\mathscr{J}_{ij}=(m_{{\rm Q}})_{ij}-(m_{{\rm Q}})_{\ k}^{k}\delta_{ij}/3$;
it then follows from the analysis above that, for a spherical body,
$\mathscr{S}_{ij}=\mathscr{J}_{ij}=0\Rightarrow\vec{\tau}_{{\rm QG}}=0$.%
}. This means that there is no gravitational counterpart to the electromagnetic
torque $\tau_{{\rm ind}}^{\alpha}$ exerted on the spherical charged
body of Sec.~\ref{sub:Electromagnetic Torque-and-force on Spherical}
(generated, from the viewpoint of the particle's frame, by the induced
electric field). This is the result we expected from the discussion
in Sec.~\ref{sub:analogy based on tidal tensors}: $\tau_{{\rm ind}}^{\alpha}$
comes from the antisymmetric part of $E_{\alpha\beta}$, or, equivalently,
from the (time) projection along $U^{\beta}$ of $B_{\alpha\beta}$,
cf.~Eqs.~(\ref{eq:tauind})-(\ref{eq:tauindBab}). Since the gravitoelectric
tidal tensor $\mathbb{E}_{\alpha\beta}$ is symmetric, and $\mathbb{H}_{\alpha\beta}$
is spatial with respect to $U^{\beta}$, the absence of an analogous
torque in gravity is thus natural.

\subsection{Summarizing with a simple realization\label{sub:Summarizing-with-a_simple}}

\begin{figure}
\includegraphics[width=0.5\textwidth]{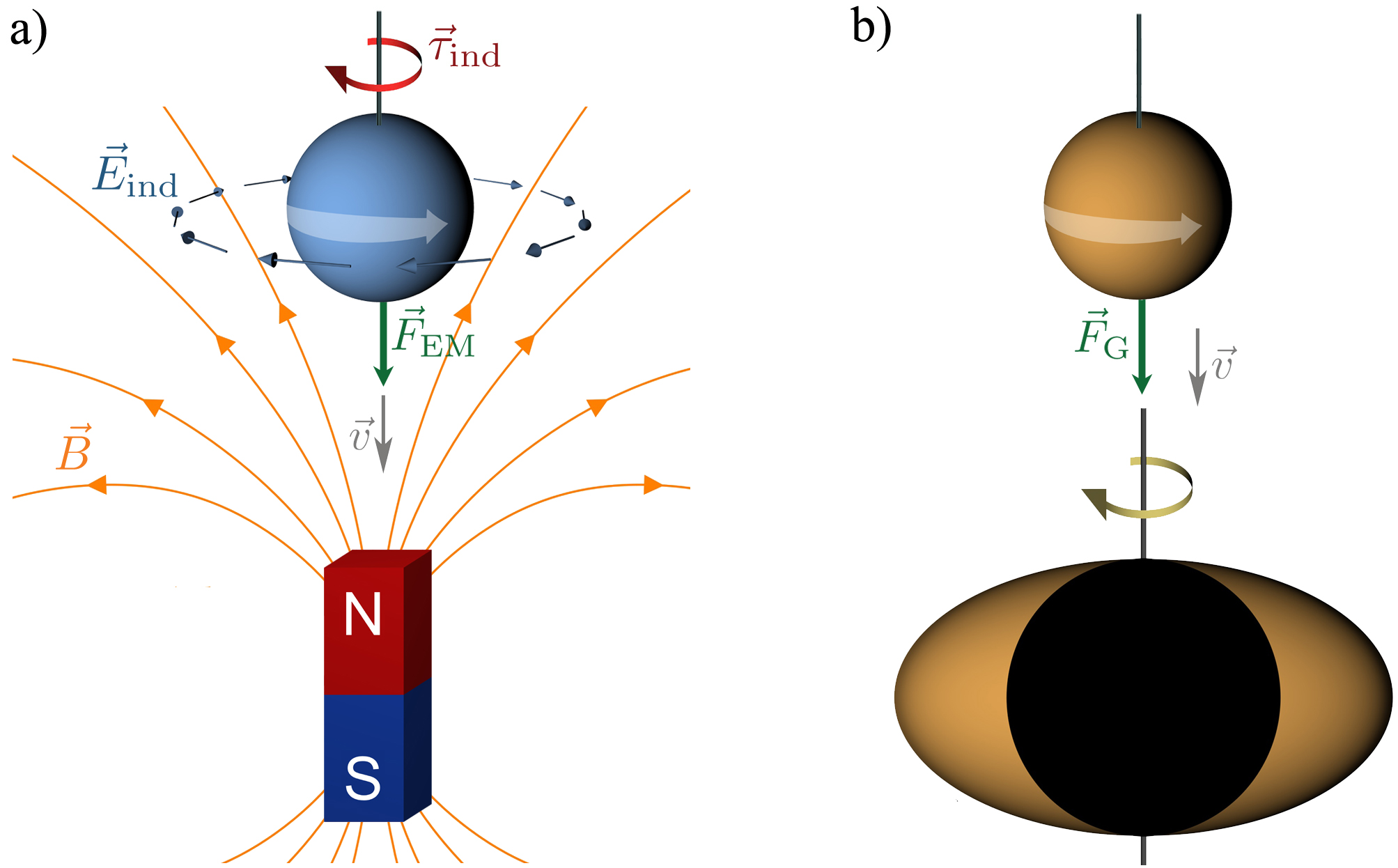}

\caption{\label{fig:Torques}a) A spinning, positively charged spherical body
being pulled by a strong magnet; $\vec{E}_{{\rm ind}}\equiv$ electric
field induced in the body's CM frame. b) A spinning spherical body
falling into a Kerr black hole. As the spinning charge moves in the
inhomogeneous magnetic field $\vec{B}$, a torque $\tau_{{\rm ind}}^{\alpha}$,
Eq.~(\ref{eq:TindSpherical}), is exerted on it due to $\vec{E}_{{\rm ind}}$,
i.e., due to the skew part $E_{[\alpha\beta]}$ of the electric tidal
tensor. This causes $S$, and the body's angular velocity $\Omega=S/I$,
to vary. The torque $\tau_{{\rm ind}}^{\alpha}$ does work at a rate
$\tau_{{\rm ind}}^{\alpha}\Omega_{\alpha}=\mathcal{P}_{{\rm ind}}$,
which exactly matches the time projection of the dipole force $F_{{\rm EM}}^{\alpha}$
it its rest frame, cf.~Eq.~(\ref{eq:WorkTauind}). This causes the
body's kinetic energy of rotation to decrease, manifest in a decrease
of proper mass $m$, and canceling out the gain in translational kinetic
energy ($\mathcal{P}_{{\rm trans}}$), so that the total work transfer,
as \emph{measured in the ``laboratory'' frame}, is \emph{zero} (cf.~Sec.~\ref{sub:Time-components-Static}).
In the gravitational case no analogous induction effects occur (as
expected, since $\mathbb{E}_{[\alpha\beta]}=0$): no torque is exerted
on the spinning particle; its angular momentum $S$, angular velocity
$\Omega$, and proper mass $m$, are constant; and there is a \emph{net}
work done on it by $F_{{\rm G}}^{\alpha}$ at a rate $\mathcal{P}_{{\rm tot}}=F_{{\rm G}}^{\alpha}v_{\alpha}$,
corresponding to an increase of translational kinetic energy.}
\end{figure}

The results in Secs.~\ref{sub:Electromagnetic-torque} and \ref{sub:Gravitational-torque}
entirely corroborate the discussion in Sec.~\ref{sub:Time-Components}
(and Sec.~\ref{sub:Mass-of-the}); namely, the manifestation of electromagnetic
induction and the absence of an analogous phenomenon in the \emph{physical}
gravitational forces and torques. In this context, it is interesting
to consider the analogous setups in Fig.~\ref{fig:Torques}: a spinning
spherical charge moving in the field of a strong magnet (or another
spinning charged body), and a spinning ``spherical'' mass moving
in Kerr spacetime.

Let us start by the electromagnetic case. A force $F_{{\rm EM}}^{\alpha}$,
Eq.~(\ref{tab:Analogy}.1a), will be exerted on the particle, causing
it to move (thereby gaining translational kinetic energy, at a rate
$\mathcal{P}_{{\rm trans}}$, Eq.~(\ref{eq:Ptrans})). As it moves
in an inhomogeneous magnetic field, a torque $\tau_{{\rm ind}}^{\alpha}$
is exerted upon it; from the viewpoint of the observer comoving with
the particle, this is due to the electric field induced by the time-varying
magnetic field. That torque will cause a variation of the particle's
angular momentum $S^{\alpha}$, and therefore of its angular velocity
$\Omega^{\alpha}=S^{\alpha}/I$ (measured with respect to the comoving
Fermi-Walker transported tetrad, cf.~Eq.~(\ref{eq:Omega})). Clearly,
$S^{2}$ \emph{is not conserved}, since $d(S^{\alpha}S_{\alpha})/d\tau=2\tau^{\alpha}S_{\alpha}\ne0$,
as we see from Eqs.~(\ref{eq:TqemSpherical})-(\ref{eq:TindSpherical})
or (\ref{eq:DfermiSSphere}). The variation of $\Omega$ also implies
a variation of the particle's rotational kinetic energy, equal to
the work of the torque $\tau_{{\rm ind}}^{\alpha}$, which in turn
is \emph{exactly} the work done by the dipole force $F_{{\rm EM}}^{\alpha}$
as measured in the \emph{frame comoving} with the particle, cf.~Eq.
\eqref{eq:KinRotTimeProj}. (This is reflected in a variation of the
particle's proper mass $m$.) From the point of view of the ``laboratory''
frame (i.e., the ``static observers'' $u^{\alpha}$), no net work
is done on the particle, $F_{{\rm EM}}^{\alpha}u_{\alpha}=0$, and
its total energy, $E=-P_{\alpha}u^{\alpha}$, is conserved, cf. \ref{sub:Time-components-Static}.
That means that the rate of variation in translational kinetic energy
$\mathcal{P}_{{\rm trans}}$ of the center of mass is exactly canceled
out by the variation of rotational kinetic energy $\mathcal{P}_{{\rm ind}}$
(the work of $\tau_{{\rm ind}}^{\alpha}$), guaranteeing that a stationary
magnetic field does not do work.

In the gravitational case, there is also a net force $F_{{\rm G}}^{\alpha}$
on the body, cf.~Eq.~(\ref{tab:Analogy}.1b) of Table \ref{tab:Analogy},
causing it to gain kinetic energy at a rate $\mathcal{P}_{{\rm trans}}=F_{{\rm G}}^{\alpha}v_{\alpha}$.
But no torque is exerted on it; up to quadrupole order we have 
\[
\frac{D_{F}S^{\alpha}}{d\tau}=0\ ;\quad S^{2}=\mbox{constant}
\]
(i.e., the spin vector of the spinning spherical mass is Fermi-Walker
transported), implying $\Omega=\mbox{constant}$. This is consistent
with the constancy of the proper mass (manifest in the fact that $F_{{\rm G}}^{\alpha}$
is orthogonal to $U^{\alpha}$), because, since there is no torque,
the kinetic energy of rotation is constant. Thus in this case the
gain in translational kinetic energy is not canceled out by a variation
of rotational kinetic energy, and therefore a stationary gravitomagnetic
field will do a net rate of work $-F_{{\rm G}}^{\alpha}u_{\alpha}=\mathcal{P}_{{\rm trans}}$
on the particle.

\textcolor{black}{We close this section with a few additional remarks.}
The application in Fig.~\ref{fig:Torques} illustrates an important
aspect of the frame dragging effect, and the contrast with the electromagnetic
analogue. For clarity, let us consider the case when the test balls
are initially non-spinning. In the electromagnetic case, Fig.~\ref{fig:Torques}a,
as the ball moves towards the magnet, it starts spinning, increasingly
faster (relative to the Fermi-Walker transported tetrad) due to the
torque $\tau_{{\rm ind}}^{\alpha}$. In the frame comoving with the
ball, $\vec{\tau}_{{\rm ind}}$ is due to the induced electric field
$\vec{E}_{{\rm ind}}$; and from the point of view of the laboratory
frame (static observers), where the field is stationary (thus there
is no induced electric field therein), $\vec{\tau}_{{\rm ind}}$ comes
from the overall effect of the Lorentz force $dq\vec{v}\times\vec{B}$
applied to each charge element $dq$ of the ball. In the gravitational
case, Fig.~\ref{fig:Torques}b, no such rotation arises. \emph{If}
initially $\Omega^{\alpha}=0$, the ball in Fig.~\ref{fig:Torques}b
will never gain any rotation \emph{relative to the local compass of
inertia}; $S^{\alpha}$ remains always zero. Indeed, an observer sitting
firmly with his tetrad on top of the ball will not detect any sign
of rotation: he will not measure any Coriolis forces acting on any
test particle that he may throw, and will see gyroscope axes fixed.
However, from the point of view of a frame adapted to the static observers
(which is anchored to the ``distant stars'', see Sec. \ref{sub:Spin-``precession''--}),
the ball indeed starts spinning increasingly faster as it approaches
the black hole. This is because, due to frame dragging, a system of
axes which is \emph{locally} \emph{non-rotating} (i.e., Fermi-Walker
transported) close to the black hole, is seen to be rotating from
a frame fixed to the distant stars. The effect is larger the closer
one gets to the black hole, and is quite analogous to the electromagnetic
situation \emph{as viewed by the static observers}: in the linear
limit, it is well known \cite{PaperIAU,PaperAnalogies,Gravitation and Inertia,Gravitation and Spacetime,Ciufolini Nature Review,Ruggiero:2002hz}
that the gravitomagnetic field $\vec{H}$ is very similar to its electromagnetic
analogue; then the gravitomagnetic ``force'' $\vec{v}\times\vec{H}$,
acting on each mass element, seemingly leads to an analogous ``torque''.
These are not, however, real forces or torques, but artifacts of the
reference frame, not measurable in any local experiment (only by locking
the frame to the distant stars, e.g.~by means of a telescope); it
is therefore no surprise that they are not manifest in the torque
equation \eqref{eq:TauQG}. For indeed \emph{it is the static observers
that rotate} relative to the local compass of inertia, which is manifest
in the fact that they have vorticity, and measure a non-zero $\vec{H}$
(causing, in their frame, test particles in geodesic motion to be
deflected by fictitious Coriolis forces $\vec{v}\times\vec{H}$, and
gyroscopes to ``precess'', cf. Sec. \ref{sub:Spin-``precession''--};
for more details, see e.g. \cite{PaperAnalogies} Secs. 3.2-3.3).
This contrasts with the situation in the electromagnetic analogue,
where $\tau_{{\rm ind}}^{\alpha}$ is a physical, covariant torque,
causing the particle to indeed have an accelerated rotation with respect
to the local compass of inertia.%
\begin{framed}%
\emph{Sec.~\ref{sec:Beyond-pole-dipole}} \emph{in brief} 
\begin{itemize}
\item The electromagnetic quadrupole torque contains the torque $\tau_{{\rm ind}}^{\alpha}$
due to Faraday's law of induction; it is a coupling of $E_{[\alpha\beta]}$
to $q_{\alpha\beta}$ (the charge ``quadrupole'').

\begin{itemize}
\item Dipole approximation ignores $q_{\alpha\beta}$; hence $\tau_{{\rm ind}}^{\alpha}$
is not manifest to dipole order; 
\item but the rate of work it does, $\tau_{{\rm ind}}^{\alpha}\Omega_{\alpha}$,
\emph{is of dipole order} ($\Omega^{\alpha}$ and $q_{\alpha\beta}$
combining into $\mu^{\alpha}$). For a rigid body, it equals the projection
of the dipole force along its worldline, $-F_{{\rm EM}}^{\alpha}U_{\alpha}$. 
\end{itemize}
\item The torque $\tau_{{\rm ind}}^{\alpha}$ has no gravitational analogue
(consistent with $\mathbb{E}_{[\alpha\beta]}=0$). 
\item A time-varying electromagnetic field torques a spherical charged body,
changing its angular momentum $S$, angular velocity $\Omega$, and
kinetic energy of rotation (manifest in $m$). The gravitational field
\emph{never} torques a ``spherical'' body; $S$, $\Omega$, and
$m$, are \emph{constant}. \end{itemize}
\end{framed}

\section{Conclusion}

In this work we studied the dynamics of spinning test particles in
general relativity, in the framework of \emph{exact} gravito-electromagnetic
analogies. A detailed summary of the main results and realizations
is given in Sec.~\ref{sec:Introduction}; herein we conclude with
some additional remarks.

Both equations of motion --- force and spin evolution --- of a spinning
particle in a gravitational field are related to their electromagnetic
counterparts by \emph{exact} analogies, valid for generic fields.
Moreover, a third analogy arises, for the so-called ``hidden momentum'',
first obtained in \cite{Wald et al 2010} as an approximate result,
and introduced herein in its exact form. All these analogies are shown
to emerge from the rigorous equations of motion for pole-dipole particles
if the Mathisson-Pirani spin condition is employed.

The first remark we want to make is that it is important to realize
that the existence of these analogies does \emph{not} mean that the
interactions are similar. These are \emph{functional} analogies: the
magnetic tidal tensor $B_{\alpha\beta}$ plays in Eq.~(\ref{tab:Analogy}.1a)
of Table \ref{tab:Analogy}, for the force exerted on a magnetic dipole,
the same role as the gravitomagnetic tidal tensor $\mathbb{H}_{\alpha\beta}$
in Eq.~(\ref{tab:Analogy}.1b) for the gravitational force exerted
on a gyroscope. The analogy extends to the Maxwell and Einstein field
equations, as manifest in Table \ref{tab:Analogy}. Moreover, in the
appropriate frame, the gravitomagnetic field $\vec{H}$ plays in the
``precession'' of the gyroscope an analogous role to $\vec{B}$
in the precession of a magnetic dipole, cf.~Eq.~(\ref{eq:Spin3+1})
(the analogy also extends, under certain conditions, to the equations
for the geodesics, for the force on the test particle, and to the
field equations, see \cite{PaperAnalogies,The many faces,Natario}).
But the analogies \emph{do} \emph{not} \emph{imply}, even in seemingly
analogous setups, that the objects are similar. First, $\vec{H}$
and $\mathbb{H}_{\alpha\beta}$, unlike their electromagnetic counterparts,
are non-linear. Second, even in the weak field regime (where the non-linearities
of the gravitational field can be neglected), the symmetries and the
time projections of the tidal tensors $B_{\alpha\beta}$ and $\mathbb{H}_{\alpha\beta}$
continue to differ crucially. The apparent similarity suggested by
the usual linear approaches in the literature, e.g.~\cite{Gravitation and Inertia,Ruggiero:2002hz,Harris1991,Black Holes},
can be misleading, as the differing terms in the force/acceleration
equations are of leading order, as shown in Sec.~\ref{sub:Weak-field-regime:}.
We have actually seen (Sec.~\ref{sub:Time-Components}) cases where
the electromagnetic and gravitational effects are\emph{ opposite}:
in a frame comoving with the test particle, the work done by the spin-curvature
force $F_{{\rm G}}^{\alpha}$ is zero ($F_{{\rm G}}^{\alpha}U_{\alpha}=0$)
whereas the work of its electromagnetic counterpart $F_{{\rm EM}}^{\alpha}$
is non-zero ($F_{{\rm EM}}^{\alpha}U_{\alpha}\ne0$); from the point
of view of ``static observers'' $u^{\alpha}$, the situation is
reversed: it is the electromagnetic force that does no work, $F_{{\rm EM}}^{\alpha}u_{\alpha}=0$
(stationary electromagnetic fields cannot do work on a magnetic dipole)
whereas the gravitational one does, $F_{{\rm G}}^{\alpha}u_{\alpha}\ne0$.

The analogies are instead suited for a comparison between the two
interactions, as this amounts to comparing mathematical objects that
play analogous dynamical roles in both theories. It is the main point
of this work that one can learn a lot (about \emph{both} of them)
from such a comparison. The differences in the structure of the gravitational
and electromagnetic tidal tensors encode fundamental differences in
the interactions, namely the phenomenon of electromagnetic induction,
and the way \emph{it manifests itself in the electromagnetic tidal
forces and torques}, which has no analogue in gravity. We have seen
in Sec.~\ref{sub:Symmetries} that $B_{\alpha\beta}$ has an antisymmetric
part, reading, in vacuum, $2B_{[\alpha\beta]}=\star F_{\alpha\beta;\gamma}U^{\gamma}$.
This equation (which encodes the Maxwell equation $\nabla\times\vec{B}=\partial\vec{E}/\partial t$)
tells us that whenever the field varies along the particle's worldline
(e.g.~when it moves in a non-uniform electric field), $B_{[\alpha\beta]}\ne0$,
hence $B_{\alpha\beta}$ is non-vanishing, and so a force $F_{{\rm EM}}^{\alpha}=B_{\beta}^{\ \alpha}\mu^{\beta}\ne0$
is exerted on the magnetic dipole (except for some special orientations
of $\vec{\mu}$). Such induction effect has no counterpart in gravity,
since, in vacuum, $\mathbb{H}_{\alpha\beta}$ is always symmetric;
indeed, it is possible for particles moving in a (non-uniform) gravitational
field to measure $\mathbb{H}_{\alpha\beta}=0$, so that no force is
exerted on them, $F_{{\rm G}}^{\alpha}=-\mathbb{H}_{\beta}^{\ \alpha}S^{\beta}=0$.
This leads to the existence of geodesic motions for spinning particles,
as exemplified in Secs.~\ref{sub:Radial-Schwa} and \ref{sub:EquatorialKerr}
by radial geodesics in Schwarzschild spacetimes, and circular geodesics
in Kerr-dS. Reinforcing the insight of the analogy, the velocity fields
for which $\mathbb{H}_{\alpha\beta}=0$ mirror the ones where, in
the electromagnetic analogue, $B_{\alpha\beta}$ reduces to its antisymmetric
part.

Likewise, the results in Sec.~\ref{sub:Time-Components}, concerning
the time components of the force, and in Sec.~\ref{sec:Beyond-pole-dipole},
concerning the torque exerted on the spinning particle, are manifestations
of the antisymmetric part of the electric tidal tensor $E_{\alpha\beta}$
(or, equivalently, to the projection of $B_{\alpha\beta}$ along $U^{\alpha}$),
and of the absence of a gravitational counterpart. The antisymmetric
part $E_{[\alpha\beta]}$ encodes the Maxwell-Faraday law $\nabla\times\vec{E}=-\partial\vec{B}/\partial t$;
the gravitoelectric tidal tensor by contrast is symmetric, $\mathbb{E}_{[\alpha\beta]}=0$,
translating in an absence of analogous induction effects in the \emph{physical}%
\footnote{In the framework of inertial forces, the fact that the time-dependent
gravitoelectric $\vec{G}$ and gravitomagnetic $\vec{H}$ fields have
a curl, in analogy with their electromagnetic counterparts, can be
interpreted as analogous to the electromagnetic induction laws, see
e.g.~\cite{Nordtvedt1988}. These, however, are reference frame artifacts;
such curls do not contribute to the tidal tensors $\mathbb{E}_{\alpha\beta}$,
$\mathbb{H}_{\alpha\beta}$ (i.e., to the tidal forces, which are
the only locally measurable forces of gravity), \emph{only the symmetrized
derivatives} of $\vec{G}$ and $\vec{H}$ do. For more details see
Secs 3.5 and 4 of \cite{PaperAnalogies}.%
} gravitational forces and torques. In this framework, we understood
the variation of proper mass $m$ of a classical particle with magnetic
dipole moment --- it arises from the work done on it by the induced
electric field (at a rate $\mathcal{P}_{{\rm ind}}=-F_{{\rm EM}}^{\alpha}U_{\alpha}$),
encoded in the projection of $B_{\alpha\beta}$ along the particle's
4-velocity $U^{\alpha}$ --- and why $m$ is conserved for a gyroscope
in a gravitational field --- it is because $\mathbb{H}^{\alpha\beta}$
is spatial with respect to $U^{\alpha}$, signaling the absence of
an analogous effect. We have also understood the contrast between
the work of these forces as measured by static observers, and the
spin dependence of Hawking's upper bound \cite{Hawking} for the energy
released when two black holes collide: if one considers a magnetic
dipole falling into a strong magnet (Fig.~\ref{fig:DipoleMagnet}b)),
there is no net gain in the particle's energy (from the point of view
of static observers); any gain in translational kinetic energy is
exactly canceled out by the work transferred to the dipole by Faraday's
induction law (i.e., by a loss in proper mass $dm/d\tau=\mathcal{P}_{{\rm ind}}$),
ensuring that the stationary magnetic field does no net work on it.
In gravity, however, since $\mathcal{P}_{{\rm ind}}$ has no counterpart
($m$ is constant), such cancellation does not occur, and therefore
a net work $-F_{{\rm G}}^{\alpha}u_{\alpha}=F_{{\rm G}}^{\alpha}v_{\alpha}$
is done on a gyroscope; there is a potential energy associated with
it, of which the Hawking-Wald spin interaction energy \cite{Wald}
is a special case. In other words: the gravitational spin interaction
energy, and the spin dependence of the black hole collision energy
(at least in the case where one black hole is much smaller than the
other, so that it can be treated as a test particle moving in a stationary
field), are justified by the fact that, \emph{unlike its electromagnetic
counterpart}, a stationary gravitational (tidal) field does work on
mass currents.

The analogies and formalism herein also provide useful tools and intuition
for practical applications, which is exemplified in Sec.~\ref{sub:Symmetries}.
From the \emph{formal} analogy between the quadratic invariants of
the Maxwell and Weyl tensors, we guessed that $\mathbb{H}_{\alpha\beta}$
should vanish for observers at rest or moving radially in the Schwarzschild
spacetime, in analogy with the situation for $B^{\alpha}$ in a Coulomb
field. The tidal tensor form of the spin-curvature force, $F_{{\rm G}}^{\alpha}=-\mathbb{H}_{\beta}^{\ \alpha}S^{\beta}$,
then tells us that no force is exerted on gyroscopes comoving with
such observers; for instance, a gyroscope dropped from rest will fall
along a geodesic towards the singularity. In the same framework, we
predicted that in the equatorial plane of the Kerr or Kerr-dS spacetimes
there should be velocity fields for which $\mathbb{H}_{\alpha\beta}=0$
(because it is so for $B^{\alpha}$ in the equatorial plane of a spinning
charge), and from that the existence of circular geodesics for spinning
particles in Kerr-dS (which were not known in the literature, to our
knowledge). Note that even the problem of the radial fall in the Schwarzschild
spacetime (the simplest in this work) could be a complex problem outside
the tidal tensor formalism/the Mathisson-Pirani spin condition (involving
possibly complicated descriptions, and difficulties in setting up
its initial conditions, see Appendix \ref{sub:Comparison-of-the-SSC-aplications}).
As for the geodesics for gyroscopes in Kerr-dS, it would be very difficult
to ever notice the effect otherwise.

In the course of this work a number of issues concerning the dynamics
of spinning particles in general relativity were clarified. First,
the problem of the equations of motion for pole-dipole particles;
the gravitational part is well established, but difficulties exist
in the electromagnetic part, as there are different versions of the
equations in the literature, and inconsistencies in their physical
interpretation, whose clarification is the purpose of Appendix \ref{sub:Dixon's-Equations}.
Moreover, the time projections of the forces, their physical content,
and relationship with the mass of the particle and the work done by
the fields, is ignored in most literature, or misunderstood (e.g.~\textcolor{black}{\cite{Dixon1970I,Dixon1965,WeyssenhoffRaabe,PomeranskiiKhriplovich,Khriplovich2008})};
they are thoroughly discussed in Sec.~\ref{sub:Time-Components}
and (for particles with electric dipole moment) in Appendix \ref{sec:The-electric-dipole}.
Another important clarification was made in Sec.~\ref{sub:Electromagnetic-torque},
concerning the quadrupole order torque according to Dixon's equations
\cite{Dixon1974III,Dixon1967,Wald et al 2010}, and the physical meaning
of the quantities involved therein. In their usual form they are equations
for the ``canonical'' angular momentum $S_{{\rm can}}^{\alpha\beta}$,
Eq.~(\ref{eq:SpinCan}), \emph{not} for the physical angular momentum
$S^{\alpha\beta}$, Eq.~(\ref{eq:Sab}); failing to notice this leads
one to overlook the torque ($\tau_{{\rm ind}}^{\alpha}$) exerted
on the body due to the curl of the electric field (i.e., to the antisymmetric
part of the electric tidal tensor), and to incorrectly conclude e.g.
that the electromagnetic field cannot torque a spherical body ---
which is known, from basic electromagnetism \textcolor{black}{\cite{Young,YoungQuestion66,Deissler},}
to be false, and would be at odds with the variation of the particle's
mass discussed in Secs. \ref{sub:Mass-of-the} and \ref{sub:Time-Components}
(which, for a rigid body, is essentially a variation of rotational
kinetic energy, cf.~Sec.~\ref{sub:Electromagnetic Torque-and-force on Spherical}).

As a future direction, we plan an investigation of the gravito-electromagnetic
analogies in the equations of motion for spinning particles to quadrupole
and higher orders in the multipole expansion.

\section*{Acknowledgments}

We thank Rui Quaresma (quaresma.rui@gmail.com) for the illustrations.
We thank João Penedones for reading the manuscript, remarks and very
helpful suggestions, and C. Herdeiro, L. Wyllemann, R. M. Wald, A.
I. Harte, I. Ciufolini, J. M. B. L. Santos, E. J. S. Lage and O. Semerák
for useful discussions. L. F. C. is funded by FCT through grant SFRH/BDP/85664/2012.
L.F.C. and J.N. were partially funded by FCT/Portugal through project
PEst-OE/EEI/LA0009/2013. M.Z. is supported by Grants MEC FPA2013-46570-C2-1-P,
MEC FPA2013-46570-C2-2-P, MDM-2014-0369 of ICCUB, 2014-SGR-104, 2014-SGR-1474,
CPAN CSD2007-00042 Consolider-Ingenio 2010, and ERC Starting Grant
HoloLHC-306605.

\begin{appendices}

\appendix

\section{The equations of motion for spinning particles in the literature\label{sec:DixonEqs}}

It is perhaps surprising that the problem of the covariant equations
describing the motion of spinning particles subject to gravitational
and electromagnetic fields is still not generally well understood,
with different methods and derivations leading to different versions
of the equations, whose relation is not always clear. Curiously, it
is the electromagnetic field that has been posing more problems (some
authors~\cite{PomeranskiiKhriplovich,Khriplovich2008} have even
concluded that such covariant description is not possible). The equations
of motion for pole-dipole particles in electromagnetic fields are
derived in unambiguous forms in~\cite{Wald et al}, for special relativity,
and in~\cite{Dixon1964} in the context of general relativity. Rigorous
derivations are also given in \cite{Dixon1967,Dixon1974III,Wald et al 2010};
in this case, however, one must be aware of the subtleties involved
in their interpretation. These equations (unlike the ones in \cite{Dixon1964,Dixon1965,Wald et al})
are symmetric with respect to electric and magnetic dipoles; this
is actually the most common form of the equations, appearing in many
other works, e.g.~\cite{BhabhaCorben,Corben,CorbenBook,Dixon1970I,BiniGemelliRuffini}.
If not properly interpreted, that would lead to physically inconsistent
predictions (given the different nature of the two dipole models),
as we shall see below and in Appendix \ref{sec:The-electric-dipole}.
Moreover, if one takes the ``angular momentum'' tensor defined in
\cite{Dixon1967,Dixon1974III} as the physical one, the ``torque''
equations therein would, at quadrupole order, seemingly contradict
well known results from elementary electromagnetism (and experimental
evidence), as discussed in Sec.~\ref{sub:Electromagnetic-torque}.
Herein we \textcolor{black}{will dissect these issues and explain
how the different versions of the equations relate to each other,
and to the ones used in this work.}

\subsection{Relation with the equations used in this work\label{sub:Relation-with-the-Eqs-Herein}}

Equations (\ref{eq:ForceDS0})-(\ref{eq:SpinDS0}) correspond to Dixon's
equations (6.31)-(6.32) of~\cite{Dixon1964} (cf.~also (3.1)-(3.2)
of~\cite{Dixon1965}), with the following simplifications in the
definitions of the moments:

\textcolor{black}{1) instead of the bitensors }in \cite{Dixon1964},
we use\textcolor{black}{{} (following \cite{Madore:1969}) the exponential
map to define the moments in curved spacetime (which amounts to using
Riemann normal coordinates $\{x^{\hat{\alpha}}\}$ in the integrals
\eqref{eq:Pgeneral}-\eqref{eq:mu_a}). The bitensor $-\sigma^{;\alpha}$
of }\cite{Dixon1964}\textcolor{black}{, which is the vector at $z^{\alpha}$
tangent to the geodesic connecting $z^{\alpha}$ to the point of integration
$x^{\alpha}$, and whose length equals that of the geodesic, has,
in the system $\{x^{\hat{\alpha}}\}$, coordinates given simply by
$-\sigma^{;\hat{\alpha}}=x^{\hat{\alpha}}$. The bitensor of geodesic
displacement $\bar{g}_{\ \alpha}^{\kappa}$ of }\cite{Dixon1964}\textcolor{black}{{}
reads, in the system $\{x^{\hat{\alpha}}\}$, $\bar{g}_{\ \hat{\alpha}}^{\hat{\kappa}}=\delta_{\ \hat{\alpha}}^{\hat{\kappa}}+\mathcal{O}(x^{2})$
(see Appendix of \cite{CostaNatario2014}); thus to dipole order (which
is linear in $x$), $\bar{g}_{\ \hat{\alpha}}^{\hat{\kappa}}\simeq\delta_{\ \hat{\alpha}}^{\hat{\kappa}}$,
and indeed} our definitions of $P^{\alpha}$, $S^{\alpha\beta}$ ($\equiv J^{\alpha\beta}$
in~\cite{Dixon1964}), and $d^{\alpha}$ ($\equiv q^{\alpha}$ in~\cite{Dixon1964})
agree with \cite{Dixon1964}.

2) The vector $w^{\gamma}$ involved in the definition of $\mu_{\alpha\beta}$
($\equiv m_{\alpha\beta}$ in \cite{Dixon1964}) via the moment $j^{\alpha\beta}$
therein, which is a vector such that displacement of every point by
$w^{\gamma}d\tau$ maps $\Sigma(\tau)$ into $\Sigma(\tau+d\tau)$,
can, to pole-dipole order, be taken as $w^{\gamma}\simeq n^{\gamma}$.
That is, $w^{\gamma}d\Sigma_{\gamma}\simeq d\Sigma$, cf.~Eq.~(\ref{eq:dSigma}).
This is easily seen in the case of flat spacetime%
\footnote{\label{fn:w}\textcolor{black}{It suffices for this purpose to work
in flat spacetime; a generalization of $w^{\alpha}$ to curved spacetime
only amounts to small corrections to something already negligible
in special relativity.}%
} \textcolor{black}{\cite{Mathisson Variational,Dixon1967}, where
we have (for $\Sigma(U)$ orthogonal to $U^{\alpha}$, and noting
that $n^{\hat{\alpha}}=U^{\hat{\alpha}}$), 
\begin{equation}
w^{\hat{\gamma}}=n^{\hat{\gamma}}\left(1-\frac{x_{\hat{\alpha}}}{n_{\hat{\beta}}n^{\hat{\beta}}}\frac{Dn^{\hat{\alpha}}}{d\tau}\right)=n^{\hat{\gamma}}(1+x_{\hat{\alpha}}a^{\hat{\alpha}})\ .\label{eq:w}
\end{equation}
} \textcolor{black}{Hence $j^{\hat{\alpha}\hat{\beta}}\equiv\int_{\Sigma(\tau,U)}x^{\hat{\alpha}}j^{\hat{\beta}}w^{\hat{\gamma}}d\Sigma_{\hat{\gamma}}$,
}Eq.~(6.8)\textcolor{black}{{} of }\cite{Dixon1964}\textcolor{black}{,
reads 
\[
j^{\hat{\alpha}\hat{\beta}}=\int_{\Sigma(\tau,U)}x^{\hat{\alpha}}j^{\hat{\beta}}d\Sigma-a_{\hat{\sigma}}\int_{\Sigma(\tau,U)}x^{\hat{\sigma}}x^{\hat{\alpha}}j^{\hat{\beta}}d\Sigma\ ,
\]
the second term being negligible to pole-dipole order.}

3)\textcolor{black}{{} The 1-form $n_{\alpha}$ normal to $\Sigma(\tau,U)$
reads, in the coordinates $\{x^{\hat{\alpha}}\}$, $n_{\hat{\alpha}}=(-1,0,0,0)(-g^{\hat{0}\hat{0}})^{-1/2}$.
Since $g^{\hat{0}\hat{0}}=-1+\mathcal{O}(x^{2})$ (see e.g.~\cite{Gravitation}),
and, at the reference worldline $z^{\alpha}$, $n_{\hat{\alpha}}=(-1,0,0,0)=U_{\hat{\alpha}}$,
we have 
\begin{equation}
n_{\hat{\alpha}}=U_{\hat{\alpha}}+\mathcal{O}(x^{2})\ ;\label{eq:n-U}
\end{equation}
hence, to dipole order, we may take (when of interest) $d\Sigma_{\hat{\delta}}\equiv-n_{\hat{\delta}}d\Sigma\simeq-U_{\hat{\delta}}d\Sigma$.
It follows that $-j^{\alpha\beta}U_{\beta}=d^{\alpha}$, cf. Eq. (\ref{eq:d_a}),
and therefore the magnetic dipole tensor $m^{\alpha\beta}$ defined
in }\cite{Dixon1964}\textcolor{black}{{} as $m^{\alpha\beta}=j^{[\alpha\beta]}-d^{[\alpha}U^{\beta]}$
matches ours: $m^{\alpha\beta}=(h^{U})_{\ \gamma}^{\alpha}(h^{U})_{\ \delta}^{\beta}j^{[\gamma\delta]}=\mu^{\alpha\beta}$,
cf. Eqs. \eqref{eq:mu_ab}, \eqref{eq:mu_a}.}

\textcolor{black}{4)} The moments are defined relative to an hypersurface
of integration $\Sigma(\tau,U)$ normal to $U^{\alpha}$ at $z^{\alpha}$,
as done in~\cite{Dixon1967,Wald et al}, whereas in e.g.~\cite{Dixon1964,Dixon1970I,Dixon1974III}
hypersurfaces $\Sigma(\tau,P)$ orthogonal to $P^{\alpha}$ are used.
That does not change the shape of the equations to dipole order, as
one can check%
\footnote{In the purely gravitational case ($F^{\alpha\beta}=0$), the integrals
(\ref{eq:Pgeneral})-(\ref{eq:Sab}), defined at $z^{\alpha}(\tau)$
over an hypersurface $\Sigma(\tau,u)$ orthogonal to $u^{\alpha}$,
are actually, to pole-dipole order, independent of $u^{\alpha}$,
see \cite{CostaNatario2014}.%
} comparing the equations in \cite{Dixon1964,Dixon1970I,Dixon1974III}
with the ones in \cite{Dixon1967} (identifying the appropriate quantities,
as explained in Sec.~\ref{sub:Dixon's-Equations} below), or in the
independent derivation in~\cite{Wald et al}.

\subsection{Dixon's ``symmetric'' equations\label{sub:Dixon's-Equations}}

In later works by Dixon~\cite{Dixon1967,Dixon1970I,Dixon1974III}
the equations of motion for spinning particles are presented in a
different form, e.g.~Eqs.~(1.33)-(1.34) of~\cite{Dixon1974III},
symmetric with respect to the electric and magnetic dipoles. Taking
into account the different signature and conventions, they read, to
dipole order, 
\begin{equation}
\frac{DP_{{\rm Dix}}^{\alpha}}{d\tau}=qF^{\alpha\beta}U_{\beta}+\frac{1}{2}F^{\mu\nu;\alpha}Q_{\mu\nu}-\frac{1}{2}R_{\ \beta\mu\nu}^{\alpha}S^{\mu\nu}U^{\beta}\ ,\label{eq:ForceCan}
\end{equation}
\begin{equation}
\frac{DS_{{\rm can}}^{\alpha\beta}}{d\tau}=2P_{{\rm Dix}}^{[\alpha}U^{\beta]}+2Q^{\theta[\beta}F_{\ \ \theta}^{\alpha]}\ ,\label{eq:SpinCan}
\end{equation}
where $Q^{\alpha\beta}$ is the \emph{electromagnetic dipole} moment
tensor about $z^{\alpha}(\tau)$ (Eq.~(5.62) of \cite{Dixon1970II},
or, for flat spacetime, Eq.~(3.44) of \cite{Dixon1967}), which reads,
in the system $\{x^{\hat{\alpha}}\}$, 
\begin{equation}
Q^{\hat{\alpha}\hat{\beta}}\equiv\int_{\Sigma(\tau,U)}x^{[\hat{\alpha}}j^{\hat{\beta}]}d\Sigma+U^{[\hat{\beta}}\int_{\Sigma(\tau,U)}x^{\hat{\alpha}]}j^{\hat{\gamma}}d\Sigma_{\hat{\gamma}}\ .\label{eq:Qab}
\end{equation}
Since $d\Sigma_{\hat{\delta}}\simeq-U_{\hat{\delta}}d\Sigma$, cf.
Eq. (\ref{eq:n-U}), this tensor embodies the intrinsic electric and
magnetic dipoles $d^{\alpha}$ and $\mu_{\alpha\beta}$, Eqs.~(\ref{eq:d_a})-(\ref{eq:mu_ab}),
as its time and space projections with respect to $U^{\alpha}$, 
\begin{equation}
d^{\alpha}=-Q^{\alpha\beta}U_{\beta},\quad\mu^{\alpha\beta}=(h^{U})_{\ \gamma}^{\alpha}(h^{U})_{\ \delta}^{\beta}Q^{\gamma\delta}\ ,\label{eq:Megnetic_Electric_Dipole}
\end{equation}
in terms of which it has the decomposition 
\[
Q^{\alpha\beta}=2d^{[\alpha}U^{\beta]}+\epsilon^{\alpha\beta\gamma\delta}\mu_{\gamma}U_{\delta}\ .
\]
It must be noted that $P_{{\rm Dix}}^{\alpha}$ and $S_{{\rm can}}^{\alpha\beta}$
($P^{\alpha}$, $S^{\alpha\beta}$ in the notation of~\cite{Dixon1967,Dixon1970I,Dixon1970II,Dixon1974III})
are \textit{not} \emph{the physical} momentum and angular momentum
given by Eqs.~(\ref{eq:Pgeneral}) and (\ref{eq:Sab}) above, but
instead contain additional electromagnetic terms, cf.~\cite{Dixon1967,Dixon1970I}.
In our framework, they can be written as 
\begin{equation}
P_{{\rm Dix}}^{\alpha}=P^{\alpha}+P'^{\alpha}\ ,\quad P'^{\hat{\alpha}}\equiv\int_{\Sigma(z,U)}\Psi^{\hat{\alpha}}j^{\hat{\beta}}d\Sigma_{\hat{\beta}}\ ,\label{eq:Pcan}
\end{equation}
\begin{equation}
S_{{\rm can}}^{\alpha\beta}=S^{\alpha\beta}+S'^{\alpha\beta}\ ,\quad S'^{\alpha\beta}\equiv2\int_{\Sigma(z,U)}x^{[\hat{\alpha}}\Phi^{\hat{\beta}]}j^{\hat{\gamma}}d\Sigma_{\hat{\gamma}}\ ,\label{eq:Scan}
\end{equation}
with 
\begin{align}
\Psi^{\hat{\alpha}}(z,x) & \equiv-\int_{0}^{1}F_{\ \hat{\beta}}^{\hat{\alpha}}(u)x^{\hat{\beta}}du\ ,\label{eq:Psi}\\
\Phi^{\hat{\alpha}}(z,x) & \equiv-\int_{0}^{1}uF_{\ \hat{\beta}}^{\hat{\alpha}}(u)x^{\hat{\beta}}du\ .\label{eq:Phi}
\end{align}
Eqs.~\eqref{eq:Psi}-\eqref{eq:Phi} are integrals along the geodesic
$\eta^{\alpha}(u)$ connecting $z^{\alpha}$ and $x^{\alpha}$, parametrized
by $u$ so that $\eta^{\alpha}(0)=z^{\alpha}$, $\eta^{\alpha}(1)=x^{\alpha}$.
In flat spacetime, these expressions are exactly%
\footnote{Therein Cartesian coordinates are used, and $F^{\alpha\beta}$ has
argument $F^{\alpha\beta}(\mathbf{z}+u\mathbf{r})$, where $r^{\alpha}=x^{\alpha}-z^{\alpha}$
is the vector connecting the reference worldline to the point $x^{\alpha}$.
Since $\eta^{\alpha}(u)$ is in this case a straightline, indeed $\eta^{\alpha}(u)=z^{\alpha}+ur^{\alpha}$.
Noting moreover that $z^{\hat{\alpha}}=0$, $r^{\hat{\alpha}}=x^{\hat{\alpha}}$
in the system $\{x^{\hat{\alpha}}\}$, one obtains \eqref{eq:Pcan}-\eqref{eq:Phi}.%
} Eqs.~(7.1)-(7.2), (7.6)-(7.7) of \cite{Dixon1967}. In curved spacetime,
they match, to the accuracy at hand, Eqs.~(3.14)-(3.15), (5.1)-(5.2)
of \cite{Dixon1970I} (corrections due to the bitensors therein are
of order $\mathcal{O}(a^{3})$ for $P'^{\alpha}$, and $\mathcal{O}(a^{4})$
for $S'^{\alpha\beta}$, where $a\equiv$ size of the body, hence
both negligible to quadrupole order, $\mathcal{O}(a^{2})$).

The lowest order approximation to these integrals is to take only
the zeroth order term in the expansion of $F^{\alpha\beta}$ around
$z^{\alpha}$, i.e., to take $F^{\alpha\beta}\approx constant$ \emph{along
the body}; this is sufficient for our purposes, as higher terms in
the expansion of $F^{\alpha\beta}$ lead to contributions of higher
multipole moments to $P'^{\alpha}$ and $S'^{\alpha\beta}$. We obtain:
\begin{equation}
P'^{\alpha}=-F_{\ \gamma}^{\alpha}d^{\gamma}\ ,\quad(i)\quad S'^{\alpha\beta}=F^{[\alpha}{}_{\sigma}q^{\beta]\sigma}\ ,\quad(ii)\label{eq:P'_S'}
\end{equation}
where $d^{\alpha}$ and $q^{\alpha\beta}$ are the charge dipole and
quadrupole moments, Eqs.~\eqref{eq:d_a} and \eqref{eq:chargequadrupole}.
As such, $S'^{\alpha\beta}$ is negligible to pole-dipole order, but
it is of crucial importance in Sec.~\ref{sec:Beyond-pole-dipole},
where terms up to quadrupole order are kept.

Note now the following: substituting (\ref{eq:ForceCan})-(\ref{eq:SpinCan}),
\eqref{eq:P'_S'} into Eqs.~\eqref{eq:ForceCan}-\eqref{eq:SpinCan}
(and noting that, to dipole order, $S^{\alpha\beta}\simeq S_{{\rm can}}^{\alpha\beta}$),
we obtain Eqs.~(\ref{eq:ForceDS0})-(\ref{eq:SpinDS0}); hence indeed
the two sets of equations \emph{are equivalent}.

As shown in~\cite{BaileyIsrael}, $P_{{\rm Dix}}^{\alpha}+qA^{\alpha}\equiv P_{{\rm can}}^{\alpha}$
and $S_{{\rm can}}^{\alpha\beta}$ have the interpretation of ``canonical
momenta'' associated to the Lagrangian of the system%
\footnote{We thank A. Harte for discussions on this point.%
}. $P_{{\rm can}}^{\alpha}$ is the quantity conserved in collisions
\cite{Israel}, and its time component $P_{{\rm can}}^{0}=-\mathbf{P}_{{\rm can}}\cdot\partial_{0}$
is the scalar conserved under stationary fields in flat spacetime,
cf.~Eq.~(\ref{eq:Conserved0}) below. The quantity $S_{{\rm can}}^{\alpha\beta}$
generalizes the canonical angular momentum of some non-relativistic
treatments \cite{Young,YoungQuestion66,Deissler}; in~\cite{Deissler},
a canonical angular momentum vector, Eq.~(31) therein, is obtained
differentiating $\partial\mathcal{L}/\partial\vec{\Omega}$ ($\mathcal{L}\equiv$
Lagrangian of the system, $\vec{\Omega}\equiv$ angular velocity of
the body). Such 3-vector is but a non-covariant form for the spatial%
\footnote{\textcolor{black}{The definition of $S_{{\rm can}}^{\gamma}$ is not
a dualization of $S_{{\rm can}}^{\alpha\beta}$, as neither $S_{{\rm can}}^{\alpha\beta}$
nor $S'^{\alpha\beta}$ are spatial with respect to $U^{\alpha}$
under the Mathisson-Pirani condition $S^{\alpha\beta}U_{\beta}=0$.
Hence $S'^{\gamma}$ and $S_{{\rm can}}^{\gamma}$ do not contain
the same information as $S_{{\rm can}}^{\alpha\beta}$ and $S'^{\alpha\beta}$
(only their spatial part).}%
} vector $S_{{\rm can}}^{\gamma}\equiv\epsilon_{\ \mu\alpha\beta}^{\gamma}S_{{\rm can}}^{\alpha\beta}U^{\mu}/2$,
as can be easily shown. From \eqref{eq:Scan}, $S_{{\rm can}}^{\gamma}=S^{\gamma}+S'^{\gamma}$,
with 
\begin{equation}
S'^{\gamma}\equiv\frac{1}{2}\epsilon_{\ \mu\alpha\beta}^{\gamma}U^{\mu}S'^{\alpha\beta}=\frac{B^{\alpha}}{2}\left[\delta_{\ \alpha}^{\gamma}q_{\sigma}^{\ \sigma}-q_{\ \alpha}^{\gamma}\right]\ ,\label{eq:S'}
\end{equation}
where we used Eq.~\eqref{eq:Fdecomp} and the orthogonality condition
$q^{\alpha\beta}U_{\alpha}=q^{\alpha\beta}U_{\beta}=0$. If the body
has uniform mass and energy density, $S'^{\gamma}=(q/2m)B^{\alpha}I_{\alpha}^{\ \gamma}$,
where $I^{\alpha\beta}$ is the moment of inertia (see Footnote \ref{fn:MomentInertia}).
In this case we have, in the particle's CM frame (where $U^{i}=0$),
$S_{{\rm can}}^{\gamma}=(0,\vec{S}_{{\rm can}})$, with $\vec{S}_{{\rm can}}=\vec{S}+\vec{S}'$
matching expression~(31) of~\cite{Deissler}.

The distinction between $P_{{\rm Dix}}^{\alpha}$ in Eqs.~(\ref{eq:ForceCan})-(\ref{eq:SpinCan})
and the physical momentum $P^{\alpha}$ should not be overlooked when
the particle possesses electric dipole moment. Since those equations
are essentially symmetric with respect to $d^{\alpha}$ and $\mu^{\alpha}$,
failing to make that distinction would lead one to believe that the
two dipoles are \emph{dynamically} similar. Given their different
nature, as defined by Eqs.~(\ref{eq:d_a})-(\ref{eq:mu_a}) (the
magnetic dipole is modeled by a current loop, the electric dipole
by a pair \textcolor{black}{of opposite charges),} that would be physically
inconsistent: i) the electric dipole would have a hidden momentum
(just like a magnetic dipole), cf.~Eq.~(\ref{eq:PdixExplicit}),
which would violate the conservation equations\textcolor{black}{;}
ii) a static electric field would do no work on the dipole (regardless
of its motion), which is well known to be false; iii) the particle's
proper mass $m$ would vary in a way consistent with a dipole arising
from from a current of magnetic monopoles, not a pair of charges;
iv) the spatial part of the force would not be consistent with the
results known from classical electromagnetism. A detailed account
of these issues is given in the next section.

At quadrupole order, it is also crucial to not confuse $S_{{\rm can}}^{\alpha\beta}$
with the physical angular momentum $S^{\alpha\beta}$ (the one which
is proportional to the angular velocity in the case of a rigid body).
Otherwise, as discussed in Sec.~\ref{sub:Electromagnetic-torque},
one would erroneously conclude that in vacuum the electromagnetic
field does not couple to the trace of $q_{\alpha\beta}$, implying
e.g.~that no torque (besides the dipole torque $\vec{\tau}=\vec{\mu}\times\vec{B}$,
if it spins) could be exerted on a spherical charged body, which is
well known, both from elementary electromagnetism and from experiment,
to be false.

\section{The electric dipole\label{sec:The-electric-dipole}}

In order to better understand some key issues in this work --- the
physical meaning of the time projection of the force on a magnetic
dipole, the variation of its proper mass, the work done on it by the
external fields, and the ``hidden momentum'' --- it is useful to
make the contrast with the case of an electric dipole.

It is clear from Eqs.~(\ref{eq:ForceDS0})-(\ref{eq:SpinDS0}) that
both the force and the spin evolution equations are different for
electric and magnetic dipoles. This is due to the intrinsic differences
of the two types of dipole: \textcolor{black}{$d^{\alpha}$, Eq.~(\ref{eq:d_a}),
is the dipole moment of the charge density, which can be modeled by
a pair of two (close) opposite charges; $\mu^{\alpha}$,} \textcolor{black}{Eq.
(\ref{eq:mu_a})}, is the dipole moment of the spatial current, modeled
by a (small) current loop. For a particle possessing only electric
dipole moment ($\mu^{\alpha\beta}=0$, $q=0$) in flat spacetime,
Eqs.~(\ref{eq:ForceDS0})-(\ref{eq:SpinDS0}) read 
\begin{align}
\frac{DP^{\alpha}}{d\tau} & \equiv F_{{\rm el}}^{\alpha}=E_{\ \beta}^{\alpha}d^{\beta}+F_{\ \beta}^{\alpha}\frac{Dd^{\beta}}{d\tau}\ ,\label{eq:Fel1}\\
\frac{DS^{\alpha\beta}}{d\tau} & =2P^{[\alpha}U^{\beta]}+2d^{[\alpha}F_{\ \ \gamma}^{\beta]}U^{\gamma}\ ,\label{eq:DS_El}
\end{align}
where $E_{\alpha\beta}\equiv F_{\alpha\gamma;\beta}U^{\gamma}$ is
the electric tidal tensor~\cite{CHPRD}.

First we note that, unlike its magnetic counterpart Eq.~(\ref{tab:Analogy}.1a)
of Table \ref{tab:Analogy}, the force on an electric dipole is not
(generically) given \textcolor{black}{by a contraction of a tidal
tensor with the dipole vector (only if $Dd^{\alpha}/d\tau=0$).} \textcolor{black}{Indeed,
it is not entirely a tidal effect, due to the extra term $F_{\ \beta}^{\alpha}Dd^{\beta}/d\tau$
(overlooked in most literature), which does not involve derivatives
of $F_{\alpha\beta}$. This term is physically interpreted as follows.
F}rom Eq.~(3.23a) of~\cite{Dixon1967} we have 
\[
\frac{Dd^{\gamma}}{d\tau}=\mathcal{J}^{\gamma}-U^{\gamma}q\ ,
\]
where $\mathcal{J}^{\hat{\alpha}}\equiv\int_{\Sigma(U,\tau)}j^{\hat{\alpha}}w^{\hat{\gamma}}d\Sigma_{\hat{\gamma}}$.
$q$ is the particle's \emph{total} charge, and $\mathcal{J}^{\alpha}$
is roughly its total current. Then $\mathcal{J}^{\gamma}-U^{\gamma}q$
is essentially the particle's spatial current with respect to $U^{\alpha}$.
For an electric dipole ($q=0$), Eq.~(\ref{eq:Fel1}) can be be re-written
as 
\begin{equation}
F_{{\rm el}}^{\alpha}=E_{\ \beta}^{\alpha}d^{\beta}+F_{\ \beta}^{\alpha}\mathcal{J}^{\beta}\ .\label{eq:Fel2}
\end{equation}
The term \textcolor{black}{$F_{\ \gamma}^{\alpha}\mathcal{J}^{\gamma}$
has a straightforward interpretation: if the dipole vector $d^{\alpha}$
varies with $\tau$ (e.g., if the dipole rotates) then it generates
a net electric current in the CM frame; therefore, a magnetic force
$F_{\ \gamma}^{\alpha}\mathcal{J}^{\gamma}$ is exerted on it, in
addition to the tidal force $E_{\ \beta}^{\alpha}d^{\beta}$. As a
simple example, consider a rotating electric dipole under a uniform
magnetic field; a net force arises from the magnetic forces (with
the same direction) that act on each of its charge poles, due to their
circular motion about the CM.}

Secondly, we note that in the term $E^{\alpha\beta}d_{\beta}$ the
indices of the tidal tensor are reversed as compared to the force
on the magnetic dipole, Eq.~(\ref{tab:Analogy}.1a)\textcolor{black}{.
In Secs.~\ref{sub:Edipole Proper-mass-and-time proj} and \ref{sub:Edipole_Time-component-as_static}
below we shall see some consequences.}

For an electric dipole at rest in an inertial frame (where $E_{ij}=\nabla_{j}E_{i}$),
the space part of (\ref{eq:Fel1}) reads $\vec{F}_{{\rm el}}=(\vec{d}\cdot\nabla)\vec{E}-\vec{B}\times D\vec{d}/d\tau$,
matching the result from classical treatments, e.g.~\cite{Vaidman}.
Note also that $D\vec{P}_{{\rm Dix}}/d\tau=\nabla(\vec{E}\cdot\vec{d})$
(analogous to the force on a magnetic dipole, $\vec{F}_{{\rm EM}}=\nabla(\vec{B}\cdot\vec{\mu})$),
which differs from the physical force $\vec{F}_{{\rm el}}=D\vec{P}/d\tau$.

\subsection{No hidden momentum for electric dipole}

Unlike the current loop, the two-charge type of dipole cannot store
hidden momentum of electromagnetic origin, see e.g.~\cite{Vaidman}.
The expression for the momentum of an electric dipole is obtained
contracting Eq.~(\ref{eq:DS_El}) with $U_{\beta}$, leading to (using
$U^{\alpha}d_{\alpha}=0$) $P^{\alpha}=mU^{\alpha}+S^{\alpha\beta}a_{\beta}$,
showing that the only hidden momentum present is the pure gauge term
$P_{{\rm hidI}}^{\alpha}=S^{\alpha\beta}a_{\beta}$ arising from the
spin condition (which exists regardless of the electromagnetic multipole
structure of the particle). This was expected from conservation arguments.
Unlike its magnetic counterpart, the electric dipole does \emph{not}
generate electromagnetic \emph{field} momentum (cross momentum $P_{\times}^{\alpha}$,
see \cite{EPAPS}) when placed in an electromagnetic field \cite{Furry}.
Now consider a \emph{stationary} configuration; in this case the conservation
equations $(T_{{\rm tot}})_{\ \ ;\beta}^{\alpha\beta}=0$ imply that
the total spatial momentum vanishes, $\vec{P}_{{\rm tot}}=0$; if
the dipole were to have any hidden momentum, it would \emph{not} be
canceled out by the field momentum, violating the conservation equations.

This shows the importance of distinguishing between the physical momentum
$P^{\alpha}$ and Dixon's momentum $P_{{\rm Dix}}^{\alpha}=P^{\alpha}+P'^{\alpha}$
of Eqs.~(\ref{eq:ForceCan})-(\ref{eq:SpinCan}); as can be seen
from (\ref{eq:Pcan}), (\ref{eq:P'_S'}), $P_{{\rm Dix}}^{\alpha}$
includes a term $-\epsilon_{\ \theta\mu\sigma}^{\alpha}d^{\theta}B^{\mu}U^{\sigma}$,
analogous to the hidden momentum $P_{{\rm hidEM}}^{\alpha}=\epsilon_{\ \theta\mu\sigma}^{\alpha}\mu^{\theta}E^{\mu}U^{\sigma}$
of the magnetic dipole (but of opposite sign): 
\begin{equation}
P_{{\rm Dix}}^{\alpha}=P^{\alpha}-E^{\beta}d_{\beta}U^{\alpha}-\epsilon_{\ \theta\mu\sigma}^{\alpha}d^{\theta}B^{\mu}U^{\sigma}\ .\label{eq:PdixExplicit}
\end{equation}
Thus, confusing $P_{{\rm Dix}}^{\alpha}$ with $P^{\alpha}$ would
lead one to believe that the electric dipole has a hidden momentum
just like a magnetic dipole, which not only would make no sense for
the dipole model at stake, as it would violate the conservation equations.

\subsection{Proper mass and time projection of the force in the CM frame\label{sub:Edipole Proper-mass-and-time proj}}

Contracting (\ref{eq:Fel1}) with $U^{\alpha}$ one obtains 
\begin{equation}
F_{{\rm el}}^{\alpha}U_{\alpha}=-E_{\gamma}\frac{Dd^{\gamma}}{d\tau}=-E_{\gamma}\mathcal{J}^{\gamma}\ ,\label{eq:TprojEl}
\end{equation}
where $E^{\alpha}\equiv F^{\alpha\beta}U_{\beta}$ is the electric
field as measured by the test particle. Hence, like the force on a
magnetic dipole ($F_{{\rm EM}}^{\alpha}$), $F_{{\rm el}}^{\alpha}$
has in general a (time) projection along the particle's worldline.
They are very different, however. As noticed above, the order of the
indices in the tidal tensor of (\ref{eq:Fel1}) is reversed compared
to \textcolor{black}{$F_{{\rm EM}}^{\alpha}=B^{\beta\alpha}\mu_{\beta}$};
\textcolor{black}{since} $E_{\alpha\beta}$ and $B_{\alpha\beta}$\textcolor{black}{{}
are spatial relative to $U^{\alpha}$ in the first, but} \textcolor{black}{\emph{not}}
\textcolor{black}{in the second index, then, }\textcolor{black}{\emph{by
contrast}} \textcolor{black}{with $F_{{\rm EM}}^{\alpha}$, the projection
of the tidal force $E^{\alpha\beta}d_{\beta}$ along $U^{\alpha}$
is zero. This means that, as measured} \textcolor{black}{\emph{in
the particle's CM frame}}\textcolor{black}{,} \textcolor{black}{the
tidal force does}\textcolor{black}{\emph{ no work}}\textcolor{black}{.
}Thus $F_{{\rm el}}^{\alpha}U_{\alpha}$\textcolor{black}{{} reduces
to the projection of the second term of }(\ref{eq:Fel1})\textcolor{black}{,
}arising from the variation of \emph{the dipole vector} $d^{\alpha}$
along the particle's worldline. This contrasts with its magnetic counterpart
$F_{{\rm EM}}^{\alpha}U_{\alpha}=U^{\beta}\mu^{\gamma}D\star\! F_{\gamma\beta}/d\tau$,
cf.~Eq.~(\ref{tab:Analogy}.1a) of Table \ref{tab:Analogy}, which
comes from the variation of \emph{the field.}

Equation~(\ref{eq:TprojEl}) makes sense: $\mathcal{J}^{\gamma}$
is \emph{essentially} the total current as measured in the dipole's
frame; when it is non-vanishing (for instance, due to a rotation of
the dipole), a non-vanishing work, in this frame, is done on the dipole
by the electric field. Noting from (\ref{eq:MomentumMP}) that $P^{\alpha}a_{\alpha}=0$,
we have 
\begin{equation}
\frac{dm}{d\tau}=-F_{{\rm el}}^{\alpha}U_{\alpha}=E_{\gamma}\frac{Dd^{\gamma}}{d\tau}\ .\label{eq:DmEl}
\end{equation}
Hence, if $Dd^{\alpha}/d\tau=0$, the particle's proper mass is constant,
which contrasts with the situation for a magnetic dipole, where $dm/d\tau$
is zero only if $DB^{\alpha}/d\tau=0$ (\emph{not} $D\mu^{\alpha}/d\tau=0$),
cf.~Eq.~\eqref{eq:dm1}.

Consider now the special case of a rigid dipole which is allowed to
rotate: $D_{F}d^{\alpha}/d\tau=\Omega_{\ \beta}^{\alpha}d^{\beta}$,
with $\Omega_{\alpha\beta}$ defined by Eqs.~(\ref{eq:Omega}). In
this case, using (\ref{eq:FermiTransport}), 
\begin{equation}
\frac{dm}{d\tau}=-F_{{\rm el}}^{\alpha}U_{\alpha}=\epsilon_{\gamma\beta\mu\nu}U^{\nu}E^{\gamma}\Omega^{\beta}d^{\mu}=\tau^{\beta}\Omega_{\beta}\ ;\label{eq:DmEl2}
\end{equation}
this is the rate of work done by the torque $\tau^{\beta}=\epsilon_{\ \mu\gamma\nu}^{\beta}U^{\nu}d^{\mu}E^{\gamma}$
exerted on the dipole by virtue of Eqs.~(\ref{eq:DS_El}), (\ref{eq:FermiTorque}).
The torque $\tau^{\beta}$ causes an accelerated rotation of the dipole;
the corresponding variation of rotational kinetic energy reflects
itself in a variation of $m$.

Note that Eqs.~(\ref{eq:DmEl})-(\ref{eq:DmEl2}) yield, e.g., the
well known work done on an electric dipole whose CM is at rest in
a static, uniform electric field, from the point of view of the rest
frame. Thus again we see the importance of not confusing $P_{{\rm Dix}}^{\alpha}$
in Eqs.~(\ref{eq:ForceCan})-(\ref{eq:SpinCan}) with the physical
momentum $P^{\alpha}$: overlooking the distinction would lead to
the conclusion that, just like for a magnetic dipole, a static field
does no work on a rotating electric dipole, which we know from basic
electromagnetism to be false.

\subsection{Time component of the force as measured by generic observers\label{sub:Edipole_Time-component-as_static}}

With respect to a congruence of observers $\mathcal{O}(u)$ of 4-velocity
$u^{\alpha}$, the time projection of the force exerted on the electric
dipole is 
\begin{equation}
-F_{{\rm el}}^{\alpha}u_{\alpha}=\gamma(E^{u})_{\beta\gamma}d^{\gamma}v^{\beta}+(E^{u})_{\alpha}\frac{Dd^{\alpha}}{d\tau}\ ,\label{eq:Fel0v2}
\end{equation}
where $(E^{u})^{\alpha}\equiv F^{\alpha\beta}u_{\beta}$ and $(E^{u})_{\beta\gamma}\equiv F_{\beta\mu;\gamma}u^{\mu}$
are, respectively, the electric field and electric tidal tensor measured
by $\mathcal{O}(u)$, and $v^{\alpha}$ (the particle's velocity relative
to $\mathcal{O}(u)$) and $\gamma$ are defined in Eqs.~(\ref{eq:U_u}).
As discussed in Sec.~\ref{sub:Time-Components}, this is the rate
of work done by the force as measured by $\mathcal{O}(u)$. \textcolor{black}{The
first term is a natural result: in a non-uniform electric field ($(E^{u})_{\alpha\beta}\ne0$),
a force is in general exerted on an electric dipole; if it is allowed
to move ($v^{\alpha}\ne0$) that force does work. The second term
contributes when $Dd^{\alpha}/d\tau\ne0$, and is non-zero even if
the fields are uniform. It is the work done by the electric field
when the dipole rotates or oscillates, discussed in the previous section.}

The power $-F_{{\rm el}}^{\alpha}u_{\alpha}$ differs significantly
from its magnetic counterpart Eq.~(\ref{eq:Pmech+Pind}). Consider
(when they exist) observers along whose worldlines the field is covariantly
constant, $F_{\ \ ;\gamma}^{\alpha\beta}u^{\gamma}=0$ (e.g. the ``static
observers'' of Sec. \ref{sub:TimeProj_Static_EM}, cf. Footnote \ref{fn:Static observers});
as we have seen in Sec.~\ref{sub:Time-components-Static}, relative
to such observers, the field does no work on a magnetic dipole, $F_{{\rm EM}}^{\alpha}u_{\alpha}=0$,
cf.~Eq.~(\ref{eq:FEMstatic}). But it \textcolor{black}{\emph{does
work}} \textcolor{black}{on an electric dipole, both terms of }(\ref{eq:Fel1})\textcolor{black}{{}
contributing to it (regarding the tidal term, the reason why $E_{\ \beta}^{\alpha}d^{\beta}$
does work, $E_{\ \beta}^{\alpha}d^{\beta}u_{\alpha}\ne0$, whereas
$F_{{\rm EM}}^{\alpha}=B_{\beta}^{\ \alpha}\mu^{\beta}$ does not,
is again due to the order of the indices in the tidal tensor). This
was to be expected given the different nature of the dipoles: in the
magnetic case, the total work is zero due to (in the simplest case
when there is no hidden momentum) a cancellation between }the variation
of translational kinetic energy and the work done on the current loop
by the electric field induced in it; the latter has no counterpart
in the electric dipole, since it does not consist of a current of
magnetic monopoles; therefore such cancellation does not occur.

\subsection{Conserved quantities, proper mass and work done by the fields\label{sub:Conserved-quantities,-proper}}

In order to better elucidate the relationship between the work done
by the fields and the variation of the proper mass, we will compare,
in a \emph{static} electromagnetic field, three different test particles:
a point monopole charge, an electric dipole, and a magnetic dipole.
Let $u^{\alpha}$ be the 4-velocity of the \emph{inertial} frame $\mathcal{O}(u)$
relative to which the fields are static. Then $u^{\alpha}$ preserves
the electromagnetic field, $\mathcal{L}_{u}F^{\alpha\beta}=0$, and,
therefore, from the constancy of expressions (5.3) of~\cite{Dixon1970I},
or (29) of~\cite{Wald et al 2010}, we have 
\begin{equation}
P_{{\rm Dix}}^{\alpha}u_{\alpha}+qA^{\alpha}u_{\alpha}=P^{\alpha}u_{\alpha}+(E^{u})^{\alpha}d_{\alpha}-q\phi=\mathrm{constant,}\label{eq:Conserved0}
\end{equation}
where $\phi\equiv-A^{\alpha}u_{\alpha}$ is the electric potential
measured in $\mathcal{O}(u)$. Using Eq.~(\ref{eq:HiddenMomentum}),
it is useful to re-write (\ref{eq:Conserved0}) as 
\begin{equation}
m+T+V+E_{{\rm hid}}=\mathrm{constant}\label{eq:Conserved}
\end{equation}
where $V=-(E^{u})^{\alpha}d_{\alpha}+q\phi$ is the potential energy
of the particle under the field, $T\equiv(\gamma-1)m$ is the kinetic
energy associated to the translation of its center of mass, $\gamma\equiv-U^{\alpha}u_{\alpha}$,
and $E_{{\rm hid}}=-P_{{\rm hid}}^{\alpha}u_{\alpha}$ the ``hidden
energy'' (i.e., the time component of the hidden momentum relative
to $\mathcal{O}(u)$, see Sec.~\ref{sub:Time-components-Static}).
In this section we shall ignore the inertial hidden momentum $P_{{\rm hidI}}^{\beta}$,
as in the applications below it either vanishes or is made negligible
by appropriate choices of the reference worldline (e.g.~Tulczyjew-Dixon,
or Mathisson-Pirani non-helical centroids). Thus, $P_{{\rm hid}}^{\beta}=P_{{\rm hidEM}}^{\beta}$
herein.

\emph{Point monopole charge} ($d^{\alpha}=P_{{\rm hid}}^{\alpha}=0$).
--- In this case, condition (\ref{eq:Conserved}) reads $m+T+q\phi=\mathrm{constant}$.
There is no exchange of energy with the proper mass of the particle,
which is a constant: 
\[
\frac{dm}{d\tau}=-\frac{DP_{\alpha}}{d\tau}U^{\alpha}=-qF_{\alpha\beta}U^{\alpha}U^{\beta}=0\,.
\]
This just tells us that, in a stationary electromagnetic field, the
``total mechanical energy'' of the particle --- kinetic energy $T$,
plus electric potential energy $V=q\phi$ --- is a constant of the
motion, as is well known. Every gain in $T$ must come from the potential
energy $V$, so there is no doubt that the field doing work, at a
rate given by the time projection of the Lorentz force $F_{{\rm L}}^{\alpha}=qF^{\alpha\beta}U_{\beta}$
relative to $\mathcal{O}(u)$, cf.~Eq.~(\ref{eq:DE/dt}): 
\[
\frac{dE}{d\tau}=-F_{{\rm L}}^{\alpha}u_{\alpha}=q\gamma(E^{u})^{\alpha}v_{\alpha}=-\frac{dV}{d\tau}=F_{{\rm L}}^{\alpha}v_{\alpha}.
\]
In vector notation, $dE/d\tau=q\gamma\vec{E}(u)\cdot\vec{v}$, with
$\vec{E}(u)=-\nabla\phi=-\nabla V/q$.

\emph{Electric dipole} ($q=P_{{\rm hid}}^{\alpha}=0$). --- Condition
(\ref{eq:Conserved}) reads $m+T-(E^{u})^{\alpha}d_{\alpha}=\mathrm{constant}$.
From Eq.~(\ref{eq:DmEl}), the proper mass $m$ is not constant;
this means that energy is exchanged between the three forms: potential
energy $V=-(E^{u})^{\alpha}d_{\alpha}$, translational kinetic energy
$T$, and $m$. Two special sub-cases are particularly enlightening: 
\begin{enumerate}
\item dipole vector covariantly constant, $Dd^{\alpha}/d\tau=0$, implying
$dm/d\tau=0$. In this case the energy exchange is similar to the
monopole charge: every gain in translational kinetic energy comes
from the potential energy $V$. It is clear that the electric tidal
field is doing work, at a rate (cf.~Eq.~(\ref{eq:Fel0v2})) 
\[
\frac{dE}{d\tau}=-F_{{\rm el}}^{\alpha}u_{\alpha}=\gamma(E^{u})_{\beta\gamma}d^{\gamma}v^{\beta}=-\frac{dV}{d\tau}=F_{{\rm el}}^{\alpha}v_{\alpha}.
\]

\item Dipole's CM at rest ($U^{\alpha}=u^{\alpha}$, $v^{\alpha}=0$), i.e.,
$T=0$. In this case, $m-E^{\alpha}d_{\alpha}=\mathrm{constant}$,
and the energy exchange occurs between the potential energy $V=-E^{\alpha}d_{\alpha}$
and proper mass $m$ (which includes rotational kinetic energy of
the particle). The work of the field thus equals the mass variation,
\[
\frac{dE}{d\tau}=-F_{{\rm el}}^{\alpha}u_{\alpha}=\frac{dm}{d\tau}=-\frac{dV}{d\tau}\ .
\]

\end{enumerate}
\emph{Magnetic dipole} ($q=d^{\alpha}=0$). --- Condition (\ref{eq:Conserved})
means in this case $m+T+E_{{\rm hid}}=\mathrm{constant}$; if we take
$\mu^{\alpha}=\sigma S^{\alpha}$, from Eq.~(\ref{eq:m}) we have
$m=m_{0}-\mu^{\alpha}B_{\alpha}$, and thus the condition becomes
$T-\mu^{\alpha}B_{\alpha}+E_{{\rm hid}}=\mathrm{constant}$. The energy
exchange is between translational kinetic energy, proper mass and
$E_{{\rm hid}}$. There is no potential energy involved (cf.~\cite{Coombes,Young,YoungQuestion66,Deissler}),
which is consistent with the fact that the static field does no work
on the magnetic dipole: $dE/d\tau=-F_{{\rm EM}}^{\alpha}u_{\alpha}=0$,
cf.~Eq.~(\ref{eq:FEMstatic}). A case of interest in the context
of this work is the one depicted in Fig.~\ref{fig:DipoleMagnet}b,
a magnetic dipole falling towards a magnet along the field's axis
of symmetry. In this case $P_{{\rm hid}}^{\alpha}=E_{{\rm hid}}=0$,
implying $T+m=\mathrm{constant}$. The energy exchange is only between
translational kinetic energy and proper mass; every gain in the former
comes at the expense of latter (which, for a rigid body, consists
essentially of a variation of \emph{rotational kinetic energy}, cf.
Sec.~\ref{sub:Electromagnetic Torque-and-force on Spherical} and
\cite{Coombes,Young,YoungQuestion66,Deissler}). Hence what the field
does is to interconvert translational kinetic energy into rotational
or other forms of internal energy.

\section{Comparison of the different spin conditions\label{sec:Spin Conditions}}

In this work we have so far been using equations of motion supplemented
by the Mathisson-Pirani (MP) spin condition, as it is the one that
makes explicit the analogies used. As we shall see below, it is also
the one that leads to the simplest description of the force/center
of mass motion in the applications in Secs.~\ref{sub:Symmetries}
and \ref{sub:Time-components-Static}. However, other spin conditions
(\ref{eq:Spin_condition}) can be used; as explained in Sec. ~\ref{sub:Center-of-mass},
the (infinite) possible choices of $u^{\alpha}$ correspond to different,
but equivalent, ways of describing the motion of a spinning body,
they differ just in the choice of its representative point. Below
we compare some best known spin conditions in the applications in
this work, and explore the gravito-electromagnetic analogies that
emerge using them.

\subsection{Comparison of the spin conditions in the applications in this work\label{sub:Comparison-of-the-SSC-aplications}}

We start with the problem of the falling motion along the symmetry
axis ($\theta=0$ in Boyer-Lindquist coordinates, hereafter the ``$z$-axis'')
of a gyroscope in a Kerr spacetime, discussed in Sec.~\ref{sub:Time-components-Static}.
Setting its initial position, velocity $\vec{U}$ and spin $\vec{S}$
all along the axis, one expects, at first sight, from symmetry arguments,
an axial fall. It turns out, however, that such naive prescription
of initial conditions does not completely determine the problem, nor
does it ensure its axial symmetry. One needs also to prescribe the
field of unit time-like vectors $u^{\alpha}$ relative to which the
CM is computed (i.e., the field entering the spin condition $S^{\alpha\beta}u_{\beta}=0$),
which, for an arbitrary choice, breaks the axial symmetry. The momentum-velocity
relation also depends on this choice, cf.~Eq.~(\ref{eq:Momentum}),
implying that $\vec{U}$ will not in general be parallel to $\vec{P}$
(hidden momentum), so that they do not both lie along the $z$-axis.
Note that, as explained in Sec.~\ref{sub:Momentum-of-the-Particle},
the acceleration of the CM does not originate solely from the force,
but also from the variation of field $u^{\alpha}$ along the CM worldline.

In order to prescribe an axisymmetric problem, we start by demanding,
as initial conditions, $\vec{U}_{{\rm in}}=U^{z}\vec{e}_{z}$ ($\vec{e}_{z}\equiv\vec{e}_{r}=\partial/\partial r$
in Boyer-Lindquist coordinates, for $\theta=0$), $\vec{u}=u^{z}\vec{e}_{z}$,
and an initial CM position $z_{{\rm in}}^{\alpha}=x_{{\rm CM}}^{\alpha}(u)|_{{\rm in}}$
also along the $z$-axis. The MP condition, $u^{\alpha}=U^{\alpha}$,
clearly allows for these initial conditions, so let us start with
it. The momentum reads, cf.~Eq.~\eqref{eq:MomentumMP}, 
\begin{equation}
P^{\alpha}=mU^{\alpha}-\epsilon_{\ \beta\gamma\delta}^{\alpha}S^{\beta}a^{\gamma}U^{\delta}\ ,\label{eq:P_U_Pirani}
\end{equation}
and the spatial part of the equation of motion $F_{{\rm G}}^{\beta}\equiv DP^{\beta}/d\tau=-\mathbb{H}^{\alpha\beta}S_{\alpha}$
(cf.~Eq.~(\ref{tab:Analogy}.1b)) reads 
\begin{equation}
m\vec{a}-\frac{D(\vec{S}\times_{U}\!\vec{a})}{d\tau}=\vec{F}_{{\rm G}}=-\mathbb{H}^{i\alpha}S_{\alpha}\vec{e}_{i}\ ,\label{eq:AxialPirani}
\end{equation}
where $\vec{S}\times_{U}\!\vec{a}$ denotes the space components of
$\epsilon_{\ \beta\gamma\delta}^{\alpha}S^{\beta}a^{\gamma}U^{\delta}$.
Initially, with $\vec{U}_{{\rm in}}=U^{z}\vec{e}_{z}$, one obtains
$\vec{F}_{{\rm G}}|_{{\rm in}}=-\mathbb{H}^{z\alpha}S_{\alpha}\vec{e}_{z}$
(it is straightforward to check that along the axis we have $\mathbb{H}^{i\alpha}=0$
if $i\ne z$); thus the force is along $z$, as expected from symmetry
arguments, given the axial symmetry of the initial setup and the fact
that $\mathbb{H}_{\alpha\beta}\equiv\star R_{\alpha\mu\beta\nu}U^{\mu}U^{\nu}$
depends only on $U^{\alpha}$. It is clear from the equation above
that one%
\footnote{Other solutions are possible, because the set of initial conditions
$\{z^{\alpha},S^{\alpha\beta},U^{\alpha},m\}|_{{\rm in}}$ is not
sufficient to uniquely specify a solution under the MP condition,
see \cite{CostaNatario2014}. Note however that, since $U_{{\rm in}}^{\alpha}$
is fixed, such solutions correspond to different values of $P_{{\rm in}}^{\alpha}$,
therefore they are \emph{not} representations of the same \emph{physical
motion} (i.e., those will be ``helical'' representations but of
different motions). %
} of the possible solutions of (\ref{eq:AxialPirani}) is the most
natural result, namely motion along the $z$-axis, with the body accelerating
in the direction of the force (and of $\vec{S}$): $\vec{a}=a^{z}\vec{e}_{z}\Rightarrow\vec{S}\times_{U}\!\vec{a}=0$,
implying $P^{\alpha}=mU^{\alpha}$, and $F_{{\rm G}}^{\alpha}=ma^{\alpha}$.
It is a ``non-helical'' solution (since it is a straightline), and
therefore the description we seek. Hence we have solved the axial
fall problem, and a unique relation between $P^{\alpha}$ and $U^{\alpha}$
was naturally established (for this solution) in the course of the
analysis.

Now let us compare with the equivalent descriptions for this problem
given by other spin conditions. For a generic field $u^{\alpha}$
with $\vec{u}$ not lying along the $z$ axis, we no longer have axial
symmetry, therefore we should not expect to obtain a centroid moving
in straightline along the axis; what we expect, in general, is a different
(possibly exotic) but equivalent description of the same physical
motion, using a different representative worldline. The problem, however,
is how to prescribe its initial conditions. If one naively sets up
an initial position $z_{{\rm in}}^{\alpha}=x_{{\rm CM}}^{\alpha}(u)|_{{\rm in}}$
lying on the $z$-axis, and then $\vec{P}$ or $\vec{U}$ (there is
an ambiguity on this choice, as they are not parallel in general,
cf.~Eq.~(\ref{eq:Momentum})) also along the $z$-axis, the solution
in general will not be an axial fall; in fact, it will not even be
a different description for it, but a different physical motion.

So first we must establish how we make sure that we are dealing with
the same particle. A pole-dipole particle is characterized by its
two moments: $P^{\alpha}$ and $S^{\alpha\beta}$. These are defined
with respect to a reference worldline $z^{\alpha}(\tau)$ and a hypersurface
of integration $\Sigma(\tau,u)$, cf.~Eqs.~(\ref{eq:Pgeneral})-(\ref{eq:Sab});
different representations of the same particle must yield the same
moments with respect to the \emph{same }point and $\Sigma(\tau,u)$.
To dipole order, $P^{\alpha}$ is independent of the spin condition
(see \cite{CostaNatario2014}), but $S^{\alpha\beta}\equiv S^{\alpha\beta}(z)$
depends on it. Let $S^{\alpha\beta}$ and $\bar{S}^{\alpha\beta}$
be the angular momentum taken about, respectively, the centroids $z^{\alpha}=x_{{\rm CM}}^{\alpha}(u)$
and $\bar{z}^{\alpha}=x_{{\rm CM}}^{\alpha}(\bar{u})$; i.e., $S^{\alpha\beta}u_{\beta}=0$,
and $\bar{S}^{\alpha\beta}\bar{u}_{\beta}=0$, cf.~Sec.~\ref{sub:Center-of-mass}.
The integral expressions for $S^{\hat{\alpha}\hat{\beta}}$ and $\bar{S}^{\hat{\alpha}\hat{\beta}}$,
in normal coordinates $\{x^{\hat{\alpha}}\}$ originating \emph{at}
$z^{\alpha}$, are given, to dipole order%
\footnote{This is because both the dependence of $S^{\alpha\beta}$ on the argument
$u^{\alpha}$ of $\Sigma$ (see \cite{CostaNatario2014}), and the
non-linearity, due to the curvature, of the transformation between
normal coordinates originating at $z^{\alpha}$ and $\bar{z}^{\alpha}$
(denote the latter by $\{x^{\tilde{\alpha}}\}$), are negligible to
dipole order: $x^{\tilde{\alpha}}=x^{\hat{\alpha}}-\bar{z}^{\hat{\alpha}}+\mathcal{O}(\|x^{\hat{\alpha}}-\bar{z}^{\hat{\alpha}}\|^{2}\Delta x)$,
cf.~e.g.~Eq.~(11.12) of \cite{Brewin}; hence, in the computation
of $\bar{S}^{\alpha\beta}$, one can use $x^{\tilde{\alpha}}\simeq x^{\hat{\alpha}}-\bar{z}^{\hat{\alpha}}$,
as the correction is of order $\mathcal{O}(a^{4})$, whereas to dipole
order only terms of $\mathcal{O}(a)$ are kept ($a\equiv$ size of
the body).%
}, by Eq.~(\ref{eq:Sab}) (in the case of $\bar{S}^{\hat{\alpha}\hat{\beta}}$,
replacing therein $x^{\hat{\alpha}}$ by $x^{\hat{\alpha}}-\bar{z}^{\hat{\alpha}}$,
so that it is taken about the point $\bar{z}^{\hat{\alpha}}$). We
obtain 
\begin{equation}
\bar{S}^{\hat{\alpha}\hat{\beta}}=S^{\hat{\alpha}\hat{\beta}}+2P^{[\hat{\alpha}}\Delta x^{\hat{\beta}]}\ ,\label{eq:SbarS}
\end{equation}
where $\Delta x^{\hat{\alpha}}=\bar{z}^{\hat{\alpha}}-z^{\hat{\alpha}}=\bar{z}^{\hat{\alpha}}$;
this is similar to the flat spacetime transformation (e.g.~\cite{Gravitation,Helical}).
Hence, to obtain a solution corresponding to the same physical motion
above, we must prescribe the same momentum $\vec{P}=P^{z}\vec{e}_{z}$,
and correct the spin tensor and initial position of the centroid using
Eq.~(\ref{eq:SbarS}). As can be seen contracting (\ref{eq:SbarS})
with $\bar{u}_{\hat{\beta}}$ (taking $u^{\alpha}=U^{\alpha}$), the
condition $\bar{S}^{\alpha\beta}\bar{u}_{\beta}=0$ yields, in general,
a centroid $\bar{z}^{\alpha}=x_{{\rm CM}}^{\alpha}(\bar{u})$ at a
different point compared to the MP centroid $z^{\alpha}=x_{{\rm CM}}^{\alpha}(U)$,
not on the $z$-axis, manifesting that the problem is no longer axisymmetric.
Since, in general, $U^{\alpha}\nparallel P^{\alpha}$, cf.~Eq.~(\ref{eq:Momentum}),
the centroid $\bar{z}^{\alpha}$ does not even move parallel to the
axis. Writing $\bar{S}_{\alpha\beta}=\epsilon_{\alpha\beta\mu\nu}\bar{S}^{\mu}\bar{u}^{\nu}$,
where $\bar{S}^{\alpha}$ denotes the new spin vector, the force now
reads 
\begin{equation}
\frac{DP^{\alpha}}{d\tau}=-\frac{1}{2}R_{\ \mu\nu\lambda}^{\alpha}U^{\mu}\bar{S}^{\nu\lambda}=-\star R_{\ \ \ \ \mu}^{\sigma\tau\alpha}U^{\mu}\bar{u}_{\tau}\bar{S}_{\sigma}\,,\label{eq:ForceSpinGeneral}
\end{equation}
which depends both on $U^{\alpha}$ and $\bar{u}^{\alpha}$, and,
in general, will also not be parallel to the axis. This clearly leads
to a more complicated description of the same problem.

The case of the Tulczyjew-Dixon (TD) condition, $\bar{u}^{\alpha}=P^{\alpha}/M$,
exemplifies some of these difficulties. First, we face the complicated
equation relating $P^{\alpha}$ and $U^{\alpha}$ \cite{Semerak I,Kunzle,TodFelice,EhlersRudoplh},
\begin{equation}
U^{\alpha}=\frac{m}{M^{2}}\left(P^{\alpha}+\frac{2\bar{S}^{\alpha\nu}R_{\nu\tau\kappa\lambda}\bar{S}^{\kappa\lambda}P^{\tau}}{4M^{2}+R_{\alpha\beta\gamma\delta}\bar{S}^{\alpha\beta}\bar{S}^{\gamma\delta}}\right)\ ,\label{eq:P_U_Dixon}
\end{equation}
which in general are not parallel; and to obtain the force, given
by Eq.~(\ref{eq:ForceSpinGeneral}), one needs to know both (not
just $U^{\alpha}$, as with the MP condition). Based only on these
equations, it would not be clear that an axial fall (of the physical
body) is possible, what kind of solution represents it in this gauge,
and how to set up its initial conditions. Using the knowledge of the
MP solution (which is an axial fall), we know that, for this problem,
$\vec{P}$ is parallel to $\vec{e}_{z}$; then, tentatively setting
$\bar{\mathbf{S}}=\bar{S}^{0}\mathbf{e}_{0}+\bar{S}^{z}\mathbf{e}_{z}$,
and $\bar{z}^{\alpha}$ along the $z$-axis, it can eventually be
shown from~(\ref{eq:P_U_Dixon}) (see e.g.~\cite{ShibataPRD96})
that, for such setup, $P^{\alpha}=mU^{\alpha}$, and therefore the
solution coincides with the one obtained using the MP condition. We
thus end up (in this case) with the same solution, but taking a more
complicated route.

In Sec.~\ref{sub:Kerr-dS} we concluded that in the equatorial plane
of Kerr-dS, for suitable $r$ and $\vec{v}$, spinning particles move
in prograde circular geodesics; we were able to do it only because
we used the MP condition. With this condition, the force is given
by a contraction of $\mathbb{H}_{\alpha\beta}$ with $S^{\alpha}$,
cf.~Eq.~(\ref{tab:Analogy}.1b). From the curvature invariants,
we deduced that in the equatorial plane there is a velocity field
for which $\mathbb{H}_{\alpha\beta}=0$, Eq.~(\ref{velocityKerr});
for certain $r=r_{{\rm geo}}$ (solution of Eq.~(\ref{eq:circ-geo-eq})),
it matches the velocity of a circular geodesic. Along such circle,
the equation of motion reduces to 
\begin{equation}
\frac{DP^{\alpha}}{d\tau}=0\Leftrightarrow ma^{\alpha}-\epsilon_{\ \beta\gamma\delta}^{\alpha}U^{\delta}\frac{D(S^{\beta}a^{\gamma})}{d\tau}=0\ ,\label{eq:Force Ker-dS}
\end{equation}
admitting $a^{\alpha}=0$ as trivial solution (obviously a ``non-helical''
one); the spinning particle will thus move along the circular geodesic.
We would not be able to reach this conclusion using other spin conditions:
for $\bar{u}^{\alpha}\ne U^{\alpha}$, the force is no longer governed
by the magnetic part of the Riemann tensor $\mathbb{H}_{\alpha\beta}$
(but instead by a tensor $\mathcal{H}_{\alpha\beta}=\star R_{\alpha\mu\beta\nu}\bar{u}^{\mu}U^{\nu}$
involving \emph{both} $\bar{u}^{\beta}$ and $U^{\beta}$, cf.~Eq.~(\ref{eq:ForceSpinGeneral})),
and therefore a similar analysis in terms of curvature invariants
is not possible. In particular, in the framework of the TD condition
$\bar{u}^{\alpha}=P^{\alpha}/M$, we doubt that it would ever be possible
to notice this effect using the system formed by Eqs.~(\ref{eq:ForceSpinGeneral})
and (\ref{eq:SpinDS0}), coupled with the momentum-velocity relation
(\ref{eq:P_U_Dixon}).

As for the application in Sec.~\ref{sub:Radial-Schwa}, the motion
in the Schwarzschild spacetime of a particle with radial initial velocity,
first notice that, for a particle with \emph{generic} spin $S^{\alpha}$,
the problem does not have spherical symmetry (regardless of the spin
condition; indeed, a force orthogonal to $\vec{e}_{r}$ arises in
the analogous electromagnetic setup, cf.~Eq.~(\ref{eq:ForceCoulombRadial})).
Using the MP condition, setting $\vec{U}=U^{r}\vec{e}_{r}$, we have,
cf.~Eqs.~(\ref{eq:HabSchwa}), $\mathbb{H}_{\alpha\beta}=0\Rightarrow DP^{\alpha}/d\tau=0$.
Hence we have (\ref{eq:Force Ker-dS}) as the equation of motion,
with trivial solution $a^{\alpha}=0\Rightarrow P^{\alpha}=mU^{\alpha}$,
i.e., the gyroscope moves along a radial geodesic. In the case of
the TD condition, again we face the complicated Eqs.~(\ref{eq:ForceSpinGeneral})-(\ref{eq:P_U_Dixon}),
not being transparent what occurs if one sets initially $\vec{U}|_{{\rm in}}=U^{r}\vec{e}_{r}$,
or if the solution thereby obtained corresponds to the same physical
motion above (a radial fall; in this framework it is not even obvious
that it occurs). From the analysis with the MP condition, we know
that, in order to represent the same problem, $\vec{P}=P^{z}\vec{e}_{z}=$\emph{constant}.
It is useful to re-write Eq.~(\ref{eq:P_U_Dixon}) in terms of tidal
tensors, 
\begin{equation}
U^{\alpha}=\frac{m}{M^{2}}\left(P^{\alpha}+\frac{\epsilon_{\ \ \tau\delta}^{\alpha\gamma}\bar{S}^{\tau}P^{\delta}(\mathbb{H}^{P})_{\sigma\gamma}\bar{S}^{\sigma}}{M^{2}+(\mathbb{F}^{P})^{\lambda\sigma}\bar{S}_{\lambda}\bar{S}_{\sigma}}\right)\ ,\label{eq:P_U_Dixon_Tidal}
\end{equation}
where $(\mathbb{H}^{P})_{\alpha\gamma}\equiv\star R_{\alpha\beta\gamma\delta}P^{\beta}P^{\delta}/M^{2}$
and $(\mathbb{F}^{P})_{\alpha\gamma}\equiv\star R\!\star_{\alpha\beta\gamma\delta}P^{\beta}P^{\delta}/M^{2}$
are, respectively, the gravitomagnetic tidal tensor and the ``$\mathbb{F}$
tensor'' \cite{BelDecomp,PaperAnalogies} measured by an observer
of 4-velocity $\bar{u}^{\alpha}=P^{\alpha}/M$. Noting, from Eqs.~(\ref{eq:HabSchwa}),
that, for radial $\vec{P}$, $(\mathbb{H}^{P})_{\alpha\beta}=0$,
Eq.~(\ref{eq:P_U_Dixon_Tidal}) yields $P^{\alpha}=mU^{\alpha}$,
and Eq.~(\ref{eq:ForceSpinGeneral}) gives $DP^{\alpha}/d\tau=0$;
i.e., we end up with the same solution obtained with the MP condition.
Other spin conditions, in general, will lead to $DP^{\alpha}/d\tau\ne0$,
and $U^{\alpha}\nparallel P^{\alpha}$ (see Fig. 6c-d of \cite{CostaNatario2014}),
thus more complicated descriptions for this motion.

In the case of the analogous electromagnetic problem, a magnetic dipole
with initial radial velocity in the Coulomb field, first we note that,
due to the electromagnetic hidden momentum $P_{{\rm hidEM}}^{\alpha}$,
in general $P^{\alpha}$ cannot be parallel to $U^{\alpha}$. Furthermore,
since $F_{{\rm EM}}^{\alpha}\ne0$ and $a^{\alpha}\ne0$, it is not
trivial to (exactly) prescribe the initial conditions for the MP non-helical
solution (which in the previous examples was ensured by $a^{\alpha}=0$).
To first order in $S$, we can impose it by taking $S^{\alpha\beta}a_{\beta}\approx0$,
see \cite{EPAPS}. With the TD condition, we face again a complicated
equation relating $P^{\alpha}$ with $U^{\alpha}$ (and therefore
$F_{{\rm EM}}^{\alpha}$ with $a^{\alpha}$), Eq.~(35) of~\cite{Wald et al 2010}.
An interesting choice for this system is the Corinaldesi-Papapetrou
condition \cite{Corinaldesi Papapetrou} $\bar{S}^{\alpha\beta}\bar{u}_{\beta}=0$,
where $\bar{\mathbf{u}}=\partial/\partial t$ corresponds to the static
observers. In this case $\bar{S}^{\alpha\beta}D\bar{u}_{\beta}/d\tau=0$,
thus $P_{{\rm hidI}}^{\alpha}=0$, cf.~Eq.~(\ref{eq:HiddenInertial}),
leading to $P^{\alpha}=mU^{\alpha}+P_{{\rm hidEM}}^{\alpha}$, which
is the simplest momentum-velocity relation possible for this problem.

More generally, in arbitrarily curved spacetimes, the inertial hidden
momentum $P_{{\rm hidI}}^{\alpha}$ can \emph{always} be made to vanish
by choosing a $\bar{u}^{\alpha}$ parallel transported along the reference
worldline, cf.~Eq.~(\ref{eq:HiddenInertial}). This choice may actually
be cast as a spin supplementary condition \cite{Semerak II} (for
its detailed discussion, see \cite{Semerak II,CostaNatario2014,SemerakPetrovTypes}).
It is especially favored for pole-dipole particles in purely gravitational
systems, because it leads to particularly simple equations: the momentum-velocity
relation is simply $P^{\alpha}=mU^{\alpha}$, and $\bar{S}^{\alpha\beta}$
is parallel transported, $D\bar{S}^{\alpha\beta}/d\tau=0$, cf.~Eq.~(\ref{eq:SpinDS0}).
On the other hand, in some treatments spin conditions for which $P_{{\rm hid}}^{\alpha}\ne0$
are preferred; that is the case of the Newton-Wigner \cite{Newton-Wigner,Pryce}
condition $\bar{u}^{\alpha}\propto P^{\alpha}/M+u_{{\rm lab}}^{\alpha}$,
where $u_{{\rm lab}}^{\alpha}$ is the 4-velocity of some ``laboratory
observer'' \cite{Hanson-Regge} (it may thus be cast as a combination
of the Tulczyjew-Dixon and Corinaldesi-Papapetrou conditions). It
is of advantage in some Hamiltonian and effective field theory approaches
\cite{Buonanno2009,Hanson-Regge,Steinhoffr,HergtSteinhoffSchaefer2010,Kunst2014,Porto,BakerOConnel19741975}
(see also \cite{Porto2006,SteinhoffHergtSchaeffer}) because it leads
to canonical Dirac brackets (to linear order in the spin, in the case
of curved spacetime \cite{Buonanno2009,Kunst2014}). The bottom line
is that the spin condition is \emph{gauge freedom}, and as such one
should choose, in each application, the one that suits it the most.
For the ones in this work (where we have been exploring exact analogies
that rely on it), it is the MP condition that is of clear advantage,
as explained above.

\subsection{Analogies under other spin conditions}

The \emph{exact} gravito-electromagnetic analogies studied so far
in this work were obtained by employing, in the equations of motion,
the Mathisson-Pirani (MP) spin condition. In this section we will
study how the situation changes by choosing other spin conditions.

\subsubsection{Analogy based on tidal tensors\label{sub:Impact Analogy-Tidal}}

For an arbitrary spin condition $\bar{S}^{\alpha\beta}\bar{u}_{\beta}=0$,
it is natural to define, as above, the spin vector $\bar{S}^{\mu}$
by $\bar{S}_{\alpha\beta}=\epsilon_{\alpha\beta\mu\nu}\bar{S}^{\mu}\bar{u}^{\nu}$,
in terms of which the spin-curvature force reads $DP^{\alpha}/d\tau=-\mathcal{H}_{\gamma}^{\ \alpha}\bar{S}^{\gamma}$,
where $\mathcal{H}_{\alpha\beta}\equiv\star R_{\alpha\mu\beta\nu}\bar{u}^{\mu}U^{\nu}$,
cf. Eq.~(\ref{eq:ForceSpinGeneral}). Thus the force is still given
by a contraction of a rank 2 tensor $\mathcal{H}_{\alpha\beta}$ with
$\bar{S}^{\alpha}$; this new tensor, however, does not coincide with
the magnetic part of the Riemann tensor $(\mathbb{H}^{u})_{\alpha\beta}=\star R_{\alpha\mu\beta\nu}u^{\mu}u^{\nu}$
as measured by any observer $u^{\alpha}$, because it results from
a contraction of $\star R_{\alpha\mu\beta\nu}$ with \emph{two} \emph{different}
vectors ($\bar{u}^{\mu}$ and $U^{\nu}$). It does not obey the field
equations in Table \ref{tab:Analogy}, since the trace and antisymmetric
parts of $\mathcal{H}_{\alpha\beta}$ no longer yield projections
of the Einstein field equations, nor equations of the type (\ref{tab:Analogy}.2b)-(\ref{tab:Analogy}.3b)
of Table \ref{tab:Analogy}.%
\footnote{Namely those will not be equations involving only tidal tensors and
sources, by contrast with both their magnetic counterparts (\ref{tab:Analogy}.2a)-(\ref{tab:Analogy}.3a)
of Table \ref{tab:Analogy}, and also with the gravitoelectric counterparts
Eqs.~(1.3b), (1.7b) of Table 1 of~\cite{PaperAnalogies}. Moreover,
the tensorial structure of $\mathcal{H}_{\alpha\beta}$ (unlike $\mathbb{H}_{\alpha\beta}$)
is not similar to its gravitoelectric counterpart $\mathbb{E}_{\alpha\beta}$,
i.e., it is not spatial in both indices with respect to the same time-like
vector, nor does it have to be symmetric in vacuum.%
} Instead, another analogy can be drawn here. First note that by choosing,
as reference worldline, the centroid $x_{{\rm CM}}^{\alpha}(\bar{u})$
given by the condition $\bar{S}^{\alpha\beta}\bar{u}_{\beta}=0$,
that generates a mass dipole $d_{{\rm G}}^{\alpha}=-\bar{S}^{\alpha\beta}U_{\beta}$
in the centroid rest frame, cf. Eq. (\ref{eq:Massdipole}). Decomposing
$\bar{S}^{\alpha\beta}$ into its time and space projections relative
to the centroid 4-velocity $U^{\alpha}=dx_{{\rm CM}}^{\alpha}(\bar{u})/d\tau$,
we have 
\begin{equation}
\bar{S}^{\alpha\beta}=2d_{{\rm G}}^{[\alpha}U^{\beta]}+\epsilon_{\ \ \mu\lambda}^{\alpha\beta}U^{\lambda}(\bar{S}^{U})^{\mu}\label{eq:Sbardecomp}
\end{equation}
where we used Eq. (4) of \cite{PaperAnalogies}, and the vector 
\begin{equation}
(\bar{S}^{U})^{\mu}\equiv\frac{1}{2}\epsilon_{\ \alpha\beta\gamma}^{\mu}\bar{S}^{\alpha\beta}U^{\gamma}\label{eq:SbarU}
\end{equation}
encodes the components of $\bar{S}^{\alpha\beta}$ spatial with respect
to $U^{\alpha}$, that is, what one would physically interpret as
the classical angular momentum 3-vector (cf. e.g. \cite{Schiff})
about $x_{{\rm CM}}^{\alpha}(\bar{u})$, as measured\emph{ in the
centroid frame} (i.e., as measured by the observer of 4-velocity $U^{\alpha}$).
Substituting Eq. (\ref{eq:Sbardecomp}) into the second member of
Eq. (\ref{eq:ForceSpinGeneral}) yields 
\begin{equation}
\frac{DP^{\alpha}}{d\tau}=-\mathbb{H}_{\beta}^{\ \alpha}(\bar{S}^{U})^{\beta}-\mathbb{E}_{\beta}^{\ \alpha}d_{{\rm G}}^{\beta}\ .\label{eq:ForceSpinGeneral2}
\end{equation}
This resembles the electromagnetic force exerted on a particle possessing
both magnetic and electric dipole moments (as measured in the centroid
frame). Indeed, the right-hand member of Eq. (\ref{eq:ForceSpinGeneral2})
is formally analogous to the the second and third terms of Eq. (\ref{eq:ForcePirani});
however the last term of (\ref{eq:ForcePirani}) (which is also part
of the force on an electric dipole), has no counterpart in (\ref{eq:ForceSpinGeneral2}).
Since this term is not a tidal term, it is natural that it has no
gravitational counterpart. An exact analogy exists however between
Eq. (\ref{eq:ForceSpinGeneral2}) and the ``canonical'' electromagnetic
force on a particle with electric and magnetic dipole moments (and
zero charge), 
\begin{equation}
\frac{DP_{{\rm Dix}}^{\alpha}}{d\tau}=B_{\beta}^{\ \alpha}\mu^{\beta}+E_{\beta}^{\ \alpha}d^{\beta}\label{eq:CanonicalForce}
\end{equation}
obtained by substituting Eq. (\ref{eq:Qab}) into (\ref{eq:ForceCan}).

\emph{Tulczyjew-Dixon (TD) condition }$(\bar{u}^{\alpha}=P^{\alpha}/M)$\emph{.---
}Noting that $U^{\alpha}=(P^{\alpha}-P_{{\rm hid}}^{\alpha})/m$,
we have in this case $d_{{\rm G}}^{\alpha}=-\bar{S}^{\alpha\beta}U_{\beta}=\bar{S}_{\ \beta}^{\alpha}P_{{\rm hid}}^{\beta}/m=\mathcal{O}$,
and $(\bar{S}^{U})^{\mu}=\bar{S}^{\mu}+\mathcal{O}$, where $\mathcal{O}$
is of order $\mathcal{O}(S^{2})$ if electromagnetic hidden momentum
is present ($P_{{\rm hid}{\rm EM}}^{\alpha}\ne0$), or $\mathcal{O}(S^{3})$
otherwise. Therefore, to a good approximation (in particular in a
pole-dipole approximation), Eq. (\ref{eq:ForceSpinGeneral2}) becomes
$F_{{\rm G}}^{\alpha}=-\mathbb{H}_{\beta}^{\ \alpha}\bar{S}^{\beta}$,
and the analogy in Table \ref{tab:Analogy} holds.

\emph{Corinaldesi-Papapetrou (CP) condition} ($\bar{u}_{\alpha}=u_{{\rm lab}}^{\alpha}$).---
This condition was introduced, for the case of Schwarzschild spacetime,
in the non-covariant form $\bar{S}^{i0}=0$ \cite{Corinaldesi Papapetrou},
where it states that the reference worldline is the centroid as measured
by the observers at rest in Schwarzschild coordinates. It can be generalized
\cite{Semerak II,CostaNatario2014} to arbitrary coordinate systems
in arbitrary spacetimes in the covariant form $\bar{S}_{\ \beta}^{\alpha}u_{{\rm lab}}^{\beta}=0$,
where $u_{{\rm lab}}^{\beta}$ is the 4-velocity of the observers
at rest in the chosen coordinate system (the ``laboratory'' observers
$u_{{\rm lab}}^{i}=0$ \cite{CostaNatario2014,Wald et al 2010}).
In this case $d_{{\rm G}}^{\alpha}=-\bar{S}^{\alpha\beta}U_{\beta}=-\bar{S}^{\alpha\beta}v_{\beta}(U,u_{{\rm lab}})\gamma(U,u_{{\rm lab}})$,
where $v^{\beta}(U,u_{{\rm lab}})$ is the velocity of the centroid
relative to the laboratory observers, cf. decomposition (\ref{eq:U_u}).
Therefore the second term of (\ref{eq:ForceSpinGeneral2}) is of first
order in $S$ and cannot in general be neglected (for instance, in
the Schwarzschild spacetime, the two terms are typically of the same
magnitude, see Sec. 3.4.2 of \cite{CostaNatario2014}). So the analogy
that holds is between Eqs. (\ref{eq:ForceSpinGeneral2}) and (\ref{eq:CanonicalForce})
(not the one in Table \ref{tab:Analogy}, between the spin curvature
and the force on a magnetic dipole).

\emph{Newton-Wigner (NW) condition} ($\bar{u}^{\alpha}\propto u_{{\rm lab}}^{\alpha}+P^{\alpha}/M$).---
In this case the reference worldline is chosen as the centroid $x_{{\rm CM}}^{\alpha}(u_{{\rm NW}})$
as measured by observers of 4-velocity (cf. \cite{Hanson-Regge,Buonanno2009,Kunst2014,SemerakPetrovTypes})
\begin{equation}
u_{{\rm NW}}^{\alpha}=K\left(u_{{\rm lab}}^{\alpha}+\frac{P^{\alpha}}{M}\right);\qquad K\equiv\sqrt{\frac{M}{2(M+m_{{\rm lab}})}}\label{eq:u_NW}
\end{equation}
($m_{{\rm lab}}\equiv-u_{{\rm lab}}^{\alpha}P_{\alpha}$), that is,
an even-weighted combination of the 4-velocity of the laboratory and
the zero 3-momentum observers. Due to that, the situation with this
spin condition is essentially similar to with the CP condition; it
resembles more the electromagnetic force on a particle possessing
both electric and magnetic moments (as measured in the centroid frame),
and is closely analogous to the ``canonical'' electromagnetic force
on such particle (except that the mass dipole $d_{{\rm G}}^{\alpha}=-\bar{S}^{\alpha\beta}U_{\beta}$
is different from the CP one, as $\bar{S}^{\alpha\beta}$ is now a
different tensor, obeying $\bar{S}_{\ \beta}^{\alpha}u_{{\rm NW}}^{\beta}=0$).

\emph{``Parallel''} \emph{condition }($D\bar{u}^{\alpha}/d\tau=0$).---
This condition chooses as reference worldline some time-like vector
$\bar{u}^{\alpha}$ parallel transported along the reference worldline
$\bar{z}^{\alpha}$. Since the initial vector $\bar{u}_{{\rm in}}^{\alpha}$
is arbitrary \cite{Semerak II}, we may choose it as $\bar{u}_{{\rm in}}^{\alpha}=U^{\alpha}$,
so that initially one obtains exactly the analogy in Table \ref{tab:Analogy},
just like for the MP condition. Since, in general, the motion is non-geodesic,
$\bar{u}^{\alpha}$ will progressively diverge from $U^{\alpha}$,
so at later instants that analogy will be only approximate, whilst
the analogy between Eqs. (\ref{eq:ForceSpinGeneral2}) and (\ref{eq:CanonicalForce})
remains exact.

\subsubsection{Spin precession}

The analogy found in Eq. (\ref{eq:Spin3+1}) of Sec. \ref{sub:Spin-``precession''--}
using the MP condition holds in an orthonormal frame comoving with
the centroid, for a spin vector $S^{\alpha}$ which represents the
angular momentum, as measured in the centroid frame, taken about the
centroid $x_{{\rm CM}}^{\alpha}(U)$ measured, again, in is own rest
frame. Other spin conditions $\bar{S}^{\alpha\beta}\bar{u}_{\beta}=0$
correspond to different angular momentum tensors $\bar{S}^{\alpha\beta}$,
taken about the centroids $x_{{\rm CM}}^{\alpha}(\bar{u})$ measured
by the observer of 4-velocity $\bar{u}^{\alpha}$ (not $U^{\alpha}$).
The vector which encodes the angular momentum about $x_{{\rm CM}}^{\alpha}(\bar{u})$,
and \emph{as measured in the centroid frame} is, as explained above,
$(\bar{S}^{U})^{\alpha}$, see Eqs. (\ref{eq:Sbardecomp})-(\ref{eq:SbarU}).
To compute its evolution equation, one first notes that $\epsilon_{\alpha\beta\gamma\delta}U^{\delta}D\bar{S}^{\alpha\beta}/d\tau=2D(\bar{S}^{U})_{\gamma}/d\tau-\epsilon_{\alpha\beta\gamma\delta}a^{\delta}\bar{S}^{\alpha\beta}$;
then, using (\ref{eq:DSabdtHidden}) (with $S^{\alpha\beta}=\bar{S}^{\alpha\beta}$)
and (\ref{eq:Sbardecomp}), we have 
\[
\frac{D(\bar{S}^{U})^{\gamma}}{d\tau}=(\bar{S}^{U})_{\mu}a^{\mu}U^{\gamma}+\epsilon_{\ \alpha\beta\delta}^{\gamma}U^{\delta}\left[d_{{\rm G}}^{\beta}a^{\alpha}+\frac{1}{2}\tau^{\alpha\beta}\right]\ 
\]
In an orthonormal tetrad $\mathbf{e}_{\hat{\alpha}}$ comoving with
the centroid, this equation reads, using (\ref{eq:taudip}) (see Sec.~\ref{sub:Spin-``precession''--}),
\begin{equation}
\frac{d\vec{\bar{S}}^{U}}{d\tau}=\vec{\bar{S}}^{U}\times\vec{\Omega}+\vec{d}_{{\rm G}}\times\vec{G}+\vec{\mu}\times\vec{B}+\vec{d}\times\vec{E}\ ,\label{eq:AnalogySpinGeneral}
\end{equation}
where $G^{\alpha}=-a^{\alpha}$ is the ``gravitoelectric field''
as measured in the centroid frame, cf Sec. \ref{sub:Momentum-of-the-Particle}.
This equation manifests that, for \emph{an arbitrary} spin condition,
an exact analogy always exists, with $\{\vec{\bar{S}}^{U},\vec{d}_{{\rm G}}\}$
playing a role analogous to the magnetic and electric dipole moment
vectors $\{\vec{\mu},\vec{d}\}$, and the inertial fields $\{\vec{\Omega},\vec{G}\}$
playing a role analogous to the electromagnetic fields $\{\vec{B},\vec{E}\}$
(all quantities measured in \emph{the centroid} \emph{frame}). As
discussed in Sec.~\ref{sub:Spin-``precession''--}, if $\mathbf{e}_{\hat{\alpha}}$
is adapted to a congruence of observers, then $\vec{\Omega}=\vec{H}/2$,
and the analogy deepens. The term $\vec{d}_{{\rm G}}\times\vec{G}\equiv-\vec{d}_{{\rm G}}\times\vec{a}$
is the exact version of the ``instrumental torque'' discussed in
\cite{Schiff}%
\footnote{\label{fn:Schiff}To make the connection with \cite{Schiff}, we note
that: therein the CP condition is considered, so $d_{{\rm G}}^{\alpha}=-\bar{S}_{\ \beta}^{\alpha}U^{\beta}=\epsilon_{\ \gamma\delta\beta}^{\alpha}U^{\beta}v^{\delta}\bar{S}^{\gamma}$
with $v^{\delta}\equiv v^{\delta}(U,u_{{\rm lab}})$, reading, in
the centroid frame, $\vec{d}_{{\rm G}}=\vec{\bar{S}}\times\vec{v}$;
$\vec{\bar{S}}\equiv\mathbf{S}$, $\vec{\bar{S}}^{U}\equiv\mathbf{S}_{0}$
in their notation; and $\vec{a}=\vec{F}/m+\mathcal{O}(S)$.%
} in the weak field and slow motion regime. If one chooses $\bar{u}^{\alpha}=U^{\alpha}$
(MP condition) then $\vec{d}_{{\rm G}}=0$, $\vec{\bar{S}}^{U}=\vec{S}$,
and, taking also particles with no electric dipole moment in the centroid
frame ($\vec{d}=0$), Eq. (\ref{eq:AnalogySpinGeneral}) reduces to
Eq. (\ref{eq:Spin3+1}). Under the TD condition $\bar{S}^{\alpha\beta}P_{\beta}=0$
the situation is similar to a good approximation: as we have seen
in Sec. \ref{sub:Impact Analogy-Tidal}, $d_{{\rm G}}^{\alpha}$ is
of order $\mathcal{O}(S^{2})$ if $P_{{\rm hid}{\rm EM}}^{\alpha}\ne0$,
or $\mathcal{O}(S^{3})$ otherwise.

Under the CP condition ($\bar{S}_{\ \beta}^{\alpha}u_{{\rm lab}}^{\beta}=0$),
$\vec{d}_{{\rm G}}$ is of order $\mathcal{O}(S)$ (cf. Footnote \ref{fn:Schiff}),
hence the situation depends on the type of force applied on the body.
If $q=\vec{d}=0$, and only gravitational and electromagnetic forces
are present, $\vec{d}_{{\rm G}}\times\vec{a}\sim\mathcal{O}(S^{2})$,
and one recovers, to a good approximation, the analogy in Eq. (\ref{eq:Spin3+1})
(with $\vec{\bar{S}}^{U}$ in the place of $\vec{S}$). Otherwise,
for a generic force (or if $q\ne0$), $\vec{d}_{{\rm G}}\times\vec{a}\sim\mathcal{O}(S)$,
non-negligible in pole-dipole, nor in weak-field slow motion approximations
\cite{Schiff}, thus in this case it is only the analogy in Eq. (\ref{eq:AnalogySpinGeneral})
that holds. With the NW condition, $\bar{S}_{\ \beta}^{\alpha}u_{{\rm NW}}^{\beta}=0$,
the situation is very similar, due to the contribution of $u_{{\rm lab}}^{\alpha}$
to $u_{{\rm NW}}^{\alpha}$ in Eq. (\ref{eq:u_NW}).

\subsubsection{Hidden momentum\label{sub:Impact_condtions_Hidden-momentum}}

Under an arbitrary spin condition neither $P_{{\rm hidI}}^{\alpha}$
nor $P_{{\rm hid}{\rm EM}}^{\alpha}$ take the forms (\ref{eq:PhidDecompMP}),
and there is no longer a close analogy between the two. For instance,
under the ``parallel'' condition $D\bar{u}^{\alpha}/d\tau=0$, one
has simply $P_{{\rm hidI}}^{\alpha}=0$; moreover $P_{{\rm hid}{\rm EM}}^{\alpha}$
(Eq. (\ref{eq:PhidEM-0})) takes in general a complicated form, encoding
not only the hidden momentum modeled in e.g. Fig. 9 of \cite{GriffithsAmJPhys}
(which is physical), but also a pure gauge part that is due solely
to the choice of centroid, see Sec. 3.5.1 of \cite{CostaNatario2014}.
An exception is the TD condition, under which Eqs. (\ref{eq:PhidDecompMP})
are still obtained to a good approximation (namely by neglecting terms
of order $\mathcal{O}(S^{2})$ and $\mathcal{O}(Sd)$, consistent
with a dipole approximation); it was actually in such approximate
form that this analogy was first introduced in \cite{Wald et al 2010}.

\end{appendices}


\begin{thebibliography}{100}
\bibitem{Wald}Robert M. Wald, Phys. Rev. D \textbf{6}, 406 (1972).

\bibitem{Ruggiero:2002hz} M.~L.~Ruggiero and A.~Tartaglia, Il
Nuovo Cimento B,\ \textbf{117}, 743 (2002). \href{http://arxiv.org/abs/gr-qc/0207065}{[arXiv:gr-qc/0207065]}.

\bibitem{Gravitation and Inertia}I. Ciufolini, J. A. Wheeler, \emph{Gravitation
and Inertia}, Princeton Series in Physics (Princeton, NJ, 1995).

\bibitem{Black Holes}Kip S. Thorne, R. H. Price, D. A. Macdonald,
\emph{Black Holes, the Membrane Paradigm} (Yale Univ. Press, 1986).

\bibitem{Harris1991} Edward G. Harris, Am. J. Phys. \textbf{59} (5)
(1991) 421

\bibitem{Natario}José Natário, Gen. Rel. Grav. \textbf{39}, 1477
(2007).


\bibitem{CHPRD} L. F. Costa, C. A. R. Herdeiro, Phys. Rev. D \textbf{78},
024021 (2008).

\bibitem{Near Zero}Kip S. Thorne in \emph{Near Zero: New Frontiers
of Physics}, Eds. J. D. Fairbank, B. S. Deaver Jr., C. W. F. Everitt,
P. F. Michelson (W. H. Freeman and Co., NY, 1988).

\bibitem{O'Connel Spin Rotation 1974}R. F. O'Connell, ``Spin, Rotation
and C, P, and T Effects in the Gravitational Interaction and Related
Experiments,'' in Experimental Gravitation: Proc. of Course 56 of
the Int. School ``Enrico Fermi'' (Academic Press, 1974), p.496.

\bibitem{lense}B. Mashhoon, F. W. Hehl, D. S. Theiss, Gen. Rel. Grav.
\textbf{16}, 711 (1984).

\bibitem{Tucker Clark}S. J. Clark, R. W. Tucker, Class. Quant. Grav.
\textbf{17}, 4125 (2000).

\bibitem{The many faces} R. T. Jantzen, P. Carini, D. Bini, Ann.
Phys. \textbf{215}, 1 (1992). \href{http://arxiv.org/abs/gr-qc/0106043}{[arXiv:gr-qc/0106043]}

\bibitem{GEM User Manual}R. T. Jantzen, P. Carini, D. Bini, ``GEM:
the User Manual'' (2004); \url{http://www34.homepage.villanova.edu/ robert.jantzen/gem/gem_grqc.pdf}

\bibitem{PaperAnalogies}L. F. Costa, J. Natário, Gen. Rel. Grav.
\textbf{46}, 1792 (2014).

\bibitem{matte} A.~Matte, Canadian J. Math. \textbf{5}, 1 (1953).

\bibitem{bel} L.~Bel, Cahiers de Physique, \textbf{16}, 59 (1962);
Eng. trans.: Gen. Rel. Grav., \textbf{32}, 2047 (2000).

\bibitem{Maartens:1997fg} R.~Maartens, B.~Bassett, Class.\ Q.\ Grav.\ \textbf{15},
705 (1998).

\bibitem{PaperInvariantes}L. F. O. Costa, L. Wylleman, J. Natário,
\href{http://arxiv.org/abs/1603.03143}{arXiv:1603.03143}

\bibitem{Semerak I}O. Semerák, Mon. Not. R. Soc. \textbf{308}, 863
(1999).

\bibitem{Wald et al 2010}S. E. Gralla, A. I. Harte, R. M. Wald, Phys.
Rev. D \textbf{81,} 104012 (2010).

\bibitem{Helical}L. F. Costa, C. Herdeiro, J. Natário, M. Zilhão,
Phys. Rev. D \textbf{85}, 024001 (2012).

\bibitem{Hawking}S. W. Hawking, Phys Rev. Lett. \textbf{26}, 1344
(1971).

\bibitem{Gravitation}C. W. Misner, Kip. S. Thorne and J. A. Wheeler,
\emph{Gravitation} (W. H Freeman and Co., San Francisco, 1973).

\bibitem{BiniStaticObs}D. Bini, A. Geralico, R. Jantzen, Class. Quant.
Grav. \textbf{22}, 4729 (2005).

\bibitem{WyllemanBeke}L. Wylleman, D. Beke, Phys. Rev. D \textbf{81,}
104038 (2010).

\bibitem{EPAPS}See supplemental material provided in ancillary file
``Supplement.PDF'' enclosed in this submission.

\bibitem{Dixon1970I} W. G. Dixon, Proc. Roy. Soc. Lond. A. \textbf{314},
499 (1970).

\bibitem{Wald et al}S. E. Gralla, A. I. Harte, R. M. Wald, Phys.
Rev. D \textbf{80}, 024031 (2009).

\bibitem{MathissonNeueMechanik}M. Mathisson, Acta Phys. Pol. \textbf{6},
163 (1937);\\
 Eng. Trans.: Gen. Rel. Grav. \textbf{42}, 1011 (2010).

\bibitem{Dixon1967}W. G. Dixon, J. Math. Phys. \textbf{8}, 1591 (1967).

\bibitem{Madore:1969}J. Madore, Ann. Inst. Henri Poincare \textbf{11},
221 (1969).

\bibitem{Jackson}J. D. Jackson, \emph{Classical Electrodynamics},
3rd Edition (John Wiley \& Sons, NY, 1999).

\bibitem{CostaNatario2014}L. F. Costa, J. Natário, in \emph{Equations
of motion in Relativistic Gravity}, D. Puetzfeld et al. (eds), Fundamental
Theories of Physics \textbf{179}, 215-258 (Springer, 2015). \href{http://arxiv.org/abs/1410.6443}{[arXiv:1410.6443]} 

\bibitem{Dixon1964}W. G. Dixon, Il Nuovo Cimento, \textbf{34}, 317
(1964).

\bibitem{Dixon1970II}W. G. Dixon, Proc. Roy. Soc. Lond. A. \textbf{319},
509 (1970).

\bibitem{Dixon1974III}W. G. Dixon, Phil. Trans. R. Soc. Lond. A \textbf{277},
59 (1974).

\bibitem{Papapetrou I}A. Papapetrou, Proc. R. Soc. London A \textbf{209},
248 (1951).

\bibitem{Dixon1965}W. G. Dixon, Il Nuovo Cimento, \textbf{38}, 1616
(1965).

\bibitem{Semerak II}K. Kyrian, O. Semerák, Mon. Not. R. Soc. \textbf{382},
1922 (2007).

\bibitem{MollerAIP}C. Møller, Ann. Inst. Henri Poincaré \textbf{11},
251 (1949).

\bibitem{MollerBook}C. Møller, \emph{The theory of relativity} (Clarendon
Press, Oxford, 1960).

\bibitem{Frenkel}J. Frenkel, Z. Phys. \textbf{37}, 243 (1926); \emph{idem}
Nature \textbf{117,} 653 (1926).

\bibitem{Mathisson Zitterbewegung}M. Mathisson, Acta Phys. Pol. \textbf{6},
218 (1937).

\bibitem{Pirani 1956}F.A.E. Pirani, Acta Phys. Pol. \textbf{15} (1956)
389; Gen. Rel. Grav. \textbf{41}, 1215 (2009).

\bibitem{TulczyjewII}W. Tulczyjew, Acta Phys. Pol. \textbf{18,} 393
(1959).

\bibitem{Taub}A. H. Taub, J. Math. Phys. \textbf{5} 112 (1964).

\bibitem{EulerTop}José Natário, Commun. Math. Phys. \textbf{281},
387 (2008).

\bibitem{Plyatsko Non-Oscillatory}R. Plyatsko, O. Stephanyshin, Acta
Phys. Pol. B \textbf{39}, 23 (2008).

\bibitem{BaylinMassless}M. Baylin, S. Ragusa, Phys. Rev. D \textbf{15},
3543 (1977).

\bibitem{BaylinMassless II}M. Bailyn, S. Ragusa, Phys. Rev. D \textbf{23},
1258 (1981).

\bibitem{MashhoonMassless}B. Mashhoon, Ann. Phys. \textbf{89}, 254
(1975).

\bibitem{Beigblock}W. Beiglböck, Commun. Math. Phys. \textbf{5},
106 (1967).

\bibitem{DixonGRG1973}W. G. Dixon, Gen. Rel. Grav. \textbf{4}, 199
(1973).

\bibitem{Corinaldesi Papapetrou}E. Corinaldesi, A. Papapetrou, Proc.
R. Soc. London A \textbf{209}, 259 (1951).

\bibitem{Newton-Wigner} T. D. Newton, E. P. Wigner, Rev. Mod. Phys.
\textbf{21}, 400 (1949).

\bibitem{Pryce}M. H. L. Pryce, Proc. Roy. Soc. Lond. A \textbf{195},
62 (1948).

\bibitem{Buonanno2009}E. Barausse, E. Racine, A. Buonanno, Phys.
Rev. D \textbf{80}, 104025 (2009).

\bibitem{Hanson-Regge}A. J. Hanson, T. Regge, Ann. Phys. \textbf{87},
498 (1974).

\bibitem{Steinhoffr}J. Steinhoff, Ann. Phys. (Berlin) \textbf{523},
296 (2011).

\bibitem{HergtSteinhoffSchaefer2010}S. Hergt, J. Steinhoff, G. Schäfer,
Class. Quant. Grav. \textbf{27}, 135007 (2010).

\bibitem{Kunst2014}G. Lukes-Gerakopoulos, J. Seyrich, D. Kunst, Phys.
Rev. D \textbf{90}, 104019 (2014).

\bibitem{Porto}R. A. Porto, I. Z. Rothstein, Phys. Rev. Lett. \textbf{97},
021101 (2006); idem Phys. Rev. D \textbf{78}, 044012 (2008); R. A.
Porto, Class. Quant. Grav. \textbf{27}, 205001 (2010).

\bibitem{BakerOConnel19741975}B. Barker, R. F. O'Connell, Gen. Rel.
Grav. \textbf{11} 149 (1979); \emph{idem} Gen. Rel. Grav. \textbf{5},
539 (1974).

\bibitem{Stephani} H.~Stephani, \emph{Relativity} , 3rd ed. (Cambridge
Univ. Press, 2004).

\bibitem{MassaZordan}E. Massa, C. Zordan, Meccanica \textbf{10},
27 (1975).

\bibitem{Massa}E. Massa., Gen. Rel. Grav. \textbf{5}, 573 (1974);\\
 \emph{idem} Gen. Rel. Grav. \textbf{5}, 555 (1974).

\bibitem{General Relativity}Robert M. Wald, \emph{General Relativity}
(The University of Chicago Press, 1984).

\bibitem{Ciufolini Nature Review} I. Ciufolini, \textit{Nature} \textbf{449},
41 (2007).

\bibitem{PaperIAU}L. F. Costa, C. Herdeiro, in \emph{Relativity in
fundamental astronomy: dynamics, reference frames, and data analysis},
Proceedings of the International Astronomical Union, edited by S.
A. Klioner, P. K. Seidelmann and M. H. Soffel (Cambridge Univ. Press,
2010), Vol. \textbf{5}, S261, pp. 31\textendash{}39. \href{http://arxiv.org/abs/0912.2146}{[arXiv:0912.2146]}

\bibitem{Shockley}W. Shockley, R. P. James, Phys. Rev Lett \textbf{18},
876 (1967).

\bibitem{GriffithsAmJPhys}D. Babson, S. P. Reynolds, R. Bjorquist,
D. J. Griffiths, Am. J. Phys. \textbf{77,} 826 (2009).

\bibitem{Vaidman}Lev Vaidman, Am. J. Phys. \textbf{58}, 978 (1990).

\bibitem{HnizdoFluid}V. Hnizdo, Am. J. Phys \textbf{65}, 92 (1997).

\bibitem{Van Vleck}S. Coleman, J. H. Van Vleck, Phys. Review \textbf{171},
1370 (1968).

\bibitem{ProcERE2011}L. F. Costa, J. Natário, M. Zilhão, Proc. Spanish.
Rel. Meeting 2011, AIP Conf. Proc. \textbf{1458}, 367 (2011). \href{http://arxiv.org/abs/1206.7093}{[arXiv:1206.7093]}

\bibitem{Hartle2009}A. Harte, Class. Quant. Grav. \textbf{26}, 155015
(2009).

\bibitem{GrallaHerrmannHidden}S. Gralla, F. Herrmann, Class. Quant.
Grav. \textbf{30}, 205009 (2013).

\bibitem{GriffithsBook}D. J. Griffiths, \emph{Introduction to electrodynamics},
(Pearson Benjamim Cummings, San Francisco, 2008).




\bibitem{CorbenBook}H. C. Corben, \emph{Classical and Quantum Theories
of Spinning Particles} (Holden Day, Inc., San Francisco, 1968).

\bibitem{Corben}H. C. Corben, Physical Review \textbf{121}, 1833
(1961).

\bibitem{WeyssenhoffRaabe}J. Weyssenhoff, A. Raabe, Acta Phys. Pol.
\textbf{9}, 7 (1947).

\bibitem{Coombes}C. A. Coombes, Am. J. Phys. \textbf{47}, 915 (1979).

\bibitem{Young}R. H. Young, Am. J. Phys. \textbf{44}, 581 (1976).

\bibitem{YoungQuestion66}R. H. Young, Am. J. Phys. \textbf{66}, 1043
(1998).

\bibitem{Deissler}R. J. Deissler, Phys. Rev. E \textbf{77} (2008)
036609.

\bibitem{WyllePRD}L. Wylleman, N. Van den Bergh, \textit{\emph{Phys.
Rev. D}}\emph{.} \textbf{74} (2006) 084001.

\bibitem{LandauLifshitz}L. Landau, E. Lifshitz, \emph{The classical
theory of fields}, $4^{th}$ ed. (BH-Elsevier; Amsterdam, 1975).

\bibitem{McIntosh et al 1994}C. McIntosh, R. Arianhod, S. T. Wade,
C. Hoenselaers, Class. Quant. Grav. \textbf{11}, 1555 (1994).

\bibitem{BarnesConjecture}A. Barnes, in \emph{Proceedings of the
XXVIIth Spanish Relativity Meeting}, eds. J. Miralles, J. Font, J.
Pons (Universidad de Alicante, 2004). \href{http://arxiv.org/abs/gr-qc/0401068}{[arXiv:gr-qc/0401068]}

\bibitem{VBergh2003}N.~Van den Bergh, Class.\ Quant.\ Grav.\ \textbf{20},
L165 (2003).

\bibitem{cherubini:02} C.~Cherubini, D.~Bini, S.~Capozziello,
R.~Ruffini, Int. J. Mod. Phys. D \textbf{11}, 827 (2002).

\bibitem{SemerakStationaryFrames}O. Semerák, Gen. Rel. Grav. \textbf{25},
1041 (1993).

\bibitem{VelocityKerr}Kjell Rosquist, Gen. Rel. Grav. \textbf{41},
2619 (2009).

\bibitem{RindlerPerlick}W. Rindler, V. Perlick, Gen. Rel. Grav. \textbf{22},
1067 (1990).

\bibitem{StuchlikKovar}Z. Stuchlik, J. Kovar, Class. Quant. Grav.
\textbf{23}, 3935 (2006).

\bibitem{ProcEre2009}L Filipe Costa, C. Herdeiro, J. Phys.: Conf.
Ser. \textbf{229} 012031 (2010).

\bibitem{Hartl}M. D. Hartl, Phys. Rev. D \textbf{67,} 024005 (2003).


\bibitem{Ciufolini Lageos} I. Ciufolini, E. Pavlis, \textit{\emph{Nature}}
\textbf{431}, 958 (2004); I. Ciufolini, E. Pavlis, R. Peron, \textit{\emph{New
Astron.}} \textbf{11}, 527 (2006).

\bibitem{WillPoissonBook}E. Poisson, C. Will, \emph{Gravity: Newtonian,
Post-Newtonian, Relativistic} (Cambridge Univ. Press, 2014).

\bibitem{WillBook}C. Will, \emph{Theory and Experiment in Gravitational
Physics} (Cambridge Univ. Press, 1993).

\bibitem{Faye2006}G. Faye, L. Blanchet, A. Buonanno, Phys. Rev. D
\textbf{74}, 104033 (2006).

\bibitem{KidderWill}L. E. Kidder, C. M. Will, A. G. Wiseman, Phys.
Rev. D \textbf{47}, R4183 (1993).

\bibitem{TagoshiOhashiOwen}H. Tagoshi, A. Ohashi, B. J. Owen, Phys.
Rev. D \textbf{63}, 044006 (2001).

\bibitem{DSX}T. Damour, M. Soffel, C.~Xu, Phys. Rev. D\emph{.}\ \textbf{43},
3273 (1991).

\bibitem{Kaplan}J. D. Kaplan, D. A. Nichols, Kip S. Thorne , Phys.
Rev. D \textbf{80}, 124014 (2009).

\bibitem{PlyatskoCQG2011}R. Plyatsko, O. Stephanyshin, M. Fenyk,
Class. Quant. Grav. \textbf{28,} 195025 (2011).

\bibitem{BaileyIsrael}I. Bailey, W. Israel, Annals of Physics \textbf{130},
188 (1980).

\bibitem{EhlersRudoplh}J. Ehlers, R. Rudolph, Gen. Rel.Grav. \textbf{8},
197 (1977).

\bibitem{ThorneHartle85}K. S. Thorne, J. B. Hartle, Phys. Rev. D
\textbf{31,}1815 (1985).

\bibitem{Frehland1971}E. Frehland, Int. J. Theor. Phys. \textbf{4},
33 (1971).

\bibitem{Gravitation and Spacetime}Hans C. Ohanian, Remo Ruffini,
\emph{Gravitation and Spacetime} (W.W. Norton \& Company, 2nd Ed.,
1994)

\bibitem{Nordtvedt1988} K. Nordtvedt, \textit{\emph{Int. J. Theo.
Phys.}} \textbf{\textit{\emph{27}}}, 1395 (1988).

\bibitem{PomeranskiiKhriplovich}A. A. Pomeranskii, I. B. Khriplovich,
JETP \textbf{86,} 839 (1998); A. A. Pomeranskii, R. A. Sen'Kov, I.
B. Khriplovich, Phys. Usp. \textbf{43} 1055 (2000).

\bibitem{Khriplovich2008}I. B. Khriplovich, Acta Phys. Pol. B, proceedings
supplement, \textbf{1,} 197 (2008). \href{http://arxiv.org/abs/0801.1881}{[arXiv:0801.1881]}

\bibitem{BhabhaCorben}H. J. Bhabha, H. C. Corben, Proc. R. Soc. London
A \textbf{178,} 273 (1941).

\bibitem{BiniGemelliRuffini}D. Bini, G. Gemelli, R. Ruffini, Phys.
Rev. D \textbf{61}, 064013 (2000).

\bibitem{Mathisson Variational}M. Mathisson, Proc. Cambrige Phil.
Soc. \textbf{36}, 331 (1940).

\bibitem{Israel}W. Israel, Gen. Rel. Grav. \textbf{9}, 451 (1978).

\bibitem{Furry}W. H. Furry, Am. J. Phys. \textbf{37}, 621 (1969).

\bibitem{Brewin}L. Brewin, Class. Quant. Grav. \textbf{26}, 175017
(2009).

\bibitem{Kunzle}H. P. Künzle, J. Math. Phys. \textbf{13}, 739 (1972).

\bibitem{TodFelice}K. P. Tod, F. de Felice, M. Calvani, Il Nuovo
Cim. \textbf{34}, 365 (1976).

\bibitem{ShibataPRD96}Y. Mino, M. Shibata, T. Tanaka, Phys. Rev.
D \textbf{53}, 622 (1996).

\bibitem{BelDecomp}L. Bel, C. R. Acad. Sci. Paris \textbf{246}, 3015
(1958); \emph{idem}, Annales de l' I. H. P. \textbf{17}, 37 (1961).

\bibitem{SemerakPetrovTypes}O. Semerák, M. \v{S}rámek, Phys. Rev.
D \textbf{92}, 064032 (2015).

\bibitem{Porto2006}R. A. Porto, Phys. Rev. D \textbf{73}, 104031
(2006).

\bibitem{SteinhoffHergtSchaeffer}S. Hergt, J. Steinhoff, G. Schäfer,
Ann. Phys. \textbf{327}, 1494 (2012).

\bibitem{Schiff}L. Schiff, Proc. N. A. S. \textbf{46}, 871 (1960).\end{thebibliography}
\end{document}